%% file: main_PRL.tex
\documentclass[%
 reprint,
superscriptaddress,
 amsmath,amssymb,
 aps,
 prl,
]{revtex4-2}

\usepackage{graphicx}
\usepackage{dcolumn}
\usepackage{bm}
\usepackage{hyperref}
\usepackage{braket}
\usepackage{xcolor}
\usepackage{subfigure}
\usepackage{mathtools}
\usepackage{dsfont}

\usepackage{physics}
\usepackage{soul}

\usepackage{lipsum} 

\begin{document}


\title{Optimal Quantum Information Transmission Under a Continuous-Variable Erasure Channel}

\author{Adam Taylor}
 \email{adam.taylor18@imperial.ac.uk}
 \affiliation{Blackett Laboratory, Imperial College London, London SW7 2AZ, United Kingdom}
\author{Michael Hanks}
\affiliation{Blackett Laboratory, Imperial College London, London SW7 2AZ, United Kingdom}
\author{Hyukjoon Kwon}
\affiliation{Korea Institute for Advanced Study, Seoul 02455, South Korea}
\author{M. S. Kim}
\affiliation{Blackett Laboratory, Imperial College London, London SW7 2AZ, United Kingdom}
\affiliation{Korea Institute for Advanced Study, Seoul 02455, South Korea}

\begin{abstract}
    Quantum capacity gives the fundamental limit of information transmission through a channel. 
    However, evaluating the quantum capacities of a continuous-variable bosonic quantum channel, as well as finding an optimal code to achieve the optimal information transmission rate, is in general challenging.
    In this work, we derive the the quantum capacity and entanglement-assisted quantum capacity of the bosonic continuous-variable erasure channel when subject to energy constraints. We then construct random codes based on scrambling information within the typical subspace of the encoding state and prove that these codes are asymptotically optimal up to a constant gap. Finally, using our random coding scheme we design a bosonic variation of the Hayden-Preskill protocol and find that information recovery depends on the ratio between the input and output modes. This is in contrast with the conventional discrete-variable scenario which requires only a fixed number of additional output qudits.
\end{abstract}

\maketitle

Determining the maximum rate at which information can be reliably communicated through many uses of a noisy quantum channel has been a central problem in quantum information theory since the fields foundation. This is given by the capacity of the channel and, unlike in classical information theory~\cite{shannonGOAT}, quantum channels admit an entire family of distinct capacities depending on the task and available resources~\cite{laszlo2018survey, holevo2020quantum, wilde2013quantum, waltrous2018theory}. These capacities play a fundamental role in quantum communications and networks~\cite{wehner2018quantumInternet, hu2023progress, azuma2023quantumRepeaters}, quantum cryptography~\cite{RevModPhys.74.145, pirandola2020advances, RevModPhys.92.025002}, quantum error correcting codes~\cite{shor1995scheme, aharonov1997fault, knill1998resilient, RevModPhys.87.307, brun2013error} and distributed quantum computing~\cite{caleffi2020distributed, barral2025review}. Central to the derivation of capacities are random coding arguments, which are often realised by implementing randomly selected unitaries from a unitarily invariant (i.e., Haar) measure~\cite{wilde2013quantum, waltrous2018theory, shor2002quantum, hayden2008decoupling, PhysRevA.75.062315}. This connects coding directly to notions of information scrambling and quantum chaos, where Haar random unitaries represent maximally chaotic dynamics~\cite{PRXQuantum.5.010201, roberts2017chaos, bhattacharyya2022towards, mele2024introduction}. 
Decoding random codes also corresponds with recovering information emitted from a black hole via Hawking radiation~\cite{PhysRevLett.71.1291, PhysRevLett.71.3743, hayden2007black, yoshida2017efficient, yoshida2022recovery}.

While quantum information theory for discrete-variable (DV) systems is highly developed \cite{wilde2013quantum, waltrous2018theory, nielsen2010quantum}, the corresponding theory for bosonic continuous-variable (CV) systems comparatively underexplored~\cite{RevModPhys.77.513, RevModPhys.84.621, serafini2017quantum}.
The main difficulty lies in the infinite-dimensional Hilbert space which can lead to non-physical features, such as infinite-energy states or diverging capacities. It also raises practical and conceptual difficulties around random unitaries and scrambling beyond the Gaussian regime, where notions such as Haar randomness or unitary $t$-designs are poorly understood~\cite{mollabashi2025information, bhattacharyya2022towards, PhysRevX.14.011013, PhysRevA.99.062334}. 
By analogy with DV systems, we would expect random codes to be optimal in CV systems. Despite this, there are no known examples of random codes being optimal for non-Gaussian noise channels, because no suitable ensemble of random unitaries has been constructed.
As experimental control over CV degrees of freedom continues to improve, it is essential to understand if there are any fundamental differences between or potential advantages over DV methods for quantum information processing.

One of the simplest noise models is the erasure channel, where the input state is either perfectly transmitted or replaced by a fixed output state in an orthogonal Hilbert space (``erased''). 
In the erased branch, all information about the input state is lost to the environment.
The quantum capacity of a $d$-dimensional erasure channel with erasure probability $p$ is given by~\cite{PhysRevLett.78.3217} 
\begin{equation}
\label{eq:DV_capacity_main}
    \mathcal{Q}_{\textrm{stan}}^{\textrm{(DV)}} = \max\{(1-2p) \log(d), \,0\},
\end{equation}
while the entanglement-assisted (EA) quantum capacity, where there exists a shared entanglement-resource between sender and receiver, is given by~\cite{PhysRevLett.83.3081}
\begin{equation}
\label{eq:DV_EA_capacity_main}
    \mathcal{Q}^{\textrm{(DV)}}_{\textrm{EA}} = (1 - p) \log(d).
\end{equation}
In CV systems, the channel capacities can diverge due to unphysical infinite-energy states. Therefore, a more suitable quantity to consider is the energy-constrained capacity, where constraints are placed on the energy of the state entering the channels~\cite{holevo1997quantum,holevo2003entanglement, PhysRevLett.98.130501, wilde2018energy}. For the CV erasure channel, the analogous capacity results remain unresolved, despite partial progress~\cite{Zhong2023information}.

Computing this channel capacity is now of both theoretical importance (i.e., helping CV quantum Shannon theory catch up with the corresponding DV results) and practical relevance in free-space optical communication, where particular trajectories of turbulence, beam blockages and beam misalignment lead to erasure-like noise~\cite{RevModPhys.74.145, vasylyev2016atmospheric, vasylyev2017free, liorni2019satellite, raya2024satellite}.
Since CV encodings in quantum optics offer compelling advantages over DV methods based on single photons (such as hardware efficiency, deterministic entanglement generation / transmission and compatibility with existing telecommunications networks), knowing the CV capacity will be important when designing future large scale quantum networks.

Existing works~\cite{PhysRevLett.101.130503, lassen2010continuous, villasenor2022three} have studied explicit error-correction schemes for the \emph{approximate} erasure channel, which maps the input state onto a \emph{non-orthogonal} vacuum state. The capacity of the CV erasure channel with proper orthogonality between the transmitted and erased branches was recently explored in Ref.~\cite{Zhong2023information}, where the CV erasure channel acting on $\rho^{\uparrow} := \rho_{\textrm{CV mode}} \otimes \ketbra{\uparrow}_{\textrm{classical flag}}$ can be defined as
\begin{equation}
\label{eq:CV_erasure_channel_def}
	\Lambda_p [\rho^{\uparrow}] \coloneq (1 - p) \rho^{\uparrow} + p \ketbra{0}^{\downarrow}.
\end{equation}
Here, $\uparrow$ ($\downarrow$) is the classical flag associated with the transmitted (erased) branch. By considering a random code for this channel that scrambles information within a typical subspace, it was argued that the resulting rate coincides with the energy-constrained quantum capacity of the channel.
However, since the typical subspace in the proposed approach~\cite{Zhong2023information} appears to retain classical correlations, contrary to the original claim, it remains unclear whether such a result has solid theoretical backing.

In this Letter, we directly compute quantum and entanglement-assisted quantum capacities of the erasure channel subject to energy constraints.
We then construct two random codes, one standard and one entanglement-assisted, for entanglement-transmission through the erasure channel, based on a different typical subspace than was considered in Ref.~\cite{Zhong2023information}.
We prove our codes are asymptotically optimal in the sense that the rates approach the energy-constrained capacity up to a constant gap in the many mode, high-energy limit.
Finally, our random coding scheme can be viewed as a CV analogue of the black hole toy models used to study the information paradox in the Hayden-Preskill protocol~\cite{PhysRevLett.71.1291,PhysRevLett.71.3743, hayden2007black, yoshida2017efficient, yoshida2022recovery}. In this CV setting, we recover the DV result exactly in the unassisted case \cite{PhysRevLett.71.1291, PhysRevLett.71.3743}. In the entanglement-assisted case, we find that recovery is possible when accessible output comprises of a constant \emph{fraction} more modes than the input, rather than the constant \emph{number} more in the DV case~\cite{hayden2007black}. We explain how this modified scaling arises from our typical subspace scrambling.

\textit{Energy-constrained quantum capacities.---}
The quantum capacity of the CV erasure channel subject to a uniform-energy constraint, where the average photon number across all the modes entering the erasure channel is upper bounded by $\bar{n}$, is equal to
\begin{equation}
\label{eq:capacity_main}
    \mathcal{Q}_{\textrm{stan}}(\Lambda_p; \bar{n}) = \max\{(1-2p) H_{\textrm{therm}}(\bar{n}), \, 0\}.
\end{equation}
We prove this and give details on the constraint in the End Matter.
It follows from~\cite{wilde2018energy} that this is also the private communication capacity subject to a uniform energy-constraint, and the entanglement-transmission and secret-key transmission capacities subject to a weaker \emph{average} energy-constraint.

We also derive the entanglement-assisted quantum capacity of the CV erasure subject to the uniform-energy constraint, 
\begin{equation}
\label{eq:EA_capacity_main}
    \mathcal{Q}_{\textrm{EA}}(\Lambda_p; \bar{n}) = (1 - p) H_{\textrm{therm}}(\bar{n}).
\end{equation}
The CV capacities in Eq.~\eqref{eq:capacity_main} and Eq.~\eqref{eq:EA_capacity_main} have the same form as the DV case in Eq.~\eqref{eq:DV_capacity_main} and Eq.~\eqref{eq:DV_EA_capacity_main} but with the energy-constraint now determining the scaling rather than the local dimension. In the limit $\bar{n} \rightarrow \infty$, the unconstrained CV capacities diverge as $\log(\bar{n})$.


\begin{figure}
    \centering
    \includegraphics[width=\linewidth]{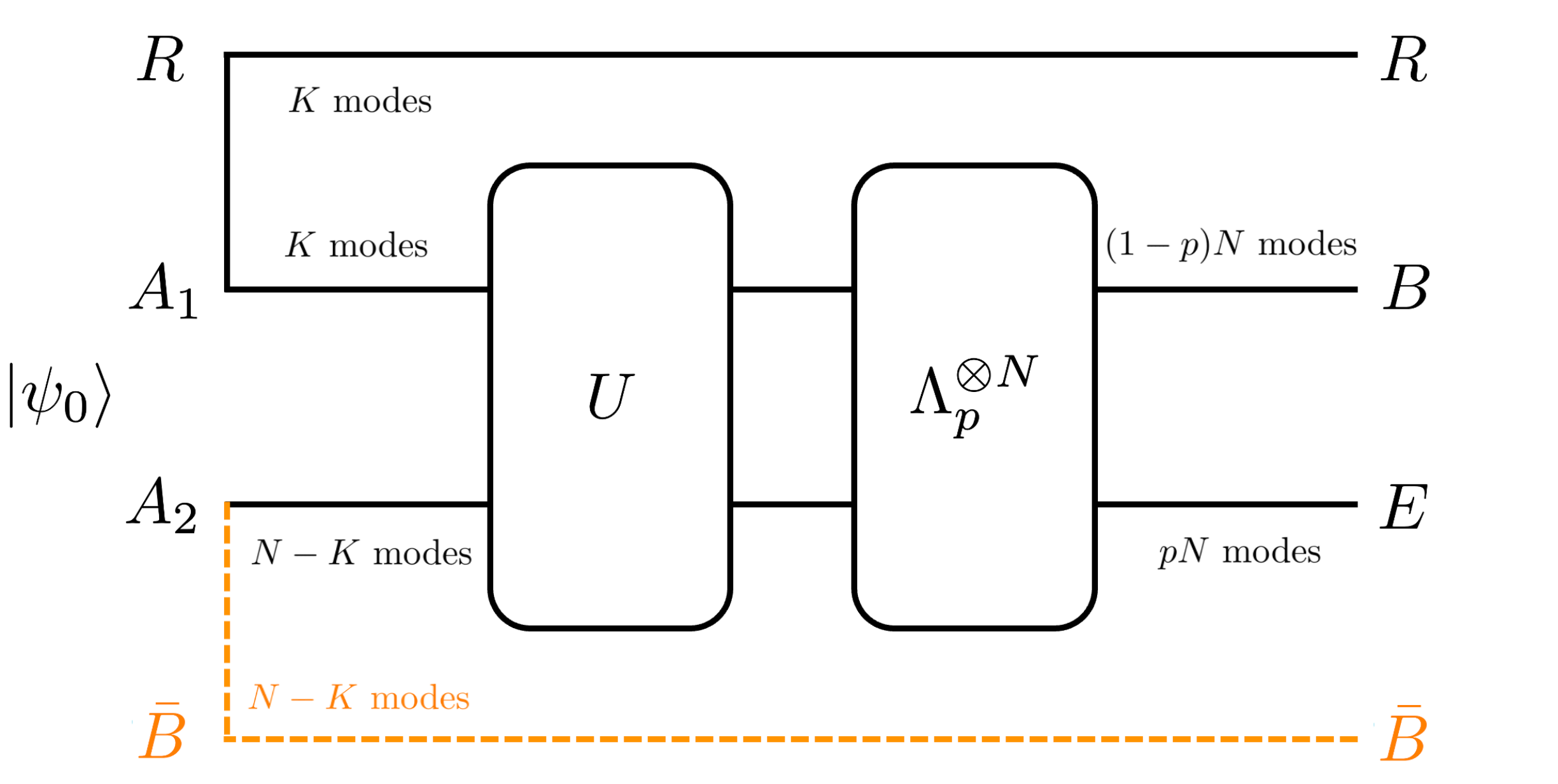}
    \caption{Random coding scheme for entanglement transmission. Alice begins with $K$ copies of TMSVs, keeping one mode per pair in the reference $R$ and placing the other in $A_1$. She introduces an additional $N-K$ modes in $A_2$ and applies a random unitary $U$ on $A=A_1A_2$. Each mode is then sent through the single-mode CV erasure channel $\Lambda_p$. Bob uses the classical flags to retain the $\approx(1-p)N$ unerased modes ($B$), while the erased modes go to the environment ($E$). In the standard scheme, $A_2$ consists of random coherent states, whereas in the entanglement-assisted scheme (dashed orange) Alice and Bob instead share $N-K$ copies of TMSVs across $A_2\bar{B}$, with Bob accessing $\bar{B}$ during decoding. When considering the black hole toy model, $A_2$ corresponds with the initial black hole, $A_1$ is the information Alice throws in, $U$ is the scrambling dynamics of the black hole, $B$ is the emitted Hawking radiation and $E$ is the residual black hole.}
    \label{fig:random_code_diagram}
\end{figure}

\begin{figure*}
  \includegraphics[width = 0.9\textwidth]{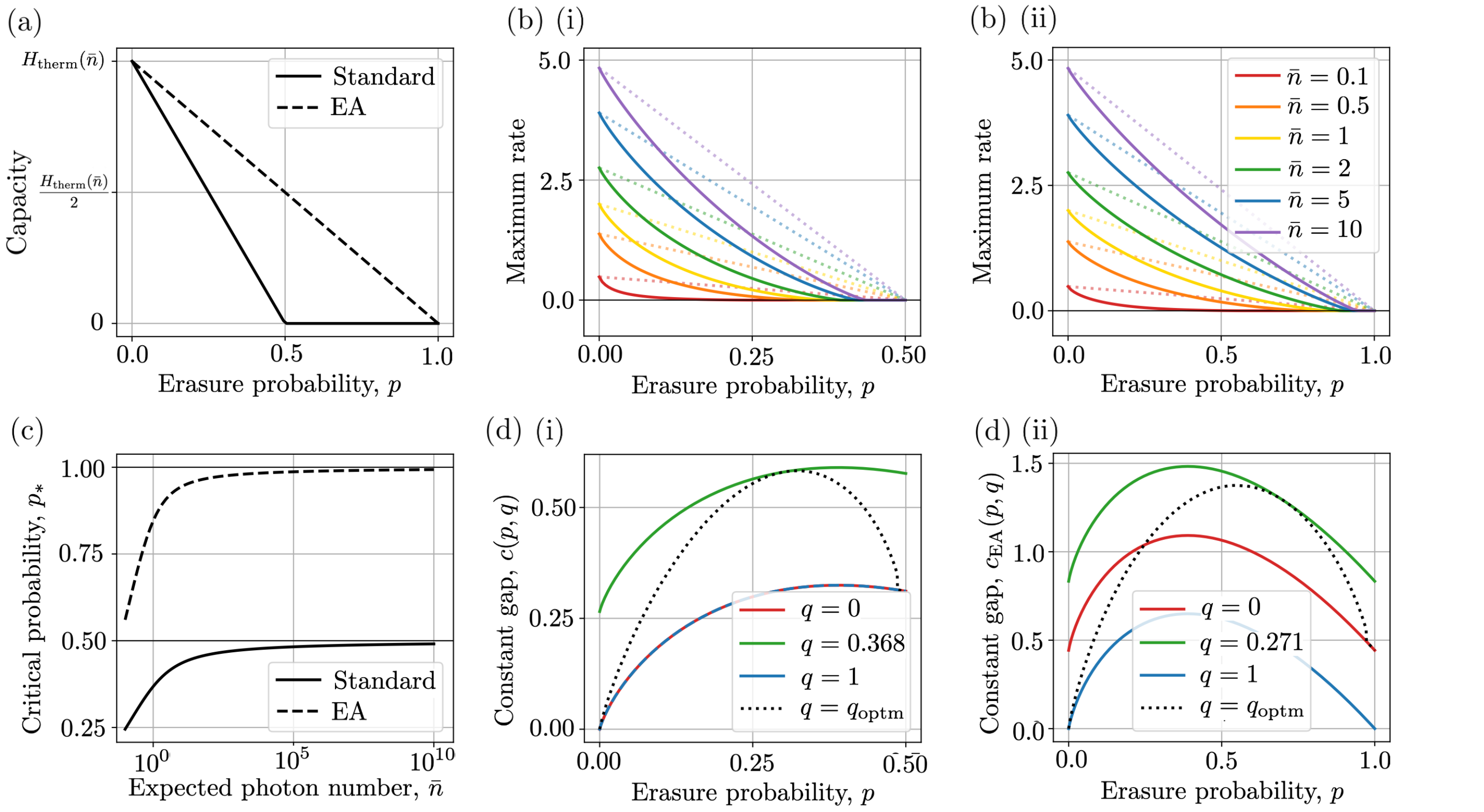}
  \caption{(a) The quantum capacities as a function of erasure probability. (b) The maximum achievable rate of the random codes in the (i) standard and (ii) entanglement-assisted cases for up to $\bar{n} = 10$. The dotted curves are the associated capacities.
  (c) The maximum erasure probability before no information transmission is possible, $p_*$ against the energy constraint $\bar{n}$ for our codes. They asymptotically approach $p_*=0.5$ and $p_* = 1$ in the standard and entanglement-assisted cases. (d) The constant functions (i) $c(p, q)$ and (ii) $c_{\textrm{EA}}(p, q)$ as a function of $p$ for different values of $q$. The $q = 0.368$ and $q = 0.271$ curves in (i) and (ii) are the  values of $q$ for which $c$ is maximised, and $q_{\textrm{optm}}$ curve is the value of $q$ that maximises the rate (i.e., the black dotted curves in (d)(i) and (d)(ii) correspond to the energy-independent gap in the plots (b)(i) and (b)(ii) respectively).}
  \label{fig:capacity_rate_constant_pcrit_plots}
\end{figure*}

\textit{Asymptotically optimal random codes.---}We now construct two random quantum codes for entanglement generation, one standard and one entanglement-assisted, whose rates achieve the corresponding energy-constrained quantum capacities (see Fig.~\ref{fig:random_code_diagram} for a diagram).  Alice initially holds $K$ copies of entangled two-mode squeezed vacuum states (TMSVs) that she wants to share with Bob \footnote{Other types of entangled states could be considered, leading to different codes based on different typical subspaces. But TMSV states maximise entanglement when subject to an energy constraint so they are our focus.}.
From each of the TMSVs, she keeps one mode in a reference space $R$ and wants to transmit the other (on Hilbert space $A_1$) through an erasure channel to Bob. She encodes the $K$ modes on $A_1$ with an additional $N-K$ modes on $A_2$ by implementing a random unitary on $A = A_1 A_2$. The modes from $A$ are then each independently sent through a single-mode CV erasure channel $\Lambda_p$ (a classical flag is attached as necessary). Bob performs a syndrome measurement on the classical flags, confirming whether he should keep the mode (in a Hilbert space labelled $B$) or discard it; the environment has access to the information originally contained in the erased modes, labelled $E$.
In the entanglement-assisted case, Alice and Bob initially share $N-K$ TMSVs on the Hilbert space $A_2 \bar{B}$. Alice uses her half during the encoding procedure, while Bob can use his half on $\bar{B}$ when decoding.

In DV systems, random codes are optimal in the large qudit number limit, where the approach is to scramble the qudits using a Haar random unitary (or even random low depth circuits~\cite{PhysRevX.11.031066}).
However, attempting to exactly extend the notion of Haar random unitaries to CV systems proves impossible; a Haar random unitary acting on an infinite Hilbert space would necessarily turn a finite-energy state into an infinite-energy state~\cite{PhysRevX.14.011013, PhysRevA.99.062334}.
In Ref.~\cite{Zhong2023information}, this problem is avoided
by considering the typical subspace of a state. For $N$ copies of a state, $\rho^{\otimes N}$, in the large $N$ limit the entire state concentrates around a typical subspace $T_\delta^N(\rho)$ with projector $\Pi^{\delta}$ such that for all $\varepsilon > 0$ and suitably large $N$ then $\tr[\Pi^{\delta} \rho^{\otimes N}] \ge 1 - \varepsilon$~\cite{wilde2013quantum, waltrous2018theory}.
The typical subspace is finite-dimensional and hence one can define the Haar ensemble of unitaries within this typical subspace. Crucially, this ensemble of unitaries does not lead to the non-physical phenomena and can also be meaningfully averaged over with Weingarten calculus~\cite{weingarten1978asymptotic, collins2006integration, collins2022weingarten}.
While Ref.~\cite{Zhong2023information} argued that passive linear optical networks induce a typical subspace on average, our analytical calculations yield a different result, showing the gap between the two results from residual classical correlations (see Supplemental Material).
To address this issue, we consider a new class of input states and explicitly show that they form a typical subspace.

We first consider the entanglement-assisted case because the reduced state on $A$ automatically has the correct tensor product structure. The initial state on $RA\bar{B}$ is an $N$-fold tensor product of TMSV states, $\ket{z}^{\otimes K}_{R A_1} \otimes \ket{z}^{\otimes (N-K)}_{A_2 \bar{B}}$, where $\ket{z}_{1,2} \coloneq (1 - |z|^2)^{\frac{1}{2}} \sum_{n = 0}^\infty z^n  \ket{n, n}_{1,2}$ for $z \in \mathbb{C}$ is a single TMSV state with expected photon number $\bar{n}_z = |z|^2 / (1 - |z|^2)$. Tracing out one of the modes from $\ket{z}_{1,2}$ yields a thermal state $\rho_z \coloneq (1-|z|^2) \sum_{n = 0}^\infty |z|^{2n} \ketbra{n}$, and hence the initial reduced state on $A$ is an $N$-fold tensor product of single-mode thermal states, $\tr_{R\bar{B}} [ (\ketbra{z})^{\otimes N}_{RA\bar{B}}] = (\rho_z)^{\otimes N}_{A}$.

In the standard case, we require the state on $A_2$ to be pure for Bob to be able to decode. The key insight is that thermal states can be written in the coherent state basis as $\int \textrm{d}p(\gamma) \, \ketbra{\gamma}$ where $\ket{\gamma} = e^{-\frac{1}{2} |\gamma|^2} \sum_{n=0}^\infty \frac{\gamma^n}{\sqrt{n!}} \ket{n}$ for $\gamma\in\mathbb{C}$ is a coherent state, and the probability density function reads $\textrm{d}p(\gamma) = e^{-\frac{1}{\bar{n}_z}|\gamma|^2} \textrm{d}^2\gamma / \pi \bar{n}_z$. To construct our typical subspace, Alice initialises her state on $A_2$ in an $(N-K)$-mode coherent state, $\ket{\bm{\gamma}} = \ket{\gamma_1, \dots, \gamma_{N-K}}$, where $\gamma_i$ in each mode is randomly sampled from $\textrm{d}p(\gamma_i)$. Therefore, the \emph{average} state on $A$ is an $N$-fold tensor product of thermal states, $\mathbb{E}_{\bm{\gamma}} [(\rho_z)^{\otimes K}_{A_1} \otimes \ketbra{\bm{\gamma}}_{A_2}] = (\rho_z)^{\otimes N}_{A}$.

The (average) initial state on $A$ is $\rho_z^{\, \otimes N}$, and hence in the large-$N$ limit we can meaningfully define the finite-dimensional typical subspace in which Haar random unitaries act, i.e., the typical subspace of the thermal state, $T_\delta^N(\rho_z)$.
Error correction is possible if and only if the environment and reference are approximately decoupled; i.e., their reduced joint density matrix satisfies $\rho_{RE} \approx \rho_R \otimes \rho_E$~\cite{PhysRevA.54.2629, schumacher2002approximate}. We next calculate when decoupling takes place, because this implies that Bob can decode the noisy output to recover $K$ pairs of TMSV states shared with Alice.

\textit{Energy constrained entanglement-transmission rate.---}In both the standard and entanglement-assisted cases, we upper bound the average trace distance to the average (and decoupled) state, $\mathbb{E}_{U, \psi_0}[\rho_{RE}(U, \psi_0)] = \rho_R \otimes \bar{\rho}_E$:
\begin{align}
\begin{split}
    \mathbb{E}_{U, \psi_0} & || \rho_{RE}(U, \psi_0) - \rho_R \otimes \bar{\rho}_E  ||_1 \\
    & \lesssim \textrm{poly}(N) \, 2^{-\frac{1}{2}N \, \xi_{\ell}(\bar{n}, p, q)}.
\end{split}
\end{align}
Here, $\ell \in \{\textrm{stan}, \textrm{EA}\}$ and $\psi_0$ label if we are in the standard or entanglement-assisted case, $q = K/N$ is the fraction of total modes Alice wishes to communicate to Bob \footnote{When decoding is possible (i.e., decoupling has taken place), this implies that the rate of entanglement-transmission of the code is given by $Q_{\textrm{rate}} = q H_{\textrm{therm}}(\bar{n})$.}, and $\xi_{\ell}$ is defined in the Supplemental Material.

In the large $N$ limit, if $\xi > 0$ then the set of possible states on $RE$ is exponentially concentrated towards the average decoupled state $\rho_R \otimes \bar{\rho}_E$, (or equivalently, the variance in state space is exponentially small). This implies that for a single-shot choice of $U$ (and of $\bm{\gamma}$ in the standard case), the output state $\rho_{RE}(U, \psi_0)$ is approximately decoupled. Therefore, if $\xi > 0$ then there exists a decoding scheme that Bob can use to recover $K$ TMSV states shared with Alice with high fidelity.

In the high-energy limit, where $\bar{n} \gg 1$, the exponent becomes, up to an energy-independent gap, the difference between the capacity and entanglement-transmission rate
\begin{align}
\begin{split}
\label{eq:result_high_energy_limit}
    \xi_{\ell}(\bar{n}, p, q) = \mathcal{Q}_{\ell} (\Lambda_p; \bar{n}) - Q_{\textrm{rate}}(\bar{n}, q) \\
    - c_{\ell}(p, q) + \textrm{O}(\bar{n}^{-1})
\end{split}
\end{align}
where $\ell \in \{\textrm{stan}, \textrm{EA}\}$ labels which case is under consideration. Here, $c_{\ell}(p, q)$ are energy-independent functions defined in the Supplemental Material, $Q_{\textrm{rate}}\coloneq q H_{\textrm{therm}}(\bar{n})$ is the entanglement-transmission rate of the code and $\mathcal{Q}_{\textrm{stan}}$ ($ \mathcal{Q}_{\textrm{EA}}$) is the energy-constrained (and entanglement-assisted) quantum capacity. Therefore, in the large $N$, high-energy limit, our random code rates can achieve the associated quantum capacities up to a constant gap. As a result, both the standard and entanglement-assisted codes are asymptotically optimal because $\lim_{N, \bar{n} \rightarrow \infty} \frac{Q_{\textrm{max rate}}}{\mathcal{Q}} = 1$.

Away from the high-energy limit, the exponent $\xi$ decreases nonlinearly with $p$ in both cases.
In Fig.~\ref{fig:capacity_rate_constant_pcrit_plots}, we plot this exponent as a function of erasure probability, alongside the energy-independent gaps $c_{\ell}$ for $\ell\in\{\textrm{stan}, \textrm{EA}\}$. As $p$ grows, we see the rate decay away from the true capacity with the magnitude of the gap controlled by $c_\ell(p, q_{\textrm{opt}})$. Since the gap is energy-independent, its impact is most significant when the energy constraint is smaller.
We also plot critical probability $p_*$ as a function of $\bar{n}$, where $p_*$ is defined as the largest erasure probability for which one can still communicate some quantum information through the channel (i.e., infimum of the set $\{p : \lim_{q \rightarrow 0} \xi(\bar{n}, p, q) \le 0\}$). These asymptotically approach the DV values of $0.5$ and $1$ in the standard and entanglement-assisted cases respectively.


Interestingly, we find that dimensions within the typical subspace are submultiplicative in the sense that $d_{XY} \le d_X d_Y$.
This means that there are no cancellations in the derivation (e.g.; $\frac{d_{X} d_{Y}}{d_{XY}^2} \ne \frac{1}{d_{XY}}$) and hence we need to introduce additional bounding techniques that are not present in the analogous DV derivation. These new techniques lead to the energy-independent gap $c_{\ell}$ that prevent capacity achieving rates when away from the high-energy limit. 
It is difficult to assess if this is consequence of our bound night being tight, or a fundamental limitation of typical subspace scrambling in bosonic systems.

\textit{Hayden-Preskill protocol for the CV erasure channel.---}The random encoding discussed above corresponds to a CV analogue of the black hole model used in the Hayden-Preskill protocol, where Alice throws information into a black hole and Bob attempts to recover it from the emitted Hawking radiation~\cite{hayden2007black, yoshida2017efficient, yoshida2022recovery}. Recent work on the Hayden-Preskill protocol has introduced conserved quantities~\cite{Nakata2023blackholesasclouded}, Clifford dynamics~\cite{yoshida2022recovery}, Hamiltonian dynamics~\cite{PhysRevResearch.6.L022021}, and chaotic and integrable unitaries~\cite{Rampp2024haydenpreskill}, whereas we extend this black hole toy model to bosonic CV modes.

In our case, from Fig.~\ref{fig:random_code_diagram} we have $A_1$ representing the infalling information, $A_2$ the pre-existing black hole, $U$ the scrambling dynamics of the black hole, $B$ the emitted Hawking radiation that Bob has access to and $E$ the residual black hole which holds the inaccessible modes. We imagine modes being emitted as Hawking radiation one at a time and therefore the initial erasure probability (i.e., fraction of modes still within the black hole) is unity. As the black hole evaporates, the erasure probability decreases to zero as all the hidden information is emitted as Hawking radiation. Alice initially deposits $K = qN$ modes in the black hole. If Bob collects all the emitted Hawking radiation, then he has access to $\tilde{q}N$ modes for $\tilde{q} = 1 - p$, while the residual black hole contains $pN$ modes. How much Hawking radiation does Bob need to collect to decode Alice's infallen information?

In the high-energy limit, Eq.~\eqref{eq:result_high_energy_limit} implies that Bob can recover Alice's information when his fraction of the output mode satisfies 
\begin{equation}
    \tilde{q} > 
    \begin{cases}
        \frac{1}{2} + \frac{1}{2}q + \varepsilon_{\textrm{stan}} & \textrm{standard case} \\
        q + \varepsilon_{\textrm{EA}} & \textrm{entanglement-assisted case}
    \end{cases}
\end{equation}
where $\varepsilon_\ell = \frac{c_\ell}{H_{\textrm{therm}}(n)} + \textrm{O}(\frac{1}{\bar{n} \, H_{\textrm{therm}}(\bar{n})})$ for $\ell\in\{\textrm{stan}, \textrm{EA}\}$. 
In the standard case, Bob requires the black hole to evaporate by $>50\%$ before he can recover any information, reproducing the DV result from Page~\cite{PhysRevLett.71.1291, PhysRevLett.71.3743}. The entanglement-assisted setting is a CV analogue of the Hayden-Preskill protocol~\cite{hayden2007black} where we find Bob can recover Alice's information if his share exceeds hers by a small, black hole dependent correction. 
This resembles the original Hayden-Preskill protocol, but with a key difference: recovery now depends on a fractional gap between the output and input, rather than a fixed number of additional qudits.
This change in scaling compared with the DV case originates from the submultiplicative dimensions within the typical subspace. When analysing the case where the input and output sizes, of $A_1$ and $B$ respectively, are held constant with respect to $N$, the complexity introduced by the submultiplicitivity renders our bound ineffective. It remains unclear whether this is a limitation of the bound itself (i.e., obscured due to looseness), or whether typical subspace scrambling genuinely requires the output to scale with the black hole size for recovery. Nonetheless, this shows that it is possible to extend the Hayden-Preskill protocol to bosonic thermal states in infinite-dimensional Hilbert spaces, which are exactly the sort of photonic states that are emitted as Hawking radiation.

\textit{Remarks.---}We have derived the energy-constrained quantum capacity and the energy-constrained entanglement-assisted quantum capacity of the CV erasure channel, finding that they mirror the DV result. This enables direct comparisons between DV and CV platforms for free-space optical communication, despite differing levels of technological maturity.
For example, the DV photonic Bell state $\propto \ket{HV} + \ket{VH}$, where $\ket{H}$ and $\ket{V}$ are respectively horizontally and vertically polarised single photons, has a local dimension $d=2$ and expected photon number $\bar{n} = 1$. A CV system of the same energy has a capacity equal to $\mathcal{Q}(\bar{n} = 1) = 2 \mathcal{Q}^{(\textrm{DV})}_{d=2}$, indicating that a similarly constrained CV architecture could transmit up to $100\%$ more entanglement in free-space optics per erasure channel use.
Moreover, squeezed vacuum states with up to $\bar{n} \approx 7.5 $ have been demonstrated~\cite{vahlbruch2016detection}; at this energy, the capacity reads $\mathcal{Q}(\bar{n} = 7.5) \approx  4.4 \, \mathcal{Q}^{(\textrm{DV})}_{d=2}$, highlighting the hardware efficiency of infinite dimensional systems and their potential for establishing high rate, large scale quantum networks.

We then constructed random codes that approach these capacities up to an energy-independent gap, a result that proves that non-Gaussian random codes can be optimal for non-Gaussian CV channels.
Our derivation also highlights the subtle role of typical subspace scrambling in CV systems, where the submultiplicity of dimensions prevents this notion of CV Haar random unitaries from being exactly analogously to their DV counterparts.
Finally, we showed that our random codes implements a CV version of the Hayden-Preskill protocol, thus linking scrambling to recoverability in CV systems and extending these toy models to infinite-dimensional thermal states.

Our results open several avenues for future work. Investigating connections between typical subspace scrambling, CV-regularised unitary $k$-designs~\cite{PhysRevX.14.011013}, and OTOC-based diagnostics~\cite{PhysRevA.99.062334, bhattacharyya2022towards} could deepen our understanding of scrambling, chaos, and random unitaries in infinite-dimensional spaces. Our random code may also help analyse non-degradable bosonic channels lacking known capacities (e.g., thermal attenuator~\cite{rosati2018narrow}, loss-dephasing~\cite{Leviant2022quantumcapacity}), potentially yielding tighter bounds. Experimentally realising the erasure channel with a classical flag would allow direct validation of its noise model and support the design of new error-correcting codes, benchmarked against our capacity results. Such advances would further highlight CV architectures as hardware-efficient platforms for high-rate quantum communication under realistic energy constraints. 

\begin{acknowledgments}
    M.S.K. acknowledges funding from the UK EPSRC through EP/Z53318X/1, EP/W032643/1 and EP/Y004752/1, the KIST through the Open Innovation fund. M.S.K. and H.K. are supported by the National Research Foundation of Korea grant (No. RS-2024-00413957) funded by the Korean government (MSIT). H.K. is supported by the KIAS Individual Grant No. CG085302 at Korea Institute for Advanced Study and Institute of Information \& Communications Technology Planning \& Evaluation (IITP) grant (No. 2022-0-00463) funded by the Korea government (MSIT).
\end{acknowledgments}

\bibliography{biblio}

\onecolumngrid
\begin{appendix}
\onecolumngrid
\section*{End Matter}
\twocolumngrid
\twocolumngrid

\section*{Capacity calculations}
We consider a setup where Alice encodes some information into $N$ CV modes via an encoding procedure $\mathcal{E}_N$ on some initial state $\rho_0$. She then transmits it through $N$ independent uses of the single-mode CV erasure channel, $\Lambda_p$, to Bob, who attempts to decode the noisy information. 
The energy-constraints we consider apply to the encoded state input on the erasure channel, namely $\mathcal{E}_N[\rho_0]$. In bosonic CV systems, we take the single-mode energy-observable to be the excitation number operator $\hat{n} = \hat{a}^\dag \hat{a}$. We then define the $N$-mode energy functional as
\begin{equation}
    E_N[\bullet] \coloneq \frac{1}{N} \sum_{i = 1}^N \tr\big[\hat{n}_i \, \bullet \big]
\end{equation}
where $\hat{n}_i \coloneq \mathds{1}^{\otimes i - 1} \otimes \hat{n} \otimes \mathds{1}^{N - i}$ is the number operator on the $i^{\textrm{th}}$ modes.
The \emph{uniform energy-constraint} then requires that the encoded state satisfies 
\begin{equation}
    E_N[ \mathcal{E}_N[\rho_0]] \le \bar{n}
\end{equation}
for some constraint parameter $\bar{n}$, i.e., the average photon number across $N$ modes is upper bounded by $\bar{n}$.

The expression for the quantum capacity of a degradable channel $\mathcal{N}$ subject to the uniform energy-constraint $E_N[\mathcal{E}_N[\rho_0]] \le \bar{n}$ is given by~\cite{wilde2018energy}
\begin{equation}
\label{eq:capacity_expression}
    \mathcal{Q}(\mathcal{N}; \bar{n}) = \sup_{\rho : \tr[\hat{n} \rho] \le \bar{n}} I_c(\rho; \mathcal{N}).
\end{equation}
Here, $\rho$ is a single-mode CV state, $I_c(\rho; \mathcal{N}) \coloneq H(\mathcal{N}[\rho]) - H(\bar{\mathcal{N}} [\rho])$ is the coherent information~\cite{PhysRevA.54.2629}, $\bar{\mathcal{N}}$ is a complementary channel to $\mathcal{N}$ and $H(\bullet) \coloneq -\tr[\bullet \log(\bullet)]$ is the von-Neumann entropy.
This expression is also the capacity for private-key distribution with a uniform energy-constraint, and the capacities of entanglement-transmission and secret-transmission subject to the weaker \emph{average} energy-constraint (see Ref.~\cite{wilde2018energy} for details).

To maximise Eq.~\eqref{eq:capacity_expression}, we first note that the CV erasure channel commutes with phase rotations $\mathcal{U}_{\theta} [\bullet] \coloneq e^{-i\theta \hat{n}} \bullet e^{i\theta \hat{n}}$, which implies that $I_c(\mathcal{U}_\theta [\rho]; \Lambda_p) = I_c(\rho; \Lambda_p)$. Second, we note that the coherent information with respect to any state $\rho$ is concave for degradable channels, and hence $\int_0^{2\pi} \frac{\textrm{d} \theta}{2\pi} I_c(\mathcal{U}_\theta[\rho]; \Lambda_p) \le I_c (\int_0^{2\pi} \frac{\textrm{d} \theta}{2\pi} \mathcal{U}_\theta [\rho]; \Lambda_p)$~\cite{yard2008capacity}. It follows that for any $\rho = \sum_{n,m} \rho_{n,m} \ketbra{n}{m}$ and associated $\rho_{\textrm{diag}} = \sum_n \rho_{nn} \ketbra{n}$, the coherent information satisfies
\begin{equation}
    I_c(\rho; \Lambda_p) \le I_c(\rho_{\textrm{diag}}; \Lambda_p).
\end{equation}
Therefore, when searching for the supremum of the coherent information we just need to consider states diagonal in the number basis.

In the number basis, we directly compute the coherent information
\begin{equation}
    I_c(\rho_{\textrm{diag}}; \Lambda_p) = (1 - 2p) H(\rho_{\textrm{diag}}).
\end{equation}
By noting that thermal states maximise entropy given energy constraints~\cite{serafini2017quantum}, it follows that
\begin{equation}
    \sup_{\rho : \tr[\hat{n} \rho] \le \bar{n}} I_c(\rho_{\textrm{diag}}; \Lambda_p) = (1 - 2p) H_{\textrm{therm}}(\bar{n}),
\end{equation}
where $H_{\textrm{therm}}(\bar{n}) = (\bar{n} + 1) \log(\bar{n} + 1) - \bar{n} \log(\bar{n})$ is the entropy of a thermal state with expected photon number $\bar{n}$. Therefore, the uniformly energy-constrained quantum capacity of the CV erasure channel is given by
\begin{equation}
    \mathcal{Q}(\Lambda_p; \bar{n}) = \max \{ (1 - 2p) H_{\textrm{therm}}(\bar{n}), \, 0 \}.
\end{equation}
By choosing a state with unbounded entropy (such as $\rho_\infty \propto \lim_{z \rightarrow 1} \sum_n |z|^{2n} \ketbra{n}$), we can immediately see that the \emph{unconstrained} quantum capacity diverges for $p < 0.5$.

The generic expression for the entanglement-assisted \emph{classical} capacity of the channel $\mathcal{N}$ subject to the uniform energy constraint is~\cite{holevo2003entanglement}
\begin{equation}
\label{eq:entanglement-assisted_classical_capacity_expression}
    \mathcal{C}_{\textrm{EA}}(\mathcal{N}; \bar{n}) = \sup_{\rho : \tr[\hat{n} \rho] \le \bar{n}} I(\rho; \mathcal{N}),
\end{equation}
where $I(\rho; \mathcal{N}) = H(\rho) + I_c(\rho; \mathcal{N})$ is the quantum mutual information. Under the constraint $\tr[\hat{n} \rho] \le \bar{n}$, thermal states simultaneously maximise both the von-Neumann entropy and the coherent information. Therefore, thermal states also maximise the quantum mutual information. We can plug this into Eq.~\eqref{eq:entanglement-assisted_classical_capacity_expression} to give the energy-constrained entanglement-assisted classical capacity of the erasure channel,
\begin{align}
\begin{split}
    \mathcal{C}_{\textrm{EA}}(\Lambda_p; \bar{n}) &= \sup_{\rho : \tr[\hat{n} \rho] \le \bar{n}} \left( H(\rho) + I_c(\rho; \Lambda_p)\right) \\
    &= 2 (1 - p) H_{\textrm{therm}}(\bar{n}).
\end{split}
\end{align}
From the known relationship $\mathcal{Q}_{\textrm{EA}} = \frac{1}{2} \mathcal{C}_{\textrm{EA}}$~\cite{bennett2002entanglement}, the energy-constrained entanglement-assisted \emph{quantum} capacity is 
\begin{equation}
    \mathcal{Q}_{\textrm{EA}}(\Lambda_p; \bar{n}) = (1 - p) H_{\textrm{therm}}(\bar{n}).
\end{equation}

\section*{Comparison with previous work~\cite{Zhong2023information}}

In Ref.~\cite{Zhong2023information}, the authors constructed a CV random code by scrambling information within the typical subspace of a state averaged over the set of passive linear optical networks (PLONs), $\bar{\rho}_A = \mathbb{E}_{\hat{V}} [\hat{V} (\rho_{A_1} \otimes \ketbra{0}_{A_2}) \hat{V}^\dag]$ where $\hat{V}$ is a particular PLON. A key first step in their derivation was the ansatz $\bar{\rho}_A = (\bar{\rho}_1)^{\otimes N}$ where $\bar{\rho}_1 = \sum_{n = 0}^\infty p(n) \ketbra{n}$ is a single mode state in the number basis; this average state would clearly admit a typical subspace in the large $N$ limit, which in turn would allow typical subspace scrambling. However, in the Supplemental Material, we analytically derive $\bar{\rho}_A$ and find that it does not form such a typical subspace due to residual classical correlations that exist between modes. Furthermore, in Ref.~\cite{Zhong2023information} the submultiplicity of the dimensions was not considered despite being a generic feature of marginal spaces within typical subspace: When considering the dimension within the typical subspace of $A_1$ and $A_2$ separately, sequences that look typical on $A_1$ and $A_2$ individually may not look typical on $A_1A_2$. On the other hand, any sequence typical on $A_1A_2$ will also be typical on $A_1$ and $A_2$ individually, implying $d_{A_1} d_{A_2} \ge d_{A_1A_2}$. This has large implications when bounding quantum information theoretic quantities, especially because we find $d_{A_1} d_{A_2}$ is exponentially larger with respect to $N$ than $d_{A_1 A_2}$. These two features -- the residual classical correlations and submultiplicity -- therefore need to be properly considered in order to accurately model CV typical subspace random codes and give well-justified theoretical guarantees.  Our derivation thus shows that typical subspace scrambling is significantly more subtle than previous work realised.

Our approach in this paper was to work directly in the typical subspace determined by the resource state Alice was attempting to share with Bob (i.e., the two-mode squeezed vacuum states). For the entanglement-assisted code the typical subspace arose naturally, while for the standard code the key insight was to sample ancilla states such that on average a typical subspace was formed. If Alice had a different resource state, then a new suitable typical subspace could be chosen based on Alice's state in order to implement the same typical subspace random coding technique.
\end{appendix}

\end{document}



\title{Supplemental Material for ``Optimal Quantum Information Transmission Under a Continuous-Variable Erasure Channel''}

\author{Adam Taylor}
 \email{adam.taylor18@imperial.ac.uk}
 \affiliation{Blackett Laboratory, Imperial College London, London SW7 2AZ, United Kingdom}
\author{Michael Hanks}
\affiliation{Blackett Laboratory, Imperial College London, London SW7 2AZ, United Kingdom}
\author{Hyukjoon Kwon}
\affiliation{Korea Institute for Advanced Study, Seoul 02455, South Korea}
\author{M. S. Kim}
\affiliation{Blackett Laboratory, Imperial College London, London SW7 2AZ, United Kingdom}
\affiliation{Korea Institute for Advanced Study, Seoul 02455, South Korea}


\begin{abstract}
    This document provides Supplemental Material for ``Optimal Information Transmission Under a Continuous-Variable Erasure Channel''. We give the preliminaries necessary to understand our results, present details of all our derivations and expand upon several of the points made in the associated paper. We also calculate the average output state from passive linear optical networks, and discuss how it relates to an ansatz used in \textit{Quantum \textbf{7}, 939 (2023)}.
\end{abstract}

\maketitle

\tableofcontents


\section{Preliminaries}

\subsection{Bosonic continuous variable systems and quantum channels}
In this work, we are concerned with continuous-variable (CV) bosonic modes, with each mode corresponding to an infinite-dimensional Hilbert space $\mathcal{H} \cong L^2(\mathbb{R})$. 
We let $\mathcal{B}(\mathcal{H})$ denote the set of all bounded linear operators acting on $\mathcal{H}$, let $\mathcal{T}(\mathcal{H}) \subset \mathcal{B}(\mathcal{H})$ be the set of all trace-class operators and let $\mathcal{D}(\mathcal{H}) \subset \mathcal{T}(\mathcal{H})$ denote the set of Hermitian positive-semidefinite operators with unit trace, i.e., the set of density operators on $\mathcal{H}$.  We will often label density operators by the Hilbert space that they act on; for example, if $\rho \in \mathcal{D}(\mathcal{H})$, then we may label it as $\rho = \rho_{\mathcal{H}}$. A state $\rho \in \mathcal{D}(\mathcal{H})$ is pure if it can be written as $\rho = \ketbra{\psi}$ for some unit-vector $\ket{\psi} \in \mathcal{H}$.
A quantum channel is a linear, completely positive and trace-preserving (CPTP) map $\mathcal{N}: \mathcal{T}(A) \rightarrow \mathcal{T}(B)$ where $A$ and $B$ are separable Hilbert spaces. It can always be written in the Kraus representation via
\begin{equation}
\label{eq:kraus_representation_definition}
    \mathcal{N} [\bullet] \coloneq \sum_{j} K_j \bullet K_j^\dag .
\end{equation}
where $K_j:A \rightarrow B$ are the Kraus operators.
The trace-preserving condition is expressed as $\sum_j K_j^\dag K_j = \mathds{1}_A$.
For simple unitary evolution, a channel $\mathcal{U}:\mathcal{T}(A) \rightarrow \mathcal{T}(B)$ has only one Kraus term, $\mathcal{U} [\bullet] \coloneq U \bullet U^\dag$ where $U: A \rightarrow B$ is a unitary operator.
Stinespring's dilation theorem allows any CPTP channel to be written as the reduced action of a single-term Kraus sum acting on a larger Hilbert space.
If $V: A \rightarrow BE$ is an isometry satisfying $V^\dag V = \mathds{1}_A$, then the channel $\mathcal{N}$ can be expressed as 
\begin{equation}
\label{eq:Stinespring_dilation_representation_of_channel}
    \mathcal{N}[\rho_A] = \tr_E[ V_{A \rightarrow BE} \rho_A V^{\dag}_{A \rightarrow BE} ]  
\end{equation}
We can similarly define a complementary channel $\bar{\mathcal{N}}: A \rightarrow E$ by tracing out $B$
\begin{equation}
\label{eq:complementary_channel_definition}
    \bar{\mathcal{N}} [\rho_A] = \tr_B [ V_{A \rightarrow BE} \rho_A V^{\dag}_{A \rightarrow BE}]
\end{equation}
There is an uncountably infinite set of equivalent Strinespring dilations.

A channel $\mathcal{N}$ is said to be degradable if there exists another CPTP map $\mathcal{M}:B \rightarrow E$ that allows it to recover the complementary channel, i.e., $\mathcal{M} \circ \mathcal{N} = \bar{\mathcal{N}}$. Similarly, a channel is antidegradable if its complementary channel is degradable. The degradability of a channel is independent of the choice of Stinespring dilation \cite{wilde2013quantum, waltrous2018theory}.

\subsection{Nomenclature}
Throughout this work, we take logarithms to be in base 2 (i.e., $\log(\bullet) \coloneq \log_2(\bullet)$). We take $\exp[\bullet] \coloneq e^{\bullet}$ and analogously $2[\bullet] \coloneq 2^{\bullet}$. When working with operators acting on other operators, we may write $X \circ Y \coloneq X\, Y \, X^{\dag}$. In Sec.~\ref{sup_mat_sec:average_output_state_from_PLON}, we introduce the notation $\tilde{\circ}$ as similar to $\circ$ but diagonal in the creation/annihilation operator basis - this is defined in Eq.~\eqref{sup_mat_eq:tilde_circ_functional_definition}. Vectors are denoted with bold indices, e.g, $\bm{x} \coloneq (x_1, x_2, \dots, x_{|\bm{x}|})$.
Number states are denoted with the Latin characters, usually $n$, $m$ or $x$, so for example $\ket{n} \coloneq \hat{a}^{\dag n} / \sqrt{n!} \, \ket{0}$, where $\ket{0}$ is the vacuum.
We refer to two-mode squeezed vacuum states with $\ket{z}_{1, 2} \coloneq (1 - |z|^2)^{\frac{1}{2}} \sum_{n = 0}^\infty z^n \ket{n, n}_{1, 2}$ where $z \in \mathbb{C}, \, |z| \in [0, 1)$ characterises the phase and magnitude of squeezing and $\bar{n}_z = |z|^2 / (1 - |z|^2)$ is the expected photon number per mode
In the limit $|z| \rightarrow 1$, this approaches the non-physical, infinite-entanglement Einstein-Podolsky-Rosen state $\ket{\textrm{EPR}} \propto \sum_{n = 0}^\infty \ket{n, n}_{1, 2}$. Tracing out one of the modes from $\ket{z}$ gives a thermal state with expected photon number $\bar{n}_z$, denoted here as $\rho_z \coloneq (1 - |z|^2) \sum_{n = 0}^\infty |z|^{2n} \ketbra{n}$. We also refer to coherent states using lower case Greek characters, such as $\ket{\gamma} \coloneq e^{-\frac{1}{2}|\gamma|^2} \sum_{n = 0}^\infty \frac{\gamma^n}{\sqrt{n!}} \ket{n}$.
When the argument in a ket is a vector, this indicates a tensor product state ${\ket{\bm{x}}} = \bigotimes_{i = 1}^{|\bm{x}|} \ket{x_i}$.
The set $\bm{P}(n, N)$ is the set of all ways of distributing $n$ indistinguishable items among $N$ bins, so for example $\bm{P}(3, 2) = \{(3, 0), (2, 1), (1, 2), (0, 3)\}$ and $\bm{P}(2, 3) = \{(2, 0, 0), (0, 2, 0), (0, 0, 2), (1, 1, 0), (1, 0, 1), (0, 1, 1)\}$. The projector onto the $n$ photon subspace in $N$ modes is given by $\Pi_N(n) \coloneq \sum_{\bm{x} \in \bm{P}(n, N)} \ketbra{\bm{x}}$. Throughout this work we encounter three entropies; the entropy of a quantum state is defined as $H(\rho) \coloneq -\tr[\rho \log(\rho)]$ for quantum state $\rho$, the binary entropy function is defined as $H_2(x) = -x \log(x) - (1-x) \log(1-x)$ for $x\in[0, 1]$, and the entropy of a single-mode thermal state with expected particle number $\bar{n} \in [0, \infty)$ is $H_{\textrm{therm}}(\bar{n}) = (\bar{n} + 1) \log(\bar{n} + 1) - \bar{n} \log(\bar{n})$.

\subsection{Channel capacities}
The coherent information of the channel $\mathcal{N}$ with respect to a state $\rho$ is defined as \cite{PhysRevA.54.2629}
\begin{equation}
\label{eq:coherent_information_wrt_state}
    I_{c} (\rho; \mathcal{N}) \coloneq H(\mathcal{N}[\rho]) - H(\bar{\mathcal{N}}[\rho]) .
\end{equation}
where $H(\rho) \coloneq -\tr[\rho \log(\rho)]$ is the von-Neumann entropy of the state $\rho$ and $\bar{\mathcal{N}}$ is any complementary channel to $\mathcal{N}$. The coherent information is superadditive with respect to the states because for any $N$ mode state $\rho^{(N)}$ with reduced single mode states $\rho_j \coloneq \tr_{k \ne j}[\rho^{(N)}]$, it satisfies
\begin{equation}
    I_c(\rho^{(N)}; \mathcal{N}^{\otimes N}) \le \sum_{j = 1}^N I_c(\rho_j; \mathcal{N})
\end{equation}
For degradable channels, the coherent information is additive with respect to states and satisfies $I_c(\rho^{(N)}; \mathcal{N}^{\otimes N}) = N \, I_c(\rho; \mathcal{N})$. This has significant ramifications when it comes to calculating the capacity of degradable channels.
The coherent information of the channel $\mathcal{N}$ is defined as the supremum of $I_c(\rho; \mathcal{N})$ over the set of states $\rho$
\begin{equation}
    I_c(\mathcal{N}) = \sup_{\rho} I_c(\rho; \mathcal{N})
\end{equation}
For clarity, we continue to use the coherent information with respect to states and explicitly include the supremum in expressions.
\\
\\
In discrete-variable systems, the quantum capacity of a channel, $\mathcal{Q}(\mathcal{N})$, is the maximum rate at which quantum information (i.e., qudits) can be reliably communicated over asymptotically many independent uses of the channel $\mathcal{N}$, where rate is defined as the amount of information sent divided by the number of channel uses.
The famed Lloyd-Shor-Devetak theorem \cite{PhysRevA.57.4153, PhysRevA.55.1613, shor2002quantum, shor2003capacities, devetak2005private} states that the capacity of a channel is given by the regularised coherent information in the limit $N \rightarrow \infty$,
\begin{equation}
    \mathcal{Q}(\mathcal{N}) = \lim_{N \rightarrow \infty} \sup_{\rho^{(N)}} I_c(\rho^{(N)}; \mathcal{N}^{\otimes N}) 
\end{equation}
For generic channels, the optimisation over an unbounded number of channel uses makes calculating the capacity exactly intractable; for example, the capacity of the $d$-dimensional qudit depolarising channel, $\mathcal{N}_\lambda[\rho] \coloneq (1 - \lambda) \rho + \frac{1}{d}\mathds{1}$, remains unknown, despite being arguably the simplest and most well studied non-unitary channel \cite{fanizza2020quantum_flags}. On the other hand, the coherent information of degradable channels is additive, and hence the associated quantum capacity can be calculated via
\begin{equation}
    \mathcal{Q}(\mathcal{N})\big|_{\mathcal{N} \,\textrm{degradable}} = \sup_{\rho} I_c(\rho; \mathcal{N})
\end{equation}
This drastically simplifies calculations.

The classical capacity of a channel, $\mathcal{C}(\mathcal{N})$, is analogous to the quantum capacity but with classical information (i.e., bit strings). Its full expression is given by the Holevo–Schumacher–Westmoreland theorem \cite{PhysRevA.56.131, holevo1998classicalcapacity}. As with the quantum capacity, the classical capacity requires optimising over an unbounded number of channel uses and hence is intractable for generic channels.

The entanglement-assisted classical capacity of a channel, $\mathcal{C}_{\textrm{EA}}(\mathcal{N})$, is the classical capacity of a quantum channel if the two parties communicating share unlimited initial entanglement \cite{bennett2002entanglement_assisted}. In this setting, the sender can use their half of the entangled state during the encoding procedure, while the receiver can use their half when decoding. The expression for the entanglement-assisted classical capacity is given by \cite{bennett2002entanglement_assisted}
\begin{equation}
\label{sup_mat_eq:EA_classical_capacity}
    \mathcal{C}_{\textrm{EA}}(\mathcal{N}) = \sup_{\rho} I(\rho; \mathcal{N})
\end{equation}
where $I(\rho; \mathcal{N}) \coloneq I_c(\rho; \mathcal{N}) + H(\rho)$ is the quantum mutual information. Unlike the unassisted quantum or classical capacities, the expression for the entanglement-assisted classical capacity does not require regularisation over an unbounded number of channel uses.
One can also consider the entanglement-assisted quantum capacity, $\mathcal{Q}_{\textrm{EA}}(\mathcal{N})$. This has a numeric value equal to half the entanglement-assisted classical capacity, $\mathcal{Q}_{\textrm{EA}} = \frac{1}{2}\mathcal{C}_{\textrm{EA}}$.

\subsection{Energy-constrained channel capacities}
The unbounded Hilbert space of CV systems means that the capacity of channels can be infinite, a result associated with states that have infinite energy. The capacities are therefore of limited practical relevance. Instead, the capacity given constrained input states is a more useful quantity to consider \cite{holevo1997quantum, holevo2003entanglement}. A natural choice is the energy-constrained quantum capacity. Since we are interested in photonic platforms, we focus on the energy operator being the number operator $\hat{n} \coloneq \hat{a}^\dag \hat{a}$ (i.e., by setting $\hbar \omega = 1$).

We imagine the sender attempts to communicate through $N$ independent uses of the channel $\mathcal{N}$. This can be modeled as the sender transmitting an (encoded) $N$-mode state $\rho^{(N)}$ through the channel $\mathcal{N}^{\otimes N}$, and hence we apply the energy constraint to the state $\rho^{(N)}$.
We define the $N$-mode energy functional as
\begin{equation}
    E_N[\bullet] \coloneq \frac{1}{N} \sum_{i = 1}^N \tr[ \hat{n}_i \bullet]
\end{equation}
where $\hat{n}_i = \mathds{1}^{\otimes i - 1} \otimes \hat{n} \otimes \mathds{1}^{\otimes N - i}$ is the number operator on the $i^{\textrm{th}}$ mode. This energy functional gives the average expected particle number across all $N$ modes.

The energy-constraint we consider is called a \emph{uniform energy-constraint}; the encoded state $\rho^{(N)}$ must satisfy
\begin{equation}
    E_N[\rho^{(N)}] \le \bar{n}
\end{equation}
for some upper bound on the energy $\bar{n} \in [0, \infty)$. In other words, the averaged expected particle number entering the noise channel $\mathcal{N}^{\otimes N}$ must be upper bounded by $\bar{n}$.

The unassisted quantum capacity for an input state $\rho^{(N)}$ subject to the uniform energy-constraint was derived by Wilde and Qi \cite{wilde2018energy}. For a \emph{degradable} channel $\mathcal{N}$, it is given by
\begin{equation}
    \mathcal{Q}(\mathcal{N}; \bar{n}) = \sup_{\rho : \tr[\hat{n} \rho] \le \bar{n}} I_c(\mathcal{N}; \rho)
\end{equation}
Note that the generic expression would involve an optimisation over an unbounded number of channel uses, but we focus on the degradable channel case.
This is equal to the private key capacity with the same energy-constraint, which is the maximum rate at which private keys can be distributed through the channel.
This is \emph{also} equal to the entanglement-transmission capacity and secret-key capacity subject to an \emph{average} energy-constraint. This is a weaker constraint than the uniform energy-constraint (refer to \cite{wilde2018energy} for details).

The entanglement-assisted classical capacity subject $E_N [\rho^{(N)}] \le \bar{n}$ was derived by Holevo \cite{holevo2003entanglement}; a channel $\mathcal{N}$ satisfying $H(\mathcal{N}(\rho)) < \infty$ for any state $\rho$ satisfying $\tr[\hat{n} \rho] \le \bar{n}$ has an energy-constrained entanglement-assisted classical capacity
\begin{equation}
    \mathcal{C}_{\textrm{EA}}(\mathcal{N}; \bar{n}) = \sup_{\rho : \tr[\hat{n} \rho] \le \bar{n}}   I(\rho; \mathcal{N}) 
\end{equation}
This expression is exactly analogous to the unconstrained entanglement-assisted classical capacity in Eq.~\eqref{sup_mat_eq:EA_classical_capacity}, but with the maximisation search space restricted to those states with energy $\le \bar{n}$. Again, the energy-constrained entanglement-assisted quantum capacity is equal to half the classical capacity,
\begin{equation}
    \mathcal{Q}_{\textrm{EA}}(\mathcal{N}; \bar{n}) = \frac{1}{2} \mathcal{C}_{\textrm{EA}}(\mathcal{N}; \bar{n})
\end{equation}

\subsection{Quantum error correction and decoupling}
The coherent information gives the quantum capacity, but it has further applications in quantum error correction. Consider a state $\rho_A$ passing through a noise channel $\mathcal{N}_{A \rightarrow B}$. Perfect error correction is possible if there exists a decoding procedure $\mathcal{F}_{B \rightarrow \bar{A}}$ such that $\mathcal{F}_{B \rightarrow \bar{A}}\big[ \mathcal{N}_{A \rightarrow B} [\rho_A] \big] = \rho_{\bar{A}}$ with $\mathcal{I}_{\bar{A} \rightarrow A}[\rho_{\bar{A}}] = \rho_A$, where $\mathcal{I}$ is the identity superoperator. In Ref.~\cite{PhysRevA.54.2629}, Schumacher and Nielsen proved that such a decoding procedure exists if and only if the coherent information of $\mathcal{N}$ with respect to $\rho$ is equal to the entropy of the initial state,
\begin{equation}
\label{sup_mat_eq:coherent_info_entropy_QEC}
    I_c(\mathcal{N}; \rho) = H(\rho) \implies \textrm{error correction is possible}
\end{equation}
If the state $\rho_A$ is purified into the Hilbert space $R$ (i.e., so $\rho_A = \tr_R [ \ketbra{\psi}_{RA} ]$), and we consider a Stinespring dilation of the channel $(\mathcal{U}_{\mathcal{N}})_{A \rightarrow BE}$ (i.e., so $\mathcal{N}_{A\rightarrow B}[\rho_A] = \tr_E \big[(\mathcal{U}_{\mathcal{N}})_{A \rightarrow BE} [\rho_A] \big]$), then Eq.~\eqref{sup_mat_eq:coherent_info_entropy_QEC} implies that error correction is possible on the evolved state $\rho_{RBE}(\mathcal{N}) := \left(\mathcal{I}_{R \rightarrow R} \otimes (\mathcal{U}_{\mathcal{N}})_{A \rightarrow BE} \right) [\ketbra{\psi}_{RA}]$ if and only if the reduced state $\rho_{RE}(\mathcal{N}) = \tr_B[\rho_{RBE}(\mathcal{N})]$ is decoupled between $R$ and $E$ \cite{PhysRevA.54.2629},
\begin{equation}
    \rho_{RE}(\mathcal{N}) = \rho_R \otimes \rho_E(\mathcal{N}) \implies \textrm{error correction is possible}
\end{equation}
Intuitively, this means that the environment $E$ has no information about the initial state. Since all the operations described above are fundamentally unitary, and unitary dynamics imply information conservation, this means that all the information about the initial state must be contained in $B$, and hence a decoding procedure exists such that $\mathcal{F}_{B \rightarrow \bar{A}}[\rho_B(\mathcal{N})] = \rho_{\bar{A}}$.
Similarly, in Ref.~\cite{schumacher2002approximate}, Schumacher and Westmoreland proved that approximate decoupling between $R$ and $E$ implies the existence of an approximate error correcting procedure, where the fidelity between $\rho_{\bar{A}}$ and $\rho_A$ is bounded by the fidelity between $\rho_{RE}(\mathcal{N})$ and $\rho_R \otimes \rho_E(\mathcal{N})$.

\subsection{Typical subspace}
In quantum information theory, the typical subspace of a state arises when one considers many copies of that state. For the spectral decomposition of a state $\rho = \sum_{x} p(x) \ketbra{x}$, the $N$-fold tensor product state is given by 
\begin{align}
\begin{split}
    \rho^{\otimes N} &= \sum_{x_1, x_2, \dots, x_N} p(x_1) p(x_2) \dots 
    p(x_N) \ketbra{x_1, x_2, \dots, x_N} \\
    &= \sum_{x_1, x_2, \dots x_N} p_N(x_1, x_2, \dots, x_N) \ketbra{x_1, x_2, \dots, x_N}
\end{split}
\end{align}
Since the probability distributions $p(x_i)$ are all independent and identically distributed, there exists a typical set of sequences $\bm{x} = (x_1, x_2, \dots, x_N)$ that dominate the total probability distribution $p_N(x_1, x_2, \dots, x_N)$ in the large $N$ limit. This follows from the asymptotic equipartition property, which itself is the law of large numbers applied to probability distributions. The set of states spanned by this typical set of sequences is called the typical subspace of the state $\rho$ in the large $N$ limit. 

The weak $\delta$-typical subspace of $\rho^{\otimes N}$, denoted $T^\delta_N(\rho)$, is the Hilbert space spanned by basis states with a sample entropy, $\bar{H}_N(\bm{x})$, that is $\delta$ close to the entropy of the state $\rho$;
\begin{equation}
    T_N^\delta (\rho) \coloneq \textrm{span} \left\{ \ket{\bm{x}} : \big|\bar{H}_N(\bm{x}) - H(\rho) \big| \le \delta \right\}
\end{equation}
where the sample entropy is defined as the normalised negative logarithm of the probability associated with the sequence $\bm{x}$,
\begin{equation}
    \bar{H}_N(\bm{x}) \coloneq - \frac{1}{N} \log( p_N(\bm{x}))
\end{equation}
The projector onto the typical subspace is given by
\begin{equation}
    \Pi^\delta_N (\rho) \coloneq \sum_{\bm{x} : |\bar{H}_N(\bm{x}) - H(\rho)| \le \delta} \ketbra{\bm{x}}
\end{equation}

For any $\delta, \varepsilon > 0$, there exists an $N_0$ such that for all $N \ge N_0$ the state $\rho^{\otimes N}$ has overlap $\ge 1 - \varepsilon$ with the typical subspace;
\begin{equation}
    \tr[ \Pi_N^\delta (\rho) \, \rho^{\otimes N} ] \ge 1 - \varepsilon
\end{equation}
This means that the entire state $\rho^{\otimes N}$ is contained within the typical subspace with arbitrary accuracy for suitably large $N$. In other words, the state concentrates in the typical subspace. The typical subspace dimension scales as $2^{N \big(H(\rho) + \delta \big)}$;
\begin{equation}
    \tr[ \Pi_N^\delta (\rho) ] \le 2^{N \big(H(\rho) + \delta \big)}
\end{equation}
In our random coding scheme, we will apply the typical space to continuous-variable states that live in an infinite-dimensional Hilbert space. This shows that we can capture all the key properties of many copies of a CV state within a finite-dimensional Hilbert space in the limit that we take enough copies. When applied to CV systems, this effectively becomes a sophisticated finite-cutoff to the dimension.
\\
\\
Refer to Ref.~\cite{wilde2013quantum} for a detailed introduction to typical sets and typical subspaces.

\subsection{The Haar measure and Weingarten calculus}

Finally, we introduce the Haar measure and Weingarten calculus used for averaging over random unitaries.
The Haar measure is the unique measure satisfying both left and right invariance. If $\textrm{d}\mu(U)$ is the Haar measure associated with the group $\textrm{U}(d)$ and $f: \textrm{U}(d) \rightarrow M$ is some function that maps $\textrm{U}(d)$ onto some space $M$, then the Haar measure satisfies the following properties:
\begin{equation}
\label{eq:Haar_measure_defining_properties_1}
    \mathbb{E}_U [f(U)] \coloneq \int_{\textrm{U}(d)} \textrm{d}\mu(U) f(U) = \int_{\textrm{U}(d)} \textrm{d}\mu(U) f(VU) = \int_{\textrm{U}(d)} \textrm{d}\mu(U) f(UV')
\end{equation}
\begin{equation}
\label{eq:Haar_measure_defining_properties_2}
    \int_{\textrm{U}(d)} \textrm{d}\mu(U) = 1
\end{equation}
\begin{equation}
\label{eq:Haar_measure_defining_properties_3}
    \int_{S\subset \textrm{U}(d)} \textrm{d} \mu(U) \le 1
\end{equation}
where $V, V' \in \textrm{U}(d)$ are arbitrary constant unitary matrices.
For example, consider the group $\textrm{U}(1) = \{e^{i\theta} | \theta \in [0, 2\pi)\}$. In this case, one can explicitly write the Haar measure as $\textrm{d}\mu(U = e^{i\theta}) = \frac{\textrm{d}\theta}{2\pi}$, from which averages can be calculated via
\begin{equation}
\label{eq:Haar_measure_U(1)}
    \mathbb{E}_U [f(e^{i\theta}) ] = \int_{0}^{2\pi} \frac{\textrm{d}\theta}{2\pi} \, f(e^{i\theta})   
\end{equation}

Explicitly expressing the Haar measure becomes increasingly non-trivial as $d$ grows. Weingarten calculus \cite{weingarten1978asymptotic,collins2006integration, collins2022weingarten} describes the series of techniques used to calculate moments of the Haar measure on certain compact groups. In particular, Collins and {\'S}niady provided the explicit formulae for averages over the Haar measure for the unitary, orthogonal and symplectic groups \cite{collins2006integration}. For $n, m \in \{1, 2, 3, \dots\}$ the integral over products of matrix elements of $U \in \textrm{U}(N)$ is given by \cite{collins2006integration}
\begin{align}
\begin{split}
\label{eq:collins_sniady_weingarten_calculus_moment_result}
    \int_{\textrm{U}(N)} \textrm{d}\mu \, &U_{i_1, j_1} U_{i_2, j_2} \dots U_{i_{n}, j_{n}}  U_{i'_1, j'_1}^* U_{i'_2, j'_2}^* \dots U_{i'_{m}, j'_{m}}^*  \\
    &=\delta_{n, m} \sum_{\sigma, \tau \in \mathcal{S}_n} \delta_{i_1, i'_{\sigma(1)}} \dots \delta_{i_n, i'_{\sigma(n')}} \delta_{j_1, j'_{\tau(1)}} \dots \delta_{j_n, j'_{\tau(n')}} \textrm{Wg}_{N,n}(\tau \sigma^{-1})
\end{split}
\end{align}
where $x^*$ denotes the complex conjugate of $x$, $\delta_{i, j}$ is the Kronecker delta function equal to unity if $i = j$ and zero otherwise, $\mathcal{S}_n$ is the symmetric group of $n$ elements and $\textrm{Wg}_{N,n}(\sigma)$ is the Weingarten function associated with the unitary group $\textrm{U}(N)$ and symmetric group $\mathcal{S}_n$ - this Weingarten function is defined in Ref.~\cite{collins2006integration}. Using this, one can average over expressions of the form $U^{\otimes n} \hat{X} \, U^{\dag \, \otimes n}$. For example, the double-twirling formula (given in Eq.~\eqref{sup_mat_eq:double_twirling_formula}) is just an application of Eq.~\eqref{eq:collins_sniady_weingarten_calculus_moment_result} with $n = 2$.

\section{Continuous-variable erasure channel}

The discrete-variable quantum erasure channel has some probability $1-p$ of doing nothing and probability $p$ of mapping the input state to an output state in an orthogonal Hilbert space,
here labeled as $\ket{e}$.
\begin{equation}
\label{eq:DV_erasure_channel_definition}
    \Lambda_p^{\textrm{DV}} [\rho] \coloneq (1 - p) \rho + p \, \ketbra{e}
\end{equation}
For $p < 0.5$ this channel is degradable, while for $p > 0.5$ it is antidegradable. The capacities of the erasure channel in the non-entanglement assisted cases were mapped out in Ref.~\cite{PhysRevLett.78.3217}, while the entanglement-assisted (EA) capacities were derived in Ref.~\cite{PhysRevLett.83.3081}. For $d$-dimensional DV systems, the standard and EA quantum capacities are respectively given by
\begin{equation}
\label{eq:DV_erasure_channel_stan_EA_quantum_capacities}
    \mathcal{Q}(\Lambda_p^{\textrm{DV}}) = \max \{(1 - 2p) \log(d), \,\,0\}, \quad \quad  \mathcal{Q}_{\textrm{EA}} (\Lambda_p^{\textrm{DV}}) = (1 - p) \log(d).
\end{equation}
Random codes are known to saturate these capacities in the asymptotic limit, even with low depth random circuits \cite{PhysRevX.11.031066}. The Hayden-Preskill problem \cite{hayden2007black} can also be considered as an example of entanglement-assisted random coding scheme for the erasure channel, for which an explicit decoding scheme was constructed in Ref.~\cite{yoshida2017efficient}.

We now turn our attention to the CV erasure channel. Physically, if an erasure takes place then the state should be mapped to the vacuum. However, naively replacing $\ket{e}$ with $\ket{0}$ in Eq.~\eqref{eq:DV_erasure_channel_definition} does not allow us to perfectly distinguish if an erasure has taken place because the vacuum is not orthogonal to our CV state. We follow Ref.~\cite{Zhong2023information} and attach an additional classical flag to the CV state that flips on erasures. We are therefore interested in the Hilbert space $\mathcal{H}_{\textrm{CV}} \otimes \mathcal{H}_{\textrm{flag}}$ where $\mathcal{H}_{\textrm{CV}}$ describes a single bosonic mode and $\mathcal{H}_{\textrm{flag}} = \textrm{span}\{\ket{\uparrow}, \ket{\downarrow}\}$ describes a flag state - we choose to associated $\ket{\downarrow}$ with erasures. The CV erasure channel $\Lambda_p: \mathcal{T}(X) \rightarrow \mathcal{T}(Y)$ for $X, Y = \mathcal{H}_{\textrm{CV}} \otimes \mathcal{H}_{\textrm{flag}}$ is defined as 
\begin{equation}
\label{eq:CV_erasure_channel_definition}
    \Lambda_p[\sigma_X] \coloneq (1 - p) \sigma_Y + p \, \ketbra{0, \downarrow}_Y
\end{equation}
With this definition, we can study the erasure channel with $\sigma_Y = \rho \otimes \ketbra{\uparrow}$, the approximate erasure channel (i.e., with no orthogonality between branches) with $\sigma_Y = \rho \otimes \ketbra{\downarrow}$ and iterated applications of the erasure channel, which gives
\begin{equation}
    \Lambda_q \circ \Lambda_p [\sigma] = \Lambda_{p + q - pq} [\sigma].
\end{equation}

A Stinespring dilation of the channel can be achieved with the isometry $V: X \rightarrow YE$ with $E \cong Y$ which satisfies $V^\dag V = \mathds{1}_X$
\begin{equation}
    V_{X\rightarrow YE} = \sqrt{1 - p} \mathds{1}_{X \rightarrow Y} \otimes \ket{0,\downarrow}_E + i\sqrt{p} \ket{0,\downarrow}_Y \otimes \mathds{1}_{X\rightarrow E}
\end{equation}
where $\mathds{1}_{X \rightarrow Y} = \sum_{n = 0}^\infty \ket{n, \uparrow}_Y \bra{n, \uparrow}_X  + \ket{n, \downarrow}_Y \bra{n, \downarrow}_X$ maps the state on $X$ onto $Y$, and similarly for $\mathds{1}_{X \rightarrow E}$. In this Stinespring dilation representation, the erasure channel and complementary channel, $\bar{\Lambda}_p: \mathcal{T}(X) \rightarrow \mathcal{T}(E)$, are given by
\begin{align}
\begin{split}
    V_{X \rightarrow YE} \, \sigma_X \, (V_{X \rightarrow YE})^\dag = (1 - p) \, \sigma_Y \otimes \ketbra{0, \downarrow}_E + p \, \ketbra{0, \downarrow}_Y \otimes \sigma_E
\end{split}
\end{align}
\begin{align}
\begin{split}
    \Lambda_p[\sigma] &= \tr_E[ V_{X \rightarrow YE}  \, \sigma \, (V_{X \rightarrow YE} )^\dag ] \\
    &= (1-p) \sigma + p \ketbra{0, \downarrow}
\end{split}
\end{align}
\begin{align}
\begin{split}
    \bar{\Lambda}_p [\sigma] &= \tr_Y[ V_{X \rightarrow YE}  \, \sigma \, (V_{X \rightarrow YE} )^\dag ] \\
    &= p \, \sigma + (1 - p) \ketbra{0, \downarrow}
\end{split}
\end{align}
We see that the complementary channel is just the erasure channel with $p \rightarrow 1 - p$. For $p<0.5$, the channel is degradable with a degrading channel $\Lambda_{q}$ where $q = (1-2p) / (1-p)$, i.e., $\Lambda_q \circ \Lambda_p = \bar{\Lambda}_p$. Similarly, for $p > 0.5$, the channel is antidegradable
\\
\\
We use these preliminaries to derive the energy-constrained quantum and entanglement-assisted capacities in the main manuscript.

\input{optimal_code}

\input{PLON_averaged_state}

\bibliography{biblio_sup}

%% file: optimal_code.tex
\section{Random coding scheme}
\label{sup_mat_sec:random_coding_scheme}

\begin{figure}
    \centering
    \includegraphics[width=0.85\linewidth]{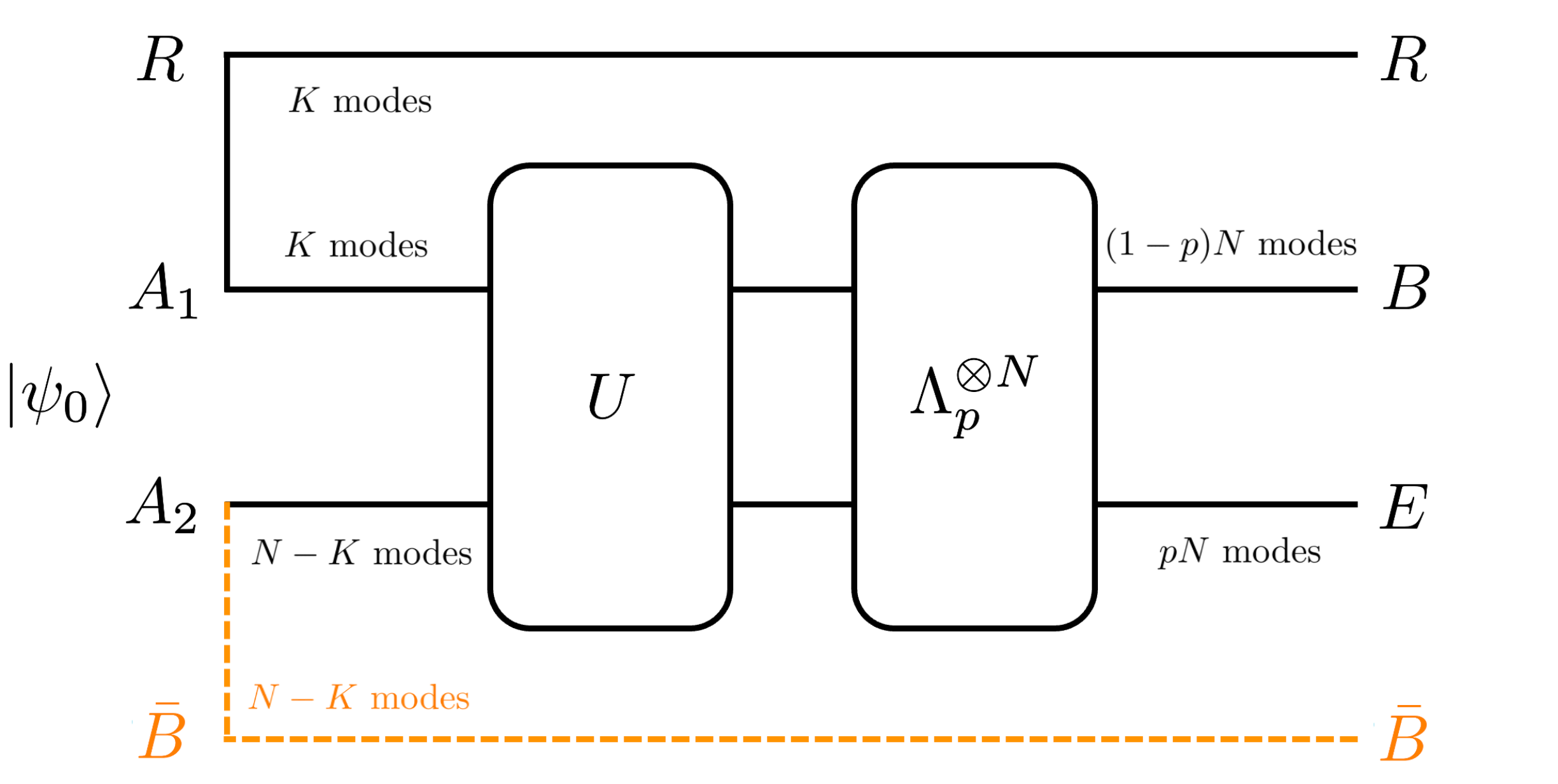}
    \caption{Random coding scheme for entanglement transmission. Alice begins with $K$ TMSVs, keeping one mode per pair in the reference $R$ and placing the other in $A_1$. She introduces an additional $N-K$ modes ($A_2$) and applies a random unitary $U$ on $A=A_1A_2$. Each mode is then sent through the single-mode CV erasure channel $\Lambda_p$. Bob uses the classical flags to retain the $\approx(1-p)N$ unerased modes ($B$), while the erased modes go to the environment ($E$). In the standard scheme, $A_2$ consists of random coherent states, whereas in the entanglement-assisted scheme (dashed orange) Alice and Bob instead share $N-K$ TMSVs across $A_2\bar{B}$, with Bob accessing $\bar{B}$ during decoding.}
    \label{sup_mat_fig:random_code_diagram}
\end{figure}

In this section, we construct asymptotically optimal random codes for entanglement-generation across the continuous-variable erasure channel. Refer to Fig.~\ref{sup_mat_fig:random_code_diagram} for a diagram of the setup, described below.
Alice initially holds $K$ entangled two-mode squeezed vacuum states (TMSVs) that she wants to share with Bob.
From each of the TMSVs, she keeps one mode in a reference space $R$ and wants to transmit the other (on Hilbert space $A_1$) through an erasure channel to Bob. She encodes the $K$ modes on $A_1$ with an additional $N-K$ modes on $A_2$ by implementing a random unitary on $A = A_1 A_2$. The modes from $A$ are then each independently sent through a single-mode CV erasure channel $\Lambda_p$ (a classical flag is attached as necessary). Bob then performs a syndrome measurement on the classical flags, confirming whether he should keep the mode (in a Hilbert space labeled $B$) or discard it; the environment has access to the information originally contained in the erased modes so we label them with $E$.
In the entanglement-assisted case, Alice and Bob initially share $N-K$ TMSVs on the Hilbert space $A_2 \bar{B}$. Alice uses her half during the encoding procedure, while Bob can use his half on $\bar{B}$ when decoding. In this case, the initial state is a tensor product of $N$ TMSV states, whereas in the standard case the initial state on $A_2$ is a tensor product of randomly sampled coherent states. Hence the initial state $\ket{\psi_0}$ reads
\begin{equation}
    \ket{\psi_0} = \begin{cases}
        (\ket{z}^{\otimes K})_{RA_1} \ket{\bm{\gamma}}_{A_2} & \textrm{standard case} \\
        (\ket{z}^{\otimes N})_{R A \bar{B}} & \textrm{entanglement-assisted case}
    \end{cases}
\end{equation}

Recall that Bob can decode Alice's information if the environment $E$ and reference $R$ are decoupled, i.e., $\rho_{RE}(U, \psi_0) \approx \rho_R \otimes \rho_E(U, \psi_0)$, where the $\psi_0$ argument distinguishes if we are in the standard or entanglement-assisted case and $U$ is the random unitary that Alice used to encode the state. We will upper bound the average trace distance of $\rho_{RE}(U, \psi_0)$ to the average (and decoupled) state $\mathbb{E}_{U, \psi_0} [\rho_{RE}(U, \psi_0)] = \rho_R \otimes \bar{\rho}_E$, namely
\begin{equation}
    \mathbb{E}_{U, \psi_0} || \rho_{RE}(U, \psi_0) - \rho_R \otimes \bar{\rho}_E ||_1.
\end{equation}
The average distance to the average state is the variance of the distribution $\rho_{RE}(U, \psi_0)$ in state space. When this is exponentially small with respect to $N$, it indicates that for a particular choice of $U$ (and $\ket{\bm{\gamma}}$ in the standard case) then the state $\rho_{RE}(U, \psi_0)$ will be exponentially close to $\rho_R \otimes \bar{\rho}_E$ in trace distance. Therefore, the state will be approximately decoupled and hence Bob can decode the $K$ TMSV states that Alice sent. 

We will show that the average trace distance can be made exponentially small with respect to $N$ for all entanglement-transmission rates up to the associated quantum capacities in the high energy limit, minus an energy-independent gap. This shows that these random codes are asymptotically optimal in the large $N$, high energy limit.
\\
\\
First we need to find an ensemble of random unitaries that suitably hide Alice's information while avoiding the diverging energies associated with maximally scrambling CV dynamics. To this end, we want the initial state on $A$ to form a typical subspace, and then consider Haar random unitaries within this subspace, an approach introduced in Ref.~\cite{Zhong2023information}. To do this, we require in the initial reduced state on $A$ to be an $N$-fold tensor product state - we show in the next section how our initial state becomes (on average) a tensor product of thermal states on which we can define a typical subspace.

\subsection{Typical subspace}
Alice initially has $K$ two-mode squeezed vacuum (TMSV) states $\ket{z}^{\otimes K}_{R A_1}$ where $\ket{z}_{1,2} \coloneq \sqrt{1 - |z|^2} \sum_{n = 0}^\infty z^{n} \ket{n,n}_{1,2}$. Tracing out one mode from a TMSV gives a thermal state $\rho_z \coloneq (1 - |z|^2) \sum_{n = 0}^\infty |z|^{2n} \ketbra{n}$. Therefore, the reduced state on $A_1$ is a tensor product of thermal states, $\tr_{X\notin A_1}[\ketbra{\psi_0}{\psi_0}] = \rho_z^{\otimes K}$.

Using this, we want the reduced state on $A$ to be, at least on average, a tensor product of thermal states. In the entanglement-assisted case, we consider Alice and Bob initially sharing an additional $N-K$ TMSV states, with Alice's half held in $A_2$ and Bob's half held in $\bar{B}$. In this setting, the reduced state on $A$ is a tensor product of thermal states, $(\psi_0)_A = \rho_z^{\otimes N}$, and hence we automatically form a typical subspace in the large $N$ limit. 

In the standard case, we can not choose the modes in $A_2$ to be mixed thermal states because then Bob would not be able to decode the information \cite{PhysRevA.54.2629}. Our approach instead is to utilise the coherent state representation of thermal states,
\begin{equation}
\label{sup_mat_eq:coherent_state_decomp_of_thermal_state}
    \rho_z = \int_{\mathbb{C}} \textrm{d}p(\gamma) \, \ketbra{\gamma}, \quad \quad 
    \quad \textrm{d}p(\gamma) = e^{-\frac{1}{\bar{n}_z}|\gamma|^2} \frac{\textrm{d}^2\gamma}{\bar{n}_z \pi}
\end{equation}
where $\ket{\gamma} = e^{-\frac{1}{2} |\gamma|^2} \sum_{n = 0}^\infty \frac{\gamma^n}{\sqrt{n!}} \ket{n}$ is a coherent state. Using this, in each mode of $A_2$ we randomly sample a coherent state $\ket{\gamma_i}$ from the probability distribution $\textrm{d}p(\gamma_i)$ to give an initial state $\psi_0 = \ket{z}_{RA_1}^{\otimes K} \ket{\bm{\gamma}}_{A_2}$ where $\ket{\bm{\gamma}} = \bigotimes_{i = 1}^{N - K} \ket{\gamma_i}$ are the randomly sampled coherent states. Crucially, when averaging over the coherent states $\bm{\gamma}$ the reduced state on $A$ is given by a tensor product of thermal states,
\begin{equation}
    \mathbb{E}_{\bm{\gamma}} [\rho_{z \, A_1}^{\otimes K} \otimes \ketbra{\bm{\gamma}}_{A_2}] = (\rho_{z}^{\otimes N})_A
\end{equation}
As such, on average a typical subspace is formed. Particular choices of $\ket{\bm{\gamma}}$ will also have a high overlap with the typical subspace (shown later) and hence it is sufficient use the thermal state typical subspace to derive our results.

\subsubsection{Typical subspace of a thermal state}
In the eigenbasis of $\rho_z^{\otimes N}$ (i.e., the Fock / number basis), the probability associated with the photon number string $\bm{x} = (x_1, x_2, \dots, x_N)$ is given by $p_N(\bm{x}) = (1 - |z|^2) |z|^{2 n_{\textrm{tot}}(\bm{x})}$ where $n_{\textrm{tot}}(\bm{x}) = \sum_{i = 1}^N x_i$ is the total number of photons in $\bm{x}$. The sample entropy $\bar{H}_N(\bm{x}) \coloneq -\frac{1}{N} \log(p_N(\bm{x}))$ is readily computed as
\begin{equation}
    \label{sup_mat_eq:sample_entropy}
    \bar{H}_N(\bm{x}) = -\log(1 - |z|^2) - \frac{n_{\textrm{tot}}(\bm{x})}{N} \log(|z|^2)
\end{equation}
The weak $\delta$-typical subspace of $\rho_z^{\otimes N}$, denoted via $T_N^\delta (\rho_z)$, is the space spanned by the eigenstates of $\rho_z^{\otimes N}$ that have a sample entropy $\delta$-close to the von Neumann entropy of $\rho_z$,
\begin{equation}
\label{sup_mat_eq:typical_subspace_def}
    T_N^\delta(\rho_z) \coloneq \textrm{span} \big\{ \ket{\bm{x}} : |\bar{H}_N(\bm{x}) - H(\rho_z) | \le \delta \big\}
\end{equation}
Inserting Eq.~\eqref{sup_mat_eq:sample_entropy} into Eq.~\eqref{sup_mat_eq:typical_subspace_def} gives a relationship between the total number of photons,  the number of modes and $\delta$
\begin{align}
\begin{split}
    |\bar{H}_N(\bm{x}) - H(\rho_z)| &= \left | \left(\bar{n}_z - \frac{n_{\textrm{tot}}(\bm{x})}{N} \right) \log(|z|^2) \right| \\ 
    &\le \delta
\end{split}
\end{align}
The $\delta \rightarrow 0$ limit is achieved when $n_{\textrm{tot}}(\bm{x}) = \bar{n}_z N$ - intuitively, this is the set of photon number strings that have the same total number of photons as the expected photon number of $\rho_z^{\otimes N}$. If we consider strings of photon numbers that deviate from the average with $n_{\textrm{tot}}(\bm{x}) = \bar{n}_z N (1 \pm \kappa)$ for $0 < \kappa < 1$, then these states are also in the $\delta$-typical subspace if
\begin{equation}
    \kappa \le \frac{\delta}{\bar{n}_z |\log(|z|^2)|}
\end{equation}
Therefore, the $\delta$-typical subspace of $N$ thermal states is given by
\begin{equation}
    T_N^\delta (\rho_z) = \textrm{span} \big\{ \ket{\bm{x}} : n_{\textrm{tot}}(\bm{x}) \in \{\tilde{n}_-, \tilde{n}_- + 1, \dots, \tilde{n}_+\}   \big\}
\end{equation}
where we have defined
\begin{align}
\begin{split}
    \tilde{n}_+ &= \floor{N (\bar{n}_z + \frac{\delta}{|\log(|z|^2)|})}, \\
    \tilde{n}_- &= \ceil{N (\bar{n}_z - \frac{\delta}{|\log(|z|^2)|})}.
\end{split}
\end{align}
In other words, it is spanned by the set of states $\ket{\bm{x}}$ that have a total number of photons within $\frac{\delta}{|\log(|z|^2)} N$ of the average photon number $\bar{n}_z N$.

The projector onto this subspace is given by
\begin{equation}
\label{sup_mat_eq:typical_subspace_projector}
    \Pi_N^\delta = \sum_{m = \tilde{n}_-}^{\tilde{n}_+} \Pi_N(m).
\end{equation}
where $\Pi_N(m)$ is the projector onto all the possible ways of distributing $m$ indistinguishable photons among $N$ modes. For example, $\Pi_2(3) = \ketbra{3, 0} + \ketbra{2, 1} + \ketbra{1, 2} + \ketbra{0, 3}$. We can decompose this over bipartitions (or arbitrary partitions) by using the identity (or iterated use of the identity)
\begin{equation}
    \Pi_N(m) = \sum_{k = 0}^m \Pi_K(k) \otimes \Pi_{N - K}(n - k).
\end{equation}

\subsubsection{States in the typical subspace}
The overlap between $\rho_z^{\otimes N}$ and its $\delta$-typical subspace $T_N^\delta(\rho_z)$, denoted with $\Delta_N \coloneq \tr [\rho_z^{\otimes N} \Pi_N^\delta]$, is given by
\begin{equation}
\label{sup_mat_eq:typical_subspace_overlap}
    \Delta_N = (1 - |z|^2)^{N} \sum_{n = \tilde{n}_-}^{\tilde{n}_+} |z|^{2n} {n + N - 1 \choose n}
\end{equation}
This can be made arbitrarily close to unity by increasing $N$ \cite{wilde2013quantum}. In the entanglement-assisted case, this quantity is the overlap between the initial state on $A$ and the typical subspace $T_N^\delta(\rho_z)$.

When we consider the standard case, the initial state includes a randomly sampled tensor product of coherent states on $A_2$, namely $\ket{z}_{RA_1}^{\otimes K} \ket{\bm{\gamma}}_{A_2}$ where each $\gamma_i$ in $\bm{\gamma} = (\gamma_1, \gamma_2, \dots, \gamma_{N-K})$ is drawn from the probability distribution $\textrm{d}p(\gamma_i)$ in Eq.~\eqref{sup_mat_eq:coherent_state_decomp_of_thermal_state}. When averaging over $\bm{\gamma}$, the average input state on $A$ is again a tensor product of thermal states $(\rho_z^{\otimes N})_A$, which forms a typical subspace on average.
In the standard case, for a \emph{particular} choice of $\bm{\gamma}$, the overlap of $\rho_{z \,A_1}^{\otimes K} \otimes \ketbra{\bm{\gamma}}_{A_2}$ is denoted $\Delta_N(\bm{\gamma})$. This clearly satisfies
\begin{equation}
    \mathbb{E}_{\bm{\gamma}} [\Delta_N(\bm{\gamma})] = \Delta_N
\end{equation}
which indicates that on \emph{average} the overlap approach unity. Since $\Delta(\bm{\gamma})$ is non-negative and upper bounded by unity, this indicates that in the large $N$ limit almost every choice of $\bm{\gamma}$ will have a high overlap with the typical subspace. This is because if there was a non-negligible fraction of $\bm{\gamma}$ choices that had a low overlap with $T_N^\delta (\rho_z)$, then it would be impossible for the average $\Delta(\bm{\gamma})$ to approach unity as $N \rightarrow \infty$. Therefore, the initial states in both the standard and entanglement-assisted cases have an overlap with the typical subspace $T_N^\delta (\rho_z)$ that can be arbitrarily large for suitably large $N$.

Finally, the overlap between $K$ single-mode thermal states and the reduced typical subspace spanning $\Pi_K^\delta$, denoted by $\Delta_K$, is
\begin{align}
\begin{split}
\label{sup_mat_eq:reduced_typical_subspace_overlap}
    \Delta_K &= \tr[\Pi_K^\delta \rho_z^{\otimes K}] \\
    &= (1 - |z|^2)^{2K} \sum_{n = 0}^{\tilde{n}_+} |z|^{2n} {n + K - 1 \choose n}
\end{split}
\end{align}
Again, this approaches unity in the large $N$ limit.

\subsubsection{Dimensions in the typical subspace}
The dimension of the typical subspace $T_N^\delta(\rho_z)$ is equal to the number of basis states that span it (henceforth we will drop the argument $\rho_z$). Since $\textrm{rank}\big( \Pi_N(m) \big) = {m + N - 1 \choose m}$, from Eq.~\eqref{sup_mat_eq:typical_subspace_projector} the dimension is given by
\begin{align}
\begin{split}
    \textrm{dim}(T_N^\delta) &= \sum_{m = \tilde{n}_-}^{\tilde{n}_+} {m + N - 1 \choose m} \\
    &= {\tilde{n}_+ + N \choose \tilde{n}_+} - {\tilde{n}_- + N - 1 \choose \tilde{n}_-}
\end{split}
\end{align}
where we used the standard result $\sum_{m = 0}^{\tilde{n}} {m + N - 1 \choose m} = {\tilde{n} + N \choose \tilde{n}}$. 

If we consider $M < N$ modes in the typical subspace in the Hilbert space $A_M$, then the reduced typical subspace on $A_M$, denoted $(T_N^\delta)_{A_M}$, is Hilbert space spanned by the reduced projection operator,
\begin{align}
\begin{split}
    (T_{N}^\delta)_{A_M} &= \textrm{span} \left\{ \tr_{A_{\bar{M}}}[\Pi_N^\delta] \right\}\\
    &= \textrm{span} \left\{ \sum_{m = \tilde{n}_-}^{\tilde{n}_+} \sum_{k = 0}^m \Pi_{A_M}(k)  \tr_{A_{\bar{M}}}[\Pi_{A_{\bar{M}}}(m - k)] \right\}
\end{split}
\end{align}
where $A_{\bar{M}}$ is the complement of $A_M$ (such that $A = A_M A_{\bar{M}}$. This is clearly equivalent to the Hilbert space spanned by the projector $\sum_{k = 0}^{\max{m} = \tilde{n}_+} \Pi_M(k)$, i.e., the set of all possible ways of distributing \emph{up to} $\tilde{n}_+$ photons among $M$ modes. The dimension of this reduced space, denoted $d_{A_M}$, is then given by
\begin{align}
\begin{split}
    d_{A_M} &= \sum_{k = 0}^{\tilde{n}_+} {k + M - 1 \choose k} \\
    &= {\tilde{n}_+ + M \choose \tilde{n}_+}
\end{split}
\end{align}
Interestingly, the dimensions within the typical subspace of $A$ are submultiplicative in following sense;
\begin{equation}
    d_{XY} \le d_X d_Y.
\end{equation}
In fact, there is an exponential gap between $d_X d_Y$ and $d_{XY}$ (proven next). This has significant ramifications when it comes to bounding quantities involving dimensions within the typical subspace. The underlying reason is that typical sequences on $XY$ will also always be typical when considering $X$ and $Y$ individually, while typical sequences on $X$ and $Y$ individually may not be typical on $XY$ when considered together. This is therefore a generic feature when considering reduced subspaces in the typical subspace.

For a simple example, imagine we had a subspace $T_2$ that had a single photon shared between two modes, so $T_2 = \textrm{span}\{\ket{0,1}, \ket{1,0}\}$. Now consider the reduced subspace obtained by tracing out one of the modes, denoted $T_1$. It can have either zero photons or one photon, and is hence given by $T_1 = \textrm{span}\{\ket{0}, \ket{1}\}$. But now we see that $T^{\otimes 2}_1 = \textrm{span}\{\ket{0,0}, \ket{0,1}, \ket{1,0}, \ket{1,1} \}$ and hence $T_2 \subset T_1^{\otimes 2}$. This means combining reduced subspaces produces a larger space than we started with, which implies submultiplicity.
\\
\\
We can employ Stirling's approximation to understand how the dimension scales with the number of modes; the binomial coefficient ${aN \choose bN}$ for $b < a$ reads
\begin{equation}
\label{sup_mat_eq:binomial_stirling}
    {aN \choose bN} \approx \sqrt{\frac{a}{2\pi b (a-b) N}} \, 2^{aN H_2(\frac{b}{a})}.
\end{equation}
where $H_2(x) = -x\log(x) - (1-x) \log(1-x)$ is the binary entropy function. We now drop the ceiling and floor functions from $\tilde{n}_{\pm}$ and assume that $N (\bar{n}_z \pm \frac{\delta}{|\log(|z|^2)|})$ is an integer - for suitably large $N$, rational $\delta$ and rational $\bar{n}_z$, this is always possible. We then define
\begin{equation}
    \tilde{m}_{\pm} = \bar{n}_z \pm \frac{\delta}{|\log(|z|^2)|}
\end{equation}
such that $\tilde{n}_{\pm} = \tilde{m}_{\pm} N$. For any given energy, a suitably small $\delta$ can be chosen such that $\frac{\delta}{|\log(|z|^2)|} \ll \bar{n}_z$ and hence in this limit $\tilde{m}_{\pm}$ effectively characterises the energy $\bar{n}_z$.

The dimension of the full typical subspace on $A$ is
\begin{equation}
    d_A = {(\tilde{m}_+ + 1) N \choose \tilde{m}_+ N} \left( 1 - \frac{\tilde{m}_- N}{(\tilde{m}_- + 1)N} \frac{{(\tilde{m}_- + 1)N \choose \tilde{m}_- N} }{{(\tilde{m}_+ + 1) N \choose \tilde{m}_+ N}}  \right)
\end{equation}
From Eq.~\eqref{sup_mat_eq:binomial_stirling}, this is approximately equal to
\begin{equation}
    d_A \approx \sqrt{\frac{\tilde{m}_+ + 1}{2\pi \tilde{m}_+ N}} 2^{N (\tilde{m}_+ + 1) H_2(\frac{\tilde{m}_+}{\tilde{m}_+ + 1})} \, \times \, \left(1 - \textrm{O}(2^{-N \big(H_{\textrm{therm}}(\tilde{m}_+) - H_{\textrm{therm}}(\tilde{m}_-) \big) } \right) 
\end{equation}
where $H_{\textrm{therm}}(\tilde{m}) = (\tilde{m} + 1) \log(\tilde{m} + 1) - \tilde{m} \log(\tilde{m})$ is the entropy of a thermal state with expected photon number $\tilde{m}$ (this follows from the relation $(x+1) H_2(\frac{x}{x+1}) = H_{\textrm{therm}}(x)$). The term of $\textrm{O}(2^{-N (...)})$ is exponentially smaller than 1 and hence can be dropped in the large $N$ limit. Hence the full dimension is approximately
\begin{equation}
\label{sup_mat_eq:dimension_in_full_space_dA}
    d_A \approx \sqrt{\frac{\tilde{m}_+ + 1}{2\pi \tilde{m}_+ N}} \, 2^{N H_{\textrm{therm}}(\tilde{m}_+)}
\end{equation}
This scales as expected from the theory of typical subspaces, namely $\textrm{dim}(T_N^\delta (\rho)) \le 2^{N (H(\rho) + \delta)}$.

Similarly, for a subsystem consisting of $X = x N$ modes for $0 < x < 1$, the reduced typical subspace dimension is
\begin{align}
\begin{split}
\label{sup_mat_eq:dimension_in_reduced_space_dX}
    d_{X} &= {(\tilde{m}_+ + x) N \choose \tilde{m}_+ N} \\
    &\approx \sqrt{\frac{\tilde{m}_+ + x}{2\pi \tilde{m}_+ N}} \, 2^{N (\tilde{m}_+ + x) H_2(\frac{\tilde{m}_+}{\tilde{m}_+ + x})}
\end{split}
\end{align}
In the high-energy limit, where $\tilde{m}_+ \gg 1$, this exponent scales as
\begin{align}
\begin{split}
\label{sup_mat_eq:high_energy_limit_reduced_dimension_exponent}
    (\tilde{m}_+ + x) H_2\left(\frac{\tilde{m}_+}{\tilde{m}_+ + x}\right) &= (\tilde{m}_+ + x) \log(\tilde{m}_+ + x) - \tilde{m}_+ \log(\tilde{m}_+) - x\log(x)\\
    &\approx x\log(\tilde{m}_+) - x\log(x) + \frac{x}{\ln(2)} + \textrm{O}(\tilde{m}_+^{-1}) \\
    &\approx x H_{\textrm{therm}}(\tilde{m}_+) - x \log(x) + \textrm{O}(\tilde{m}_+^{-1})
\end{split}
\end{align}
where in line 2 we used $\log(1 + x/\tilde{m}_+) = \frac{x}{\ln(x)\tilde{m}_+} + \textrm{O}(\tilde{m}_+^{-1})$ and in line 3 we used $H_{\textrm{therm}}(\tilde{m}_+) = \log(\tilde{m}_+) + \frac{1}{\ln(2)} + \textrm{O}(\tilde{m}_+^{-1})$.

\subsubsection{Submultiplicative dimensions in the typical subspace}

\begin{figure}
    \centering
    \includegraphics[width=0.85\linewidth]{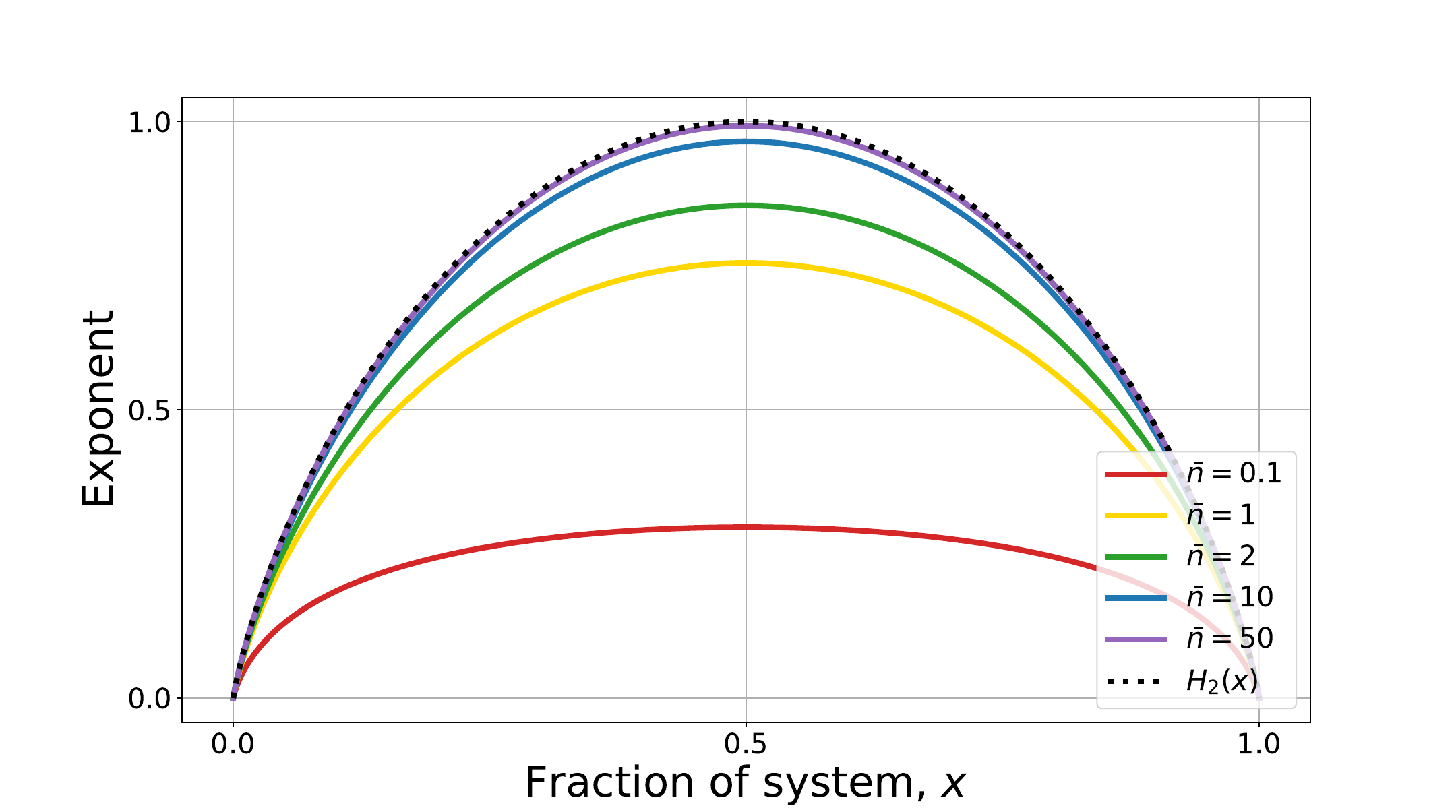}
    \caption{The exponent component of $\frac{d_{A_X}d_{A_Y}}{d_A} = \textrm{poly}(N) \, 2^{N \, \textrm{exponent}}$ where the exponent is given in Eq.~\eqref{sup_mat_eq:submultiplicative_of_dimension}. The system $X$ has $xN$ modes, while $Y$ has the other $(1-x)N$ modes. The curve is positive for every $\bar{n}$ with the exponent increasing as the energy increases, until it asymptotically approaches the binary entropy function, $H_2(x)$. The positivity of the exponent means that $d_{A_X} d_{A_Y}$ is exponentially larger than $d_A$, which in turn implies that the submultiplicativity of the dimensions has a significant impact. The exponent peaks at $x = \frac{1}{2}$ when the system is evenly bipartitioned}
    \label{sup_mat_fig:submultiplicative_dimension}
\end{figure}

Dimensions within the typical subspace are submultiplicative in the sense that $d_{XY} \le d_X d_Y$. The reason for this is that states that look typical in the spaces $X$ and $Y$ individually may not look typical on the full space $XY$ when brought together. We now employ the Stirling formula approximations of $d_A$ and the reduced $d_X$ to prove that the full dimension is exponentially smaller than the product of the reduced dimensions in the typical subspace. Let $X$ consist of $xN$ and $Y$ of a different $yN$ modes, and let $x + y = 1$ such that the Hilbert space $XY$ is the entire typical subspace on $A$. Note that an identical argument can be made where $XY$ is itself a reduced space of the typical subspace.
\begin{align}
\begin{split}
\label{sup_mat_eq:submultiplicative_of_dimension}
    \frac{d_{X} d_{Y}}{d_A} &= \textrm{poly}(N) \, 2^{N \Big( (\tilde{m}_+ + x) H_2\left(\frac{\tilde{m}_+}{\tilde{m}_+ + x}\right) + (\tilde{m}_+ + y) H_2\left(\frac{\tilde{m}_+}{\tilde{m}_+ + y}\right)  - (\tilde{m}_+ + 1) H_2 \left(\frac{\tilde{m}_+}{\tilde{m}_+ + 1}\right) \Big) } \\
    & = \textrm{poly}(N) \, 2^{N \Big( (\tilde{m}_+ + x) \log(\tilde{m}_+ + x) + (\tilde{m}_+ + y) \log(\tilde{m}_+ + y) - (\tilde{m}_+ + 1) \log(\tilde{m}_+ + 1) - x\log(x) - y \log(y)  \Big)}
\end{split}
\end{align}
where we used $(\tilde{m}_+ + z) H_2(\tilde{m}_+ / (\tilde{m}_+ + z)) = (\tilde{m}_+ + z) \log(\tilde{m}_+ + z) - \tilde{m}_+ \log(\tilde{m}_+) - z \log(z)$ in the second line. This exponent is positive for every value of $x$ and $\tilde{m}_+$, indicating that $d_X d_Y$ is exponentially larger than $d_{XY}$ in the large $N$ limit. We plot the exponent in Fig.~\ref{sup_mat_fig:submultiplicative_dimension}, whose positivity demonstrates that the submultiplicity of dimensions is a significant effect that needs to be properly accounted for when considering reduced spaces in typical subspaces.

One can directly see this by looking at the high energy limit of Eq.~\ref{sup_mat_eq:high_energy_limit_reduced_dimension_exponent}, where the exponent of $d_X d_Y / d_{XY}$ is given by
\begin{equation}
\label{sup_mat_eq:submultiplicity_high_energy_limit}
    \frac{d_X d_Y}{d_{XY}} = \textrm{poly}(N) \, 2^{N \big[(x + y - 1) H_{\textrm{therm}}(\tilde{m}_+) - x\log(x) - y\log(y) \big]}
\end{equation}
Since $x+y = 1$, this gives the residual exponent $-x\log(x) - y\log(y) = H_2(x)$ which is clearly positive for all $0< x < 1$. In Fig~\ref{sup_mat_fig:submultiplicative_dimension}, we plot the full exponent from Eq.~\eqref{sup_mat_eq:submultiplicative_of_dimension}, which is always positive and approaches the binary entropy function as the energy increases,.

\subsection{Scrambling in the typical subspace}

In standard finite-dimensional quantum systems, scrambling unitaries distribute initially localised information nonlocally across the entire Hilbert space. The archetypic example of such an ensemble of unitaries would be the Haar ensemble. Attempting to extend the notion of Haar random unitaries to continuous-variable systems presents issues; a CV state initially localised to some finite-energy subspace would necessarily be transformed into an unnormalised, infinite energy state if one attempted to implement the exact analogous Haar random unitary. Even realising a \emph{unitary} 2-design is not possible in CV systems as proven by the impossibility of having quantum \emph{state} 2-designs \cite{blume2014curious}, although approximate energy-constrained unitary 2-designs are not ruled out \cite{PhysRevX.14.011013, PhysRevA.99.062334}. This fundamentally arises from the infinite-dimensional Hilbert space for a finite number of modes.

To get around this, we consider scrambling the state within the typical subspace of the state, an idea that was introduced in Ref.~\cite{Zhong2023information}. Since the typical subspace is finite-dimensional, we can scramble information within it without getting non-physical infinite energy states, and we can also define the ensemble of Haar random unitaries within this space.
In this framework, rather than scrambling information across the entire infinite-dimensional Hilbert space, we scramble the information within the region of the Hilbert space that the state already occupies.

For our entanglement-assisted case, we choose to implement the operator $\mathds{1}_{R} \otimes U_A \otimes \mathds{1}_{\bar{B}}$ where $U_A$ is drawn from the Haar measure of unitary operators within the typical subspace of $A$. The initial state is an $N$-fold tensor product of two-mode squeezed vacua $\ket{z}^{\otimes N}_{RA\bar{B}}$ (recall that the reduced state on $A$ is a tensor product of thermal state $\tr_{R\bar{B}} [(\ketbra{z}^{\otimes N})_{RA\bar{B}}] = (\rho_z^{\otimes N})_A$. The Haar average output state of $U_A \ket{z}^{\otimes N}_{RA\bar{B}}$ can be calculated using Weingarten calculus \cite{weingarten1978asymptotic, collins2006integration}.
\begin{equation}
    \mathbb{E}_U [U_A \ketbra{z}{z}_{RA\bar{B}} U_A^\dag] = \rho_{R\bar{B}} \otimes \bar{\rho}_A, \quad \bar{\rho}_A = \frac{(\Pi_{N}^\delta)_A}{d_A}
\end{equation}
where $\bar{\rho}_A$ is the maximally mixed state on the typical subspace of $A$, $\Pi_N^\delta$ is the typical subspace projector on $N$ modes and $\rho_{R \bar{B}} = (\rho_z^{\otimes N})_{R \bar{B}}$ is the reduced state on $R\bar{B}$. Notice that this has removed all correlations between $A$ and $R \bar{B}$.

The reduced density matrix on $A_X$ consisting of $X < N$ modes (whose complement is $A_Y$ such that $A = A_X A_Y$), is given by $\bar{\rho}_{A_X} = \tr_{A_Y}[\bar{\rho}_A]$. This is computed as
\begin{align}
\begin{split}
\label{sup_mat_eq:average_state_maximially_mixed_typical_subspace}
    \bar{\rho}_{A_X} &= \tr_{A_Y} \left[\frac{(\Pi_N^\delta)_A}{d_A}\right] \\
    &= \frac{1}{d_A} \tr_{A_Y} \left[ \sum_{n = \tilde{n}_-}^{\tilde{n}_+} \sum_{m = 0}^n \Pi_{X, A_X}(m) \otimes \Pi_{N-X, A_Y}(n - m)  \right] \\
    &= \frac{1}{d_A} \sum_{n = \tilde{n}_-}^{\tilde{n}_+} \sum_{m = 0}^n {n - m + N - X - 1 \choose n - m} \Pi_{X, A_X}(m)
\end{split}
\end{align}
where $\Pi_M(m)$ is the projector onto the $m$ photon subspace in $M$ modes. Unlike for DV maximally mixed states over many qudits, this expression is \emph{not} equal to the reduced projector onto the typical subspace associated with $A_X$, i.e., $\bar{\rho}_{A_X} \ne (\Pi_{X}^\delta)_{A_X} / d_{A_X}$.

The reason for this is because the maximally mixed state over a typical subspace that obeys a global conservation rule, i.e., that the total number of photons in each term of the wavefunction must be in the set  $\{\tilde{n}_-, \tilde{n}_- + 1, \dots, \tilde{n}_+\}$. These conservation rules mean that the projector does not factorise cleanly and hence $\Pi_{N, A}^\delta \ne \Pi_{X A_X}^\delta \otimes \Pi_{Y, A_Y}^\delta$. This is also the cause of the submultiplicity.

\subsection{Bounding the average distance to a decoupled state}
In this subsection, we bound the average distance to the average, decoupled state. In both the standard and entanglement-assisted cases, the average state on $RE$ is given by
\begin{equation}
    \mathbb{E}_{U, \psi_0} \left [\tr_{X \notin RE}[U_A \ketbra{\psi_0} U_A^\dag]  \right] = \rho_R \otimes \bar{\rho}_E
\end{equation}
where $\ket{\psi_0}$ dictates which case we are considering; if $\ket{\psi_0} = \ket{z}^{\otimes N}_{RA\bar{B}}$, then we are in the entanglement-assisted case and are not averaging over the initial state, whereas if $\ket{\psi_0} = \ket{z}_{RA_1}^{\otimes K} \ket{\bm{\gamma}}_{A_2}^{N-K}$ then we are in the standard case and we average over the coherent state samples $\bm{\gamma}$.

This average state is clearly decoupled. For a particular choice of $U$ and $\ket{\psi_0}$, the reduced state on $RE$ is given by $\rho_{RE}(U, \psi_0)$. If this state is ``close'' in trace distance to the average decoupled state, then this implies $\rho_{RE}(U, \psi)$ is also approximated decoupled. We will now bound the expected trace distance between $\rho_{RE}(U, \psi)$ and $\rho_R \otimes \bar{\rho}_E$; we derive the result
\begin{equation}
\label{sup_mat_eq:average_distance_to_average_state_1}
    \mathbb{E}_{U, \psi_0} || \rho_{RE}(U, \psi_0) - \rho_R \otimes \bar{\rho}_E ||_1 \le \textrm{poly}(N) \, 2^{-N  \xi_\ell(\bar{n}, p)}
\end{equation}
Recall that $p$ is the probability of erasure and $\bar{n}$ is the expected photon number of the $K = qN$ TMSV states that Alice is trying to communicate to Bob. We derive the function $\xi$ which is labelled with $\ell \in \{\textrm{standard}, \textrm{entanglement-assisted}\}$. In the large $N$ limit, if $\xi > 0$ then this implies decoupling has approximately taken place, and hence Bob can decode Alice's $K$ TMSV states. 

Notice that Eq.~\eqref{sup_mat_eq:average_distance_to_average_state_1} is equal to the variance in state space of the random state $\rho_{RE}(U, \psi_0)$, which has a distribution centered on $\rho_R \otimes \bar{\rho}_E$;
\begin{align}
\begin{split}
     \textrm{Var} [\rho_{RE}] &= \mathbb{E}_{U, \psi_0} \big[||\rho_{RE}(U, \psi_0) - \mathbb{E}_{U, \psi_0} \left[ \rho_{RE}(U, \psi_0) \right] ||_1 \big] \\
     &= \mathbb{E}_{U, \psi_0} ||\rho_{RE}(U, \psi_0) - \rho_{R} \otimes \bar{\rho}_E ||_1
\end{split}
\end{align}
We can therefore understand Eq.~\eqref{sup_mat_eq:average_distance_to_average_state_1} as bounding how concentrated the distribution of random states is around the average decoupled state.

\subsubsection{Trace norm to Hilbert-Schmidt norm and Jensen's inequality}
In order to use the inequality $||X||_1 \le \sqrt{\textrm{rank}(X)} ||X||_2$, we require the object $X$ to be an element of a Hilbert space with finite rank. If $X$ is an infinite-dimensional operator (such as $\rho_{RE}$) and $\Pi$ is a projector of finite rank (such as $(\Pi_M^\delta)_{E}$), it follows that $||\Pi X \Pi ||_1 \le \sqrt{\textrm{rank}(\Pi)} ||\Pi X \Pi ||_2 = \sqrt{\textrm{rank}(\Pi) \tr[\Pi X^\dag \Pi X \Pi]}$. We now apply this to the left hand side of Eq.~\eqref{sup_mat_eq:average_distance_to_average_state_1}.

First, recall that we are projecting the state in $A$ onto the typical subspace $T_N^\delta (\rho_z)$. We can apply this to the initial state in the entanglement-assisted case to make the entire state finite-dimensional. While the dimension of an arbitrary state $\ket{\psi_0}_{R\bar{B}A}$ projected onto $\mathds{1}_{R\bar{B}} \otimes (\Pi_{N}^\delta)_A$ need not be finite (e.g; $\ket{\psi_0} = \ket{\sigma_1}_{RB}\ket{\sigma_2}_A$ will continue to be infinite-dimensional because of the component on $R\bar{B}$), in our case the photon numbers in $R \bar{B}$ are perfectly correlated with those in $A$. Therefore, projecting the state onto the typical subspace of $A$ ends up being equivalent to applying the typical subspace projector onto either just $R\bar{B}$, or both $A$ and $R\bar{B}$, i.e.,
\begin{equation}
    \Big(\mathds{1}_{R\bar{B}} \otimes \Pi^\delta_{N \, A} \Big) \ket{z}^{\otimes N}_{R\bar{B} A} = \Big( \Pi_{N \, R \bar{B}}^\delta \otimes \mathds{1}_A \Big) \ket{z}^{\otimes N}_{R \bar{B} A} = \Big(\Pi_{N \, R\bar{B}}^\delta \otimes \Pi_{N \, A}^\delta \Big) \ket{z}_{R\bar{B} A}^{\otimes N}
\end{equation}
The analogous results apply in the standard case, but with the state on $R$ being projected onto the \emph{reduced} typical subspace with projector $(\Pi_K^\delta)_R$. Therefore, the rank of $\Pi_{N \, A}^\delta \ket{\psi_0}$ is finite.

Since the random unitary $U_A$ scrambles information within the typical subspace, it does not mix basis states inside the typical subspace with those outside it. Therefore, we can commute $(\Pi_{N}^\delta)_A$ through $U_A$ while maintaining all information about the initial state confined to the typical subspace. Applying all this with the above trace norm to Hilbert-Schmidt norm to the left hand side of Eq.~\eqref{sup_mat_eq:average_distance_to_average_state_1} allows us to derive
\begin{align}
\begin{split}
\label{sup_mat_eq:trace_norm_to_hilbert_schmidt_norm}
    \mathbb{E}_{U, \psi_0} || \rho_{RE}&(U, \psi_0) - \rho_R \otimes \bar{\rho}_E||_1 \le \mathbb{E}_{U, \psi_0} \sqrt{\textrm{rank} \left((\Pi_{K}^\delta)_R \otimes (\Pi_{M}^\delta)_E \right)} \, ||\rho_{RE}(U, \psi_0) - \rho_R \otimes \bar{\rho}_E ||_2 \\
    &\le \sqrt{d_{RE} \mathbb{E}_{U, \psi_0} \tr_{RE} \left[\left(\Big((\Pi_K^\delta)_R \otimes (\Pi_M^\delta)_E\Big) \circ \Big(\rho_{RE}(U, \psi_0) - \rho_R \otimes \bar{\rho}_E\Big) \right)^2\right] } \\
    &= \sqrt{d_{RE} \left( \mathbb{E}_{U, \psi_0} \left[\tr_{RE} [\left(\rho_{RE}^\delta (U, \psi_0) \right)^2] \right] - \tr_{RE}\left[(\rho_R^\delta \otimes \bar{\rho}_E^\delta)^2 \right] \right)}
\end{split}
\end{align}
where in the first line we bounded the trace norm with the Hilbert-Schmidt norm, in the second line we used the fact that $d_{RE} = \textrm{rank}(\Pi_K^\delta \otimes \Pi_M^\delta)$ (recall that $M = pN$ is the number of erased modes), used $X \circ \rho = X \rho X^\dag$ we also used Jensen's inequality to bring the expectation value inside the square root, and in the third line we introduced the notation $\rho^\delta = \Pi^\delta \rho \Pi^\delta$ to indicate that $\rho$ has been projected into the typical subspace.

Notice that the term on the right in the final line of Eq.~\eqref{sup_mat_eq:trace_norm_to_hilbert_schmidt_norm} is the purity of the average state $\rho_{R} \otimes \bar{\rho}_E$ - we now proceed to calculate this.

\subsubsection{Purity of the average reduced state}
In both the standard and entanglement-assisted cases, the average state on $E$ is given in Eq.~\eqref{sup_mat_eq:average_state_maximially_mixed_typical_subspace} by setting $X = E$, while the reduced state on $R$ is a tensor product of $K$ thermal states,
\begin{equation}
    \mathbb{E}_{U, \psi_0} \left[ \tr_{X \notin RE}[ U_A \ketbra{\psi_0} U_A^\dag ]\right] = (\rho^{\otimes K}_z)_R \otimes \bar{\rho}_E, \quad \bar{\rho}_E = \frac{1}{d_A} \tr_{B}[(\Pi^\delta_{N})_A]
\end{equation}
The purity of a single-mode thermal state $\rho_z = (1 - |z|^2)\sum_{n = 0}^\infty |z|^{2n} \ketbra{n}$ is $\frac{1 - |z|^2}{1 + |z|^2}$. We denote the purity of the $K$ single-mode thermal states projected onto the typical subspace $\Pi_K^\delta$ as $\beta_K$. We can analytically calculate it by using the decomposition $\rho_z^{\otimes K} = (1 - |z|^2)^K \sum_{n = 0}^\infty |z|^{2n} \Pi_K(n)$, where recall we defined $\Pi_K(n)$ as the projector onto the $n$ photon subspace in $K$ modes.
\begin{align}
\begin{split}
\label{sup_mat_eq:thermal_state_purity_typical_subspace}
    \beta_{K} &= \tr\left[ \left((\rho_z^{\otimes K})^\delta\right)^2\right] \\
    &= (1 - |z|^2)^{2K} \sum_{n, n' = 0}^\infty |z|^{2(n + n')} \tr\left[ \Pi_K^\delta \Pi_K(n) \Pi_K^\delta \Pi_K(n') \right] \\
    &= (1 - |z|^2)^{2K} \sum_{n = 0}^{\tilde{n}_+} |z|^{4n} {n + K - 1 \choose n}
\end{split}
\end{align}
This expression holds for any $1 \le K \le N - 1$. If we took $K = N$, then $\beta$ would be obtained by changing the range of the sum from $n \in \{0, 1, \dots, \tilde{n}_+\}$ to $n \in \{\tilde{n}_-, \dots, \tilde{n}_+\}$. This has an analytic representation in terms of Gaussian hypergeometric functions. For future reference, we can use the submultiplicity and finite rank of $\Pi_K^\delta$ to upper bound $\beta_K$ by
\begin{align}
\begin{split}
    \beta_K &\le \beta_1^K \\
    &\le \left(\frac{1 - |z|^2}{1 + |z|^2}\right)^K.
\end{split}
\end{align}
But we leave it as the exact $\beta_K$ for the time being.

We now calculate the purity of the average state $\bar{\rho}_{E}$; note that this state is already confined to the typical subspace so we can drop the $\delta$ superscript. 
\begin{align}
\begin{split}
\label{sup_mat_eq:reduced_maximally_mixed_state_purity}
    \tr[\bar{\rho}_E^2] &= \frac{1}{d_A^2} \sum_{n, n' = \tilde{n}_-}^{\tilde{n}_+} \sum_{m = 0}^n \sum_{m' = 0}^{m'} \tr[\Pi_M(m) \Pi_M(m')] {n - m + N - M - 1 \choose n - m} {n' - m' + N - M - 1 \choose n' - m'} \\
    &= \frac{1}{d_A^2} \sum_{n, n' = \tilde{n}_-}^{\tilde{n}_+} \sum_{m = 0}^\infty {m + M - 1 \choose m} {n - m + N - M - 1 \choose n - m } {n' - m + N - M - 1 \choose n' - m} \\
    &= \frac{1}{d_A^2} \alpha_E
\end{split}
\end{align}
where we made use of the typical subspace projector decomposition $\Pi_N^\delta = \sum_{n = \tilde{n}_-}^{\tilde{n}_+} \sum_{m = 0}^n \Pi_M(m) \otimes \Pi_{N-M}(n - m)$ and defined $\alpha_E$ as the double sum from penultimate line (where $E$ has $M$ modes in it). We have been unable to find an analytic. In Sec.~\ref{sup_mat_sec:bounding_alpha_X_purity_reduced_state}, we bound the expression for $\alpha_X$ consisting of $X$ modes.

Despite the average state on $A$ being the maximally mixed state over the typical subspace, the reduced average state on $E$ is not maximally mixed over the reduced typical subspace. This again arises from the submultiplicity of dimensions in the typical subspace and means the standard bounding techniques for these sort of calculations (e.g, seen in \cite{hayden2008decoupling, dupuis2010decoupling, Zhong2023information}) are insufficient.

Combining the purities from Eq.~\eqref{sup_mat_eq:reduced_maximally_mixed_state_purity} and Eq.~\eqref{sup_mat_eq:thermal_state_purity_typical_subspace} gives the average state purity in the typical subspace, where $\beta_R = \beta_K$.
\begin{equation}
\label{sup_mat_eq:average_state_purity}
    \tr_{RE} \left[\big( \rho_R^\delta \otimes \bar{\rho}_E^\delta \big)^2 \right] = \frac{\alpha_E \beta_K}{d_A^2}
\end{equation}

\subsubsection{Second moment of the reduced state I}
We now calculate the second moment $\mathbb{E}_{U, \psi_0} \tr_{RE} \left[ \big( \rho_{RE}^\delta(U, \psi_0) \big)^2\right]$. To do this, we employ the swap trick \cite{mele2024introduction} $\tr[X_{\mathcal{H}}^2] = \tr[\hat{F}_{\mathcal{H}, \mathcal{H}'} \, X_{\mathcal{H}} \otimes X_{\mathcal{H}'}]$ where $\hat{F}_{\mathcal{H}, \mathcal{H}'} = \sum_{i, j} \ketbra{i, j}{j, i}_{\mathcal{H}, \mathcal{H}'}$ is the swap operator between two isomorphic Hilbert spaces $\mathcal{H}$ and $\mathcal{H}'$. We carry out this section of the derivation in the standard case, but it equally applies to the entanglement-assisted case by including an additional $\bar{B} \bar{B}'$ subspace in the final line.
\begin{align}
\begin{split}
    \label{sup_mat_eq:swap_trick_setup}
    \tr_{RE} [\big( \rho_{RE}^\delta (U, \psi_0) \big)^2] &= \tr_{RR' EE'} \left[ \hat{F}_{RR'} \otimes \hat{F}_{EE'} \, \Big( \rho_{RE}^\delta (U, \psi_0) \otimes \rho_{R'E'}^\delta (U, \psi_0)  \Big)  \right] \\
    &= \tr_{RR' EE'} \left[ \hat{F}_{RR'} \otimes \hat{F}_{EE'} \Bigg( \Big(\Pi_{RE}^\delta \tr_{B} [U \ketbra{\psi_0}U^\dag] \Pi_{RE}^\delta \Big) \otimes \Big(\Pi_{R'E'}^\delta \tr_{B'} [U \ketbra{\psi_0} U^\dag] \Pi_{R'E'}^\delta \Big) \Bigg)  \right] \\
    &= \tr_{RR' AA'} \left[ \hat{F}_{RR'}^\delta \otimes \hat{F}_{EE'}^\delta \otimes \mathds{1}_{BB'}^\delta \, \Big(U \ketbra{\psi_0^\delta} U^\dag \Big)^{\otimes 2}_{RR'AA'}   \right] \\
    &= \tr_{RR'AA'} \left[ \hat{F}^\delta_{RR'} \otimes \Big( U^{\dag \otimes 2} \big( \hat{F}_{EE'}^\delta \otimes \mathds{1}_{BB'}^\delta \big) U^{\otimes 2} \Big) \, \big( \ketbra{\psi_0^\delta}^{\otimes 2} \big)_{RR' AA'} \right]
\end{split}
\end{align}
where we used $A = BE$, the operator $U$ acts only on $A$ and $A'$, and the state $\ket{\psi_0^\delta} = (\Pi_N^\delta)_A \ket{\psi_0}$ is projected onto the typical subspace, and we used the notation $\Pi_{\mathcal{H}}^\delta$ being the projector on the typical subspace in the Hilbert space $\mathcal{H}$ (i.e., dropped the label with the number of modes, allowing us to join Hilbert spaces together such that $\Pi_{AA'}^\delta = (\Pi_N^\delta)_A \otimes (\Pi_N^\delta)_{A'}$). The entanglement-assisted case is given by transforming $\hat{F}_{RR'}^\delta \rightarrow \hat{F}_{RR'}^\delta \otimes \mathds{1}_{\bar{B} \bar{B}'}^\delta$ and having $\ket{\psi_0}_{RA}\rightarrow \ket{\psi_0}_{RA\bar{B}}$ in the final line.

We thus have an expression for $\tr_{RE}[\big(\rho^\delta_{RE}(U, \psi_0)\big)^2]$ for a specific choice of $U$ and $\psi_0$. We now average this expression over the set of unitaries $U$ using Weingarten calculus \cite{weingarten1978asymptotic, collins2006integration, collins2022weingarten}. To do this, we need to calculate $\mathbb{E}_U [U^{\otimes 2} \hat{X} \,U^{\otimes 2}]$ with operator $\hat{X} = \Pi_{AA'}^\delta \circ \big( \hat{F}_{EE'} \otimes \mathds{1}_{BB'}\big)$. The finite rank of $\Pi_{AA'}^\delta$ means we can employ the double-twirling formula \cite{collins2006integration} (which itself is derived from Eq.~\eqref{eq:collins_sniady_weingarten_calculus_moment_result}) because $\hat{X}$ is finite dimensional.
\begin{align}
\begin{split}
\label{sup_mat_eq:double_twirling_formula}
    \mathbb{E}_U [U^{\dag \otimes 2} \hat{X} \, U^{\otimes 2}] = \Big( \tr_{AA'}[\hat{X}] - \frac{1}{d_A^2} \tr_{AA'} \left[ \hat{F}_{AA'} \hat{X} \right] \Big) \frac{\mathds{1}_{AA'}^\delta}{d_A^2 - 1} \, + \, \Big( \tr_{AA'} \left[ \hat{F}_{AA'} \hat{X} \right] - \frac{1}{d_A^2} \tr_{AA'} [\hat{X}] \Big) \frac{\hat{F}^\delta_{AA'}}{d_A^2 - 1}
\end{split}
\end{align}
Next, we compute the coefficient $\tr[\hat{X}]$ by bipartitioning the projector into $\Pi^\delta_N = \sum_{n = \tilde{n}_-}^{\tilde{n}_+} \sum_{m = 0}^n \Pi_M(m) \otimes \Pi_{N-M}(n-m)$.
\begin{align}
\begin{split}
\label{sup_mat_eq:tr(X)_step1}
    \tr [\hat{X}] &= \tr_{AA'}\left[ \Pi_{AA'} \big( \hat{F}_{EE'} \otimes \mathds{1}_{BB'}\big)  \right] \\
    &= \sum_{n, n' = \tilde{n}_-}^{\tilde{n}_+} \sum_{m = 0}^n \sum_{m' = 0}^{n'} \tr_{EE'} \left[ \big((\Pi_M(m))_E \otimes (\Pi_M(m'))_{E'} \big) \hat{F}_{EE'} \right] \times \tr_{BB'} \left[(\Pi_{N-M}(n-m))_B \otimes (\Pi_{B'}(n' - m'))_{B'} \right]
\end{split}
\end{align}
Here, the trace over the $BB'$ subspace gives the rank of the projectors $\Pi_{N-M}(n-m)$, i.e., ${n - m + N - M - 1 \choose n - m} {n' - m' + N - M - 1 \choose n' - m'}$. To calculate the trace over the $EE'$ subspace, we express the $m$ photon subspace projector, $\Pi_M(m)$, in terms of sums over photon number strings $\ket{\bm{x}} = \ket{x_1, x_2, \dots, x_M}$. This can be done by using the set of all ways of distributing $m$ indistinguishable items among $M$ bins, denoted $\bm{P}(m, M)$. With this, we can write $\Pi_M(m) = \sum_{\bm{x} \in \bm{P}(m, M)} \ketbra{\bm{x}}$. Similarly, we can write the swap operator as $\hat{F}_{E E'} = \sum_{\bm{p}, \bm{q} \in \mathbb{N}_0^{M}} \ketbra{\bm{p}, \bm{q}}{\bm{q}, \bm{p}}_{EE'}$. This gives the trace over the $EE'$ subspace for a specific choice of $m$ and $m'$ as
\begin{align}
\begin{split}
    \tr_{EE'}\left[ \big((\Pi_{M}(m))_E \otimes (\Pi_{M}(m'))_{E'} \big) \hat{F}_{EE'} \right] &= \sum_{\bm{x} \in \bm{P}(m, M)} \sum_{\bm{y} \in \bm{P}(m', M)} \sum_{\bm{p}, \bm{q} \in \mathbb{N}_0^{M}} \braket{\bm{p}, \bm{q}}{\bm{x}, \bm{y}} \braket{\bm{x},\bm{y}}{\bm{q}, \bm{p}} \\
    &= \delta_{m, m'} {m + M - 1 \choose m}    
\end{split}
\end{align}
where final line follows from simply reducing the first line and noting that $|\bm{P}(m, M)| = {m + M - 1 \choose m}$.

Plugging these two expressions back into $\tr[\hat{X}]$ from Eq.~\eqref{sup_mat_eq:tr(X)_step1} and noting that ${n - m + N - M - 1 \choose n - m}$ is zero for $m > n$ yields
\begin{align}
\begin{split}
    \tr_{AA'} [\hat{X}] &= \sum_{n, n' = \tilde{n}_-}^{\tilde{n}_+} \sum_{m = 0}^{\infty} {m + M - 1 \choose m} {n - m + N - M - 1 \choose n - m} {n' - m + N - M - 1 \choose n' - m}
\end{split}
\end{align}
Perhaps surprisingly, this is proportional to $\alpha_E$ from the purity of the reduced maximally mixed state, $\alpha_E = d_A^2 \tr[\bar{\rho}_E^2]$, in Eq.~\eqref{sup_mat_eq:reduced_maximally_mixed_state_purity}. Again, this expression is upper bounded in Sec.~\ref{sup_mat_sec:bounding_alpha_X_purity_reduced_state}. Under obvious symmetry arguments, we can immediately read off that $\tr[\hat{F}_{AA'} \hat{X}] = \alpha_{B}$ where $\alpha_B$ simply relabels $M \leftrightarrow N-M$ in $\alpha_E$. 

The double-twirling formula acting on $\hat{X} = \Pi_{AA'}^\delta \circ \big( \hat{F}_{EE'} \otimes \mathds{1}_{BB'} \big)$ therefore gives
\begin{equation}
\label{sup_mat_eq:second_moment_of_X_1}
    \mathbb{E}_U \left[U^{\dag \otimes 2} \hat{X} \, U^{\otimes 2} \right] = D_1 \mathds{1}_{AA'} + D_2 \hat{F}_{AA'}, \\
\end{equation}
\begin{equation}
\label{sup_mat_eq:second_moment_of_X_2}
    D_1 = \frac{1}{d_A^2 + 1} \left(\alpha_E - \frac{\alpha_B}{d_A}\right), \quad D_2=\frac{1}{d_A^2 + 1} \left(\alpha_B - \frac{\alpha_E}{d_A}\right).
\end{equation}

\subsubsection{Second moment of the reduced state II}
Having calculated $\tr [U^{\dag \, \otimes 2} \hat{X} \, U^{\otimes 2}]$, we can now compute the second moment of the reduced state. In the standard case where $\ket{\psi_0} = \ket{z}^{\otimes K}_{RA_1} \ket{\bm{\gamma}}_{A_2}$, it gives
\begin{align}
\begin{split}
\label{sup_mat_eq:D1D2EA}
    \mathbb{E}_{U} \tr_{RE} \left[ \rho^\delta_{RE}(U, \bm{\gamma})^2 \right]
    &= D_1 \tr_{RR'AA'} \left[ \big(\hat{F}_{RR'}^\delta \otimes \mathds{1}^\delta_{AA'}\big) \big(\ketbra{\psi_0^\delta}^{\otimes 2}\big)_{RR'AA'} \right]  \\
    &\quad \quad + D_2\tr_{RR'AA'} \left[\big(\hat{F}_{RR'}^\delta \otimes \hat{F}_{AA'}^\delta\big) \big(\ketbra{\psi_0^\delta}^{\otimes 2}\big)_{RR'AA'} \right] \\
    &\le D_1 \tr_{RR'A_1A_1'} \left[ \big(\hat{F}_{RR'}^\delta \otimes \mathds{1}_{A_1A_1'}^\delta\big) \big(\ketbra{z}^{\delta \, \otimes 2K}\big)_{RR'A_1A_1'} \right] \times \tr_{A_2A_2'} \left[ \big(\ketbra{\bm{\gamma}}^{\delta \, \otimes 2}\big)_{A_2 A'_2}  \right] \\
    &\quad \quad + D_2 \tr_{RA_1} \left[ \big(\ketbra{z}^{\delta \, \otimes K}\big)^2_{RA_1} \right] \times \tr_{A_2}\left[ \big(\ketbra{\bm{\gamma}}^\delta\big)^2_{A_2}\right]
\end{split}
\end{align}
where in the final two lines we decomposed the initial state over $RA_1$ and $A_2$ and employed the submultiplicity of dimensions to bipartition the typical subspace identity, $\textrm{rank}(\mathds{1}_A) \le \textrm{rank}(\mathds{1}_{A_1}) \textrm{rank}(\mathds{1}_{A_2})$.
Similarly, in the entanglement-assisted case with $\ket{\psi_0} = \ket{z}^{\otimes N}_{RA\bar{B}}$, if we again use the submultiplicity property we derive
\begin{align}
\begin{split}
\label{sup_mat_eq:D1D2}
    \mathbb{E}_{U} \tr_{RE} \left[ \rho^\delta_{RE}(U)^2  \right] &= D_1 \tr_{RR'AA'\bar{B}\bar{B}'} \left[ \big( \hat{F}_{RR'}^\delta \otimes \mathds{1}^\delta_{AA'\bar{B}\bar{B}'} \big) \big( \ketbra{\psi_0^\delta}^{\otimes 2} \big)_{RR'AA'\bar{B}\bar{B}'}  \right] \\
    & \quad \quad + D_2 \tr_{RR' AA' \bar{B}\bar{B}'} \left[\big( \hat{F}^\delta_{RR'} \otimes \hat{F}^\delta_{AA'} \otimes \mathds{1}^\delta_{\bar{B}\bar{B}'} \big)  \big( \ketbra{\psi_0^\delta}^{\otimes 2} \big)_{RR'AA'\bar{B}\bar{B}'}  \right] \\
    &\le D_1 \tr_{RR'A_1A_1'} \left[ \big(\hat{F}_{RR'}^\delta \otimes \mathds{1}_{A_1A_1'}^\delta\big) \big(\ketbra{z}^{\delta \, \otimes 2K}\big)_{RR'A_1A_1'} \right] \times \tr_{A_2A_2'\bar{B}\bar{B}'} \left[ \big(\ketbra{z}^{\delta \, \otimes 2(N-K)}\big)_{A_2A_2' \bar{B}\bar{B}'}  \right] \\
    & \quad \quad + D_2 \tr_{RA_1} \left[\big(\ketbra{z}^{\delta \, \otimes K}\big)_{RA_1}^2\right] \times \tr_{A_2 A_2' \bar{B}\bar{B}'} \left[ \big(\hat{F}_{A_2A'_2}^\delta \otimes \mathds{1}^\delta_{\bar{B}\bar{B}'} \big) \big(\ketbra{z}^{\delta \, \otimes 2(N-K)}\big)_{A_2A_2' \bar{B}\bar{B}'} \right]
\end{split}
\end{align}

We now proceed to upper bound the various trace terms appearing in Eq.~\eqref{sup_mat_eq:D1D2EA} and Eq.~\eqref{sup_mat_eq:D1D2}. First, we take the $RR'A_1A_1'$ trace associated with the coefficient $D_1$ in both the standard and EA case.To do this, we again work in the photon number basis, where $\ket{z}^{\otimes K}_{RA_1} = (1 - |z|^2)^{\frac{K}{2}} \sum_{n = 0}^\infty z^n \sum_{\bm{x} \in \bm{P}(n, K)} \ket{\bm{x}, \bm{x}}$ and $\hat{F}_{RR'} = \sum_{\bm{p}, \bm{q} \in \mathbb{N}_0^K} \ketbra{\bm{p}, \bm{q}}{\bm{q},\bm{p}}$
\begin{align}
\begin{split}
    \tr_{RR'A_1A'_1} \Big[ \big(\hat{F}_{RR'}^\delta &\otimes \mathds{1}_{A_1 A_1'}^\delta\big) \big( \ketbra{z}^{\delta \, \otimes 2K} \big)_{RR'A_1A_1'} \Big] = (1 - |z|^2)^K \sum_{n, n', m, m' = 0}^{\tilde{n}_+} z^{n+n'} z^{* \, m + m'} \\
     &\times \sum_{\bm{x}\in\bm{P}(n, K)} \sum_{\bm{y} \in \bm{P}(n', K)} \sum_{\bm{z} \in \bm{P}(m, K)} \sum_{\bm{w} \in \bm{P}(m', K)} \sum_{\bm{p}, \bm{q} \in \mathbb{N}_0^K} \braket{\bm{p},\bm{q}}{\bm{x},\bm{y}} \braket{\bm{z}, \bm{w}}{\bm{q}, \bm{p}}_{RR'} \times \tr_{A_1 A_1'} \left[ \ketbra{\bm{x}, \bm{y}}{\bm{z}, \bm{w}}\right] \\
     &= (1 - |z|^2)^{2K} \sum_{n = 0}^{\tilde{n}_+} |z|^{4n} {n + K - 1 \choose n}
\end{split}
\end{align}
This is equal to the purity of $K$ single-mode thermal states projected into the typical subspace, which we initially encountered in Eq.~\eqref{sup_mat_eq:thermal_state_purity_typical_subspace}, and is denoted with $\beta_K$. Recall that via the submultiplicity, this is upper bounded by $\beta_K \le \left(\frac{1 - |z|^2}{1 + |z|^2}\right)^K$. 

If we look at the $A_2A_2'\bar{B}\bar{B}'$ subspace trace associated with the coefficient $D_2$ in the EA case, we can see it has the same form as above but with $K \rightarrow N-K$, and therefore
\begin{equation}
    \tr_{A_2 A'_2 \bar{B} \bar{B}'} \left[ \big( \hat{F}_{A_2A_2'}^\delta \otimes \mathds{1}^\delta_{\bar{B}\bar{B}'} \big) \big( \ketbra{z}^{\delta \, \otimes 2(N-K)}\big)_{A_2 A_2' \bar{B}\bar{B}'}\right] = \beta_{N-K}
\end{equation}

The $RA_1$ trace associated with the coefficient $D_2$ in both the standard and EA case gives
\begin{align}
\begin{split}
    \tr_{RA_1} \left[ \big( \ketbra{z}^{\delta \, \otimes K}  \big)_{RA_1}^2 \right]  &= \left| \bra{z}^{\otimes K} \left(\Pi_K^\delta \otimes \Pi_K^\delta \right) \ket{z}^{\otimes K}  \right|^2 \\
    &= \left( (1 - |z|^2)^K \sum_{n = 0}^{\tilde{n}_+} |z|^{2n} {n + K - 1 \choose n} \right)^2
\end{split}
\end{align}
which is equal to the squared overlap between $\rho_z^{\otimes K}$ and the reduced typical subspace on $K$ modes, i.e., $\Delta_K^2$ with $\Delta_K$ from Eq.~\eqref{sup_mat_eq:reduced_typical_subspace_overlap}. Via the submultiplicity, this is upper bounded by $\Delta_1^{2K}$.

Similarly, the $A_2 A_2' \bar{B} \bar{B}'$ subspace trace associated with the $D_1$ coefficient in the EA case is clearly just the squared overlap between $\ket{z}^{\otimes N-K}$ and $\Pi_{N-K}^\delta \otimes \Pi_{N-K}^\delta$, namely
\begin{align}
\begin{split}
    \tr_{A_2 A_2' \bar{B} \bar{B}'} \left[ \big(\ketbra{z}^{\delta \, \otimes 2(N-K)} \big)_{A_2A_2'\bar{B}\bar{B}'}\right] = \Delta_{N-K}^2
\end{split}
\end{align}

We now look at the 2 remaining terms which now have a $\bm{\gamma}$ dependence. The trace associated with $D_1$ is equal to the squared overlap between $\ket{\bm{\gamma}}$ and the reduced typical subspace,
\begin{align}
\begin{split}
    \tr_{A_2 A_2'} \left[ \big( \ketbra{\bm{\gamma}}^{\delta \, \otimes 2} \big)_{A_2 A_2'}  \right] &= \left| \bra{\bm{\gamma}} \Pi_{N-K}^\delta \ket{\bm{\gamma}} \right|^2
\end{split}
\end{align}
The $D_2$ trace ends up also being equal to this, 
\begin{align}
\begin{split}
    \tr_{A_2} \left[ \big( \ketbra{\bm{\gamma}}^{\delta} \big)_{A_2}^2 \right] &= \left| \bra{\bm{\gamma}} \Pi_{N-K}^\delta \ket{\bm{\gamma}} \right|^2
\end{split}
\end{align}

Bringing everything together, we find that the second moment of the squared reduced state $\rho^\delta_{RE}(U, \psi_0)$ is
\begin{align}
\begin{split}
\label{sup_mat_eq:second_moment_upper_bound_final}
    \mathbb{E}_U \tr_{RE} \left[ \rho_{RE}^\delta(U, \psi_0)^2  \right] &= \begin{cases}
        \big(D_1 \beta_K + D_2 \Delta_K^2 \big) \left|\bra{\bm{\gamma}} \Pi_{N-K}^\delta \ket{\bm{\gamma}} \right|^2 & \textrm{standard case} \\
        D_1 \beta_K \Delta_{N-K}^2 + D_2 \Delta_K^2 \beta_{N-K} & \textrm{entanglement-assisted case}
    \end{cases}
\end{split}
\end{align}
where $D_1, D_2$ are defined in Eq.~\eqref{sup_mat_eq:second_moment_of_X_2}, $\Delta_X$ is defined in Eq.~\eqref{sup_mat_eq:reduced_typical_subspace_overlap} and $\beta_X$ is defined in Eq.~\eqref{sup_mat_eq:reduced_maximally_mixed_state_purity}.

To finally finish calculating an upper bound on the second moment, we need to average over the initial states in the standard case, which is equivalent averaging over $\bm{\gamma}$ for $\left| \bra{\bm{\gamma}} \Pi_{N-K}^\delta \ket{\bm{\gamma}}\right|^2$ - we denote this average as $\zeta_{N-K}$.
\begin{align}
\begin{split}
    \zeta_{N-K} &= \mathbb{E}_{\bm{\gamma}} \left| \bra{\bm{\gamma}} \Pi_{N-K}^\delta \ket{\bm{\gamma}} \right|^2
\end{split}
\end{align}

\subsubsection{Bounding the average distance to the decoupled state}
Having upper bounded both $\mathbb{E}_{U, \psi_0} \tr_{RE}[ \rho_{RE}^\delta(U, \psi_0)^2]$ in Eq.~\eqref{sup_mat_eq:second_moment_upper_bound_final} and $\tr_{RE} [ (\rho_R \otimes \bar{\rho}_E)^2 ]$ in Eq.~\eqref{sup_mat_eq:average_state_purity}, we can now upper bound the expected trace distance to the average state from Eq.~\eqref{sup_mat_eq:trace_norm_to_hilbert_schmidt_norm}; we note that $D_1$ and $D_2$ can be written as
\begin{equation}
    D_1 = \frac{\alpha_E}{d_A^2} - \frac{\alpha_E + \alpha_B}{d_A^3} + \textrm{O}(d_A^{-4}), \quad \quad D_2 = \frac{\alpha_B}{d_A^2} - \frac{\alpha_E + \alpha_B}{d_A^3} + \textrm{O}(d_A^{-4})
\end{equation}
Using this, for the difference $\mathbb{E}_{U, \psi_0} \tr_{RE}[\rho^\delta_{RE}(U, \psi_0)^2] -\tr_{RE}[(\rho_R \otimes \bar{\rho}_E)^2)]$ in the standard case we derive
\begin{align}
\begin{split}
    \mathbb{E}_{U, \bm{\gamma}} \left[\tr_{RE}[ \rho_{RE}^\delta(U, \bm{\gamma})^2] \right] - \tr_{RE} [ (\rho_R \otimes \bar{\rho}_E)^2] &= \frac{\alpha_E \beta_K (\zeta_{N-K} - 1) + \alpha_B \zeta_{N-K} \Delta_K^2}{d_A^2} + \textrm{O}(d_A^{-3}) \\
    &\le \frac{\alpha_B \zeta_{N-K} \Delta_K^2 }{d_A^2} + \textrm{O}(d_A^{-3})
\end{split}
\end{align}
where we noted that $\zeta_{N-K} < 1$. In the entanglement-assisted case, we derive
\begin{align}
\begin{split}
    \mathbb{E}_U \left[\tr_{RE}[ \rho_{RE}^\delta(U)^2] \right] - \tr_{RE} [ (\rho_R \otimes \bar{\rho}_E)^2] &= \frac{\alpha_E \beta_K (\Delta_{N-K}^2 - 1)}{d_A^2} + \frac{\alpha_B \beta_{N-K} \Delta^2_K}{d_A^2} + \textrm{O}(d_A^{-3}) \\
    &\le \frac{\alpha_B \beta_{N-K}\Delta_K^2}{d_A^2} + \textrm{O}(d_A^{-3})
\end{split}
\end{align}
where we similarly noted $\Delta_{N-K}^2 \le 1$.

Bringing everything together, we derive the following upper bounds on the expected trace distance to the average state:
\begin{equation}
\label{sup_mat_eq:expected_distance_to_average_decoupled_state_full_result}
    \mathbb{E}_{U, \psi_0} ||\rho_{RE}(U, \psi_0) - \rho_R \otimes \bar{\rho}_E||_1 \le \sqrt{\frac{d_R d_E \alpha_B \Delta_K^2}{d_A^2}} \, \times \begin{cases}
        \sqrt{\zeta_{N-K}} & \textrm{standard case} \\
        \sqrt{\beta_{N-K}} & \textrm{entanglement-assisted case}
    \end{cases}
\end{equation}
This can be simplified a little further by simply inserting $\zeta_{N-K} \le 1$ and $\Delta_{K}^2 \le 1$, but we leave these expressions whole here.
In Sec.~\ref{sup_mat_sec:results}, we manipulate and simplify these expressions to give our final result in terms of the entanglement-transmission rate and probability of erasure. But first, we upper bound the purity of the average reduced state, $\alpha_B$, in the next section.

\subsection{Bounding the purity of the average reduced state, $\alpha_X$}
\label{sup_mat_sec:bounding_alpha_X_purity_reduced_state}
In this section, we upper bound the purity of the reduced maximally mixed state, denoted by $\alpha_X$, for the subsystem consisting of $X$ modes. The expression is given by
\begin{align}
\begin{split}
    \alpha_X = \sum_{n, n' = 0}^{\tilde{n}_+} \sum_{m = 0}^\infty {m + X - 1\choose m} {n - m + N - X - 1 \choose n - m} {n' - m + N - X - 1 \choose n' - m}.
\end{split}
\end{align}
We can use the standard result $\sum_{n = 0}^{\tilde{n}} {n - m + Y \choose n - m} = {\tilde{n} - m + Y + 1 \choose \tilde{n} - m}$ to give
\begin{align}
\begin{split}
    \alpha_X &= \sum_{m = 0}^\infty {m + X - 1 \choose m} \left( {\tilde{n}_+ - m + N - X \choose \tilde{n}_+ - m} - {\tilde{n}_- - m + N - X - 1 \choose \tilde{n}_- - m - 1} \right)^2.
\end{split}
\end{align}
By expanding the square, this can then be expressed $\alpha_X = \alpha_{\tilde{n}_+, \tilde{n}_+} - 2\alpha_{\tilde{n}_+ \tilde{n}_- - 1} + \alpha_{\tilde{n}_- - 1, \tilde{n}_- - 1}$, where we defined the functions $\alpha_{\tilde{n}, \tilde{n}'}$ as
\begin{equation}
    \alpha_{\tilde{n}, \tilde{n}'}(A, B, C) = \sum_{m = 0}^\infty {m + A \choose m} {\tilde{n} - m + B \choose \tilde{n} - m} {\tilde{n}' - m + C \choose \tilde{n}' - m}.
\end{equation}
Since $\alpha_{\tilde{n}_- - 1, \tilde{n}_- - 1} \le \alpha_{\tilde{n}_+, \tilde{n}_- - 1}$, we can upper bound $\alpha_X$ with just the leading term;
\begin{equation}
    \alpha_X \le \alpha_{\tilde{n}_+, \tilde{n}_+} (X, N - X, N - X)
\end{equation}
To upper bound $\alpha_X$, we will now upper bound $\alpha_{\tilde{n}, \tilde{n}'}(A, B, C)$.

We can view $\alpha_{\tilde{n}, \tilde{n}'}$ as a particular element of the bivariate sequence $(\alpha_{n, n'})_{n,n' \in \mathbb{N}_0}$. From this, we can define and calculate the bivariate ordinary generating function, $\mathcal{G}$, of the sequence $(\alpha_{n, n'})$
\begin{equation}
    \mathcal{G}(x, y) = \sum_{n, n' = 0}^\infty \alpha_{n, n'} x^n y^{n'}.
\end{equation}
To calculate this, we now define $a_k = {k + A \choose k}$, $b_k = {k + B \choose k}$ and $c_k = {k + C \choose k}$ such that $\alpha_{n, n'}(A, B, C) = \sum_{m = 0}^\infty a_m b_{n - m} c_{n' - m}$. The generating function of the sequence $(a_k)_{k \in \mathbb{N}_0}$, denoted $\mathcal{A}$, is
\begin{align}
\begin{split}
    \mathcal{A}(x) &= \sum_{n = 0}^\infty {n + A \choose n} x^n \\
    &= \sum_{n = 0}^\infty {-A - 1 \choose n} (-x)^n \\
    &= (1 - x)^{-(A + 1)}.
\end{split}
\end{align}
The analogous results hold for the generating functions $\mathcal{B}$ and $\mathcal{C}$ of the respective sequences $(b_k)$ and $(c_k)$. We can now calculate the bivariate generating function of $(\alpha_{n, n'})$ as
\begin{align}
\begin{split}
    \mathcal{G}(x, y) &= \sum_{n, n' = 0}^\infty \alpha_{n, n'} x^n y^{n'} \\
    &= \sum_{n, n', m = 0}^\infty a_m b_{n - m} c_{n' - m} \, x^n y^{n'} \\
    &= \sum_{p, q, m = 0}^\infty a_m (xy)^m \, b_p x^p \, c_q y^q \\
    &= \mathcal{A}(xy) \mathcal{B}(x) \mathcal{C}(y) \\
    &= (1 - xy)^{-(A + 1)} (1 - x)^{-(B + 1)} (1 - y)^{-(C + 1)},
\end{split}
\end{align}
where in the second line we used $\alpha_{n, n'} = \sum_{m = 0}^\infty a_m b_{n-m} c_{n'-m}$ and in the
third line we made the variable transformation $p = n - m$ and $q = n' - m$. From this, extracting the term with the powers $x^{\tilde{n}} y^{\tilde{n}'}$ gives the value of $\alpha_{\tilde{n}, \tilde{n}'}$. Formally, this could be calculated by taking the derivative $\lim_{x,y \rightarrow 0} \frac{1}{\tilde{n}! \tilde{n}'!} \frac{\partial^{\tilde{n}}}{\partial x^{\tilde{n}}} \frac{\partial^{\tilde{n}'}}{\partial y^{\tilde{n}'}} \mathcal{G}(x, y)$, but the resulting expression is too complicated to work with.
\\
\\
Instead, our approach is to extend the domain of $x$ and $y$ to the complex plane and then look at the Laurent series of the (now complex) function $\mathcal{G}(x, y)$. It is clear that this function is well defined without singularities on the bidisk $\mathcal{D}_2 = \{ (x, y) : x, y \in \mathbb{C}, 0 < |x|, |y| < 1 \}$.

The Laurent series of a function in some region, if it exists, is unique. Therefore, the Laurent series of $\mathcal{G}(x, y)$ centered on $(0, 0)$ in the bidisk $\mathcal{D}_2$ is given simply by the definition of the generating function, $\mathcal{G}(x, y) = \sum_{n, n' = 0}^\infty \alpha_{n, n'} x^n y^{n'}$. This means that the Laurent coefficients, $\ell_{n, n'}$, are themselves given by elements of the sequence $(\alpha_{n, n'})$;
\begin{equation}
    \mathcal{G}(x, y) = \sum_{n, n' = -\infty}^\infty \ell_{n, n'} x^n y^{n'}, \quad \ell_{n, n'} =
    \begin{cases}
        \alpha_{n, n'} & n, n' \ge 0 \\
        0 & \textrm{else}
    \end{cases}
\end{equation}
We also know that the Laurent coefficients are defined by the contour integral
\begin{equation}
    \ell_{n, n'} =  \oint_\Gamma \frac{\textrm{d}x}{2\pi i} \frac{\textrm{d}y}{2\pi i} \, \frac{\mathcal{G}(x, y)}{x^{n + 1} y^{n' + 1}}
\end{equation}
where $\Gamma$ is a counterclockwise curve around the origin of the bidisk $\mathcal{D}_2$. We now upper bound this integral.

To do this, we choose the bicircular path $\Gamma = \{ (re^{i\phi}, se^{i\theta}) : r,s,\phi,\theta \in \mathbb{R}; 0<r,s<1; \phi,\theta \in [0, 2\pi) \}$ where $r$ and $s$ are constants and $\phi$ and $\theta$ vary from $0$ to $2\pi$. Given this path, the coefficient $\ell_{n, n'}$ is given by
\begin{align}
\begin{split}
    \ell_{n, n'} &= \int_{\phi = 0}^{2\pi} ire^{i\phi} \frac{\textrm{d} \phi}{2\pi i}  \int_{\theta = 0}^{2\pi i} ise^{i\theta} \frac{\textrm{d}\theta}{2\pi i} \frac{\mathcal{G}(re^{i\phi}, se^{i\theta} )}{r^{n + 1} s^{n' + 1} e^{i \big(\phi (n + 1) + \theta (n' + 1)\big)} } \\
    &= \frac{1}{r^n s^{n'}} \int_{\phi = 0}^\infty \frac{\textrm{d}\phi}{2\pi} \int_{\theta = 0}^\infty \frac{\textrm{d}\theta}{2\pi} \, \mathcal{G}(re^{i\phi}, se^{i\theta}) e^{-i(\phi n + \theta n')}
\end{split}
\end{align}
To upper bound the magnitude of the integral $|\ell_{n, n'}|$, we employ the triangle inequality and max/min integral inequality
\begin{align}
\begin{split}
    |\ell_{n, n'}| &\le \frac{1}{r^n s^{n'} } \int_{\phi = 0}^\infty \frac{\textrm{d}\phi}{2\pi} \int_{\theta = 0}^\infty \frac{\textrm{d}\theta}{2\pi} \, \left| \mathcal{G}(re^{i\phi}, se^{i\theta}) \right| \\
    &\le \frac{\max_{\phi, \theta \in [0, 2\pi)} |\mathcal{G}(re^{i\phi}, se^{i\theta})|}{r^n s^{n'}} \int_{\phi = 0}^\infty \frac{\textrm{d}\phi}{2\pi} \int_{\theta = 0}^\infty \frac{\textrm{d}\theta}{2\pi} \\
    &= \max_{\phi, \theta \in [0, 2\pi)}\frac{|\mathcal{G}(re^{i\phi}, se^{i\theta})|}{r^n s^{n'}}
\end{split}
\end{align}
This inequality holds for any $0 < r,s < 1$. Expanding out $|\mathcal{G}(re^{i\phi}, se^{i\theta})| = |(1 - rs e^{i(\phi + \theta)})^{-(A+1)} (1 - re^{i\phi})^{-(B+1)} (1 - se^{i\theta})^{-(C+1)}|$, this expression is \emph{maximised} when $\phi = \theta = 0$; this is because $|1-z|$ for constant $|z| < 1$ is \emph{minimised} when $z$ is parallel to the real line. It follows that we can upper bound $\alpha_{\tilde{n}, \tilde{n}'} = \ell_{\tilde{n}, \tilde{n}'}$ using any $r$, $s$ satisfying $0 < r, s < 1$ via
\begin{align}
\begin{split}
    \alpha_{\tilde{n}, \tilde{n}'} \le \frac{1}{r^{\tilde{n}} s^{\tilde{n}'} (1 - rs)^{A + 1} (1 - r)^{B + 1} (1 - s)^{C + 1}}
\end{split}
\end{align}
Recalling that $\alpha_X \le \alpha_{\tilde{n}_+, \tilde{n}_+} (X, N - X, N - X)$, this implies 
\begin{equation}
\label{sup_mat_eq:full_alphaX_bound_no_approx}
    \alpha_X \le \frac{(rs)^{-\tilde{n}_+} (1 - rs)^{-X} \left((1 - r) (1 - s) \right)^{-(N - X)}}{(1-rs)(1-r)(1-s)}
\end{equation}
This holds for any $X$.
\\
\\
Now we choose to take $X = xN$ and attempt to minimise this bound by finding the optimal $r$ and $s$. Recalling that $\tilde{n}_+ = \tilde{m}_+ N$, the bound can be expressed as
\begin{align}
\begin{split}
    \alpha_X \le \frac{\left((rs)^{\tilde{m}_+} (1-rs)^{x} ((1-r)(1-s))^{1-x} \right)^{-N} }{(1-rs) (1-r)(1-s)}
\end{split}
\end{align}
In the large $N$ limit, this is minimised by maximising the term raised to the power of $-N$. Numerically we find this is maximised when $r = s$ so make the transformation $s \rightarrow r$. We then need to maximise the function
\begin{equation}
    r^{2m} (1+r)^x (1 - r)^{2 - x}
\end{equation}
where we rewrote $(1 - rs) \rightarrow (1 - r^2) = (1-r)(1+r)$. One can easily show that this is maximised on $r \in (0, 1)$ by
\begin{equation}
\label{sup_mat_eq:r_opt_definition}
    r_{\textrm{opt}} = \sqrt{\frac{\tilde{m}_+}{\tilde{m}_+ + 1} + \left(\frac{1-x}{2( \tilde{m}_+ + 1)}\right)^2} \, - \frac{1 - x}{2(\tilde{m}_+ + 1)}
\end{equation}
If we consider the high energy limit $\tilde{m}_+ \gg 1$ (recall that $\tilde{m}_+ \approx \bar{n}$), then $r_{\textrm{opt}} \approx \sqrt{\frac{\tilde{m}_+}{\tilde{m}_+ + 1}} - \frac{1-x}{2(\tilde{m}_+ + 1)} + \textrm{O}(\tilde{m}_+^{-2})$.

The $\alpha_X$ bound can be written in terms of $r_{\textrm{opt}}$ using logarithms in the following way - note that we use the notation $2[\bullet] = 2^{\bullet}$ (analogous to $\exp[\bullet] = e^{\bullet}$)
\begin{align}
\begin{split}
\label{sup_mat_eq:alpha_X_bound_exact}
    \alpha_X &\le \frac{2\left[-N \Big( \tilde{m}_+ \log(r_{\textrm{opt}}^2) + x \log(1 - r_{\textrm{opt}}^2) + 2(1 - x) \log(1 - r_{\textrm{opt}})  \Big) \right]}{(1 + r_{\textrm{opt}})(1-r_{\textrm{opt}})^3}
\end{split}
\end{align}
This is our general bound for $\alpha_X$ for $X = xN$ modes.
\\
\\
Finally, we want to find how $\alpha_X$ behaves in the high energy limit, $\tilde{m}_+ \gg 1$, where $r_{\textrm{opt}} \approx \sqrt{\frac{\tilde{m}_+}{\tilde{m}_+ + 1}} - \frac{1-x}{2(\tilde{m}_+ + 1)} + \textrm{O}(\tilde{m}_+^{-2})$. We will look at just part of the exponent of the bound from Eq.~\eqref{sup_mat_eq:alpha_X_bound_exact}, namely $-\big(\tilde{m}_+ \log(r_{\textrm{opt}}^2) + x\log(1 - r_{\textrm{opt}}^2) + 2(1 - x) \log(1 - r_{\textrm{opt}}) \big)$. Taking $r_{\textrm{opt}} \approx \sqrt{\frac{\tilde{m}_+}{\tilde{m}_+ + 1}} - \frac{1-x}{2(\tilde{m}_+ + 1)} + \textrm{O}(\tilde{m}_+^{-2})$, it can be shown by careful rearrangement that in this limit this exponent is given by
\begin{align}
\begin{split}
    -\big(\tilde{m}_+ \log(r_{\textrm{opt}}^2) + x\log(1 - r_{\textrm{opt}}^2) + 2(1 - x) \log(1 - r_{\textrm{opt}}) \big) & \\
    = (2 - x) H_{\textrm{therm}}&(\tilde{m}_+) + 2(1 - x) - (2-x)\log(2-x) +  \textrm{O}(\tilde{m}_+^{-1})
\end{split}
\end{align}
To derive this, we used $\log(1 - \sqrt{\tilde{m}_+ / \tilde{m}_+ + 1}) = \log(\sqrt{\tilde{m}_+ + 1} - \sqrt{\tilde{m}_+}) - \log(\sqrt{\tilde{m}_+ + 1})$ and also Taylor expanded the argument of $\log(\sqrt{1 + \tilde{m}_+^{-1}} - 1)$ to get $\log(\sqrt{1 + \tilde{m}_+^{-1}} - 1) = \log(\frac{1}{2}\tilde{m}_+^{-1}) + \textrm{O}(\tilde{m}_+^{-1})$. We also used the approximation $\log(\tilde{m}_+ + y) = \log(\tilde{m}_+) + \frac{y}{\tilde{m}_+ \ln(2)} + \textrm{O}(\tilde{m}_+^{-2})$ for any constant $y$ of order unity, and $\log(\tilde{m}_+) = H_{\textrm{therm}}(\tilde{m}_+) - 1$.
This finally gives our bound on the purity of the reduced maximally mixed states in the high energy limit.
\begin{equation}
\label{sup_mat_eq:alpha_X_bound_high_energy_limit}
    \alpha_X \le \textrm{poly}(N) \, \,2\left[-N \Big( (2 - x)H_{\textrm{therm}}(\tilde{m}_+) + 2(1 - x) - (2-x) \log(2 - x) + \textrm{O}(\tilde{m}_+^{-1}) \Big) \right]
\end{equation}

\subsection{Main Results}
\label{sup_mat_sec:results}
\begin{figure}
    \centering
    \includegraphics[width=0.85\linewidth]{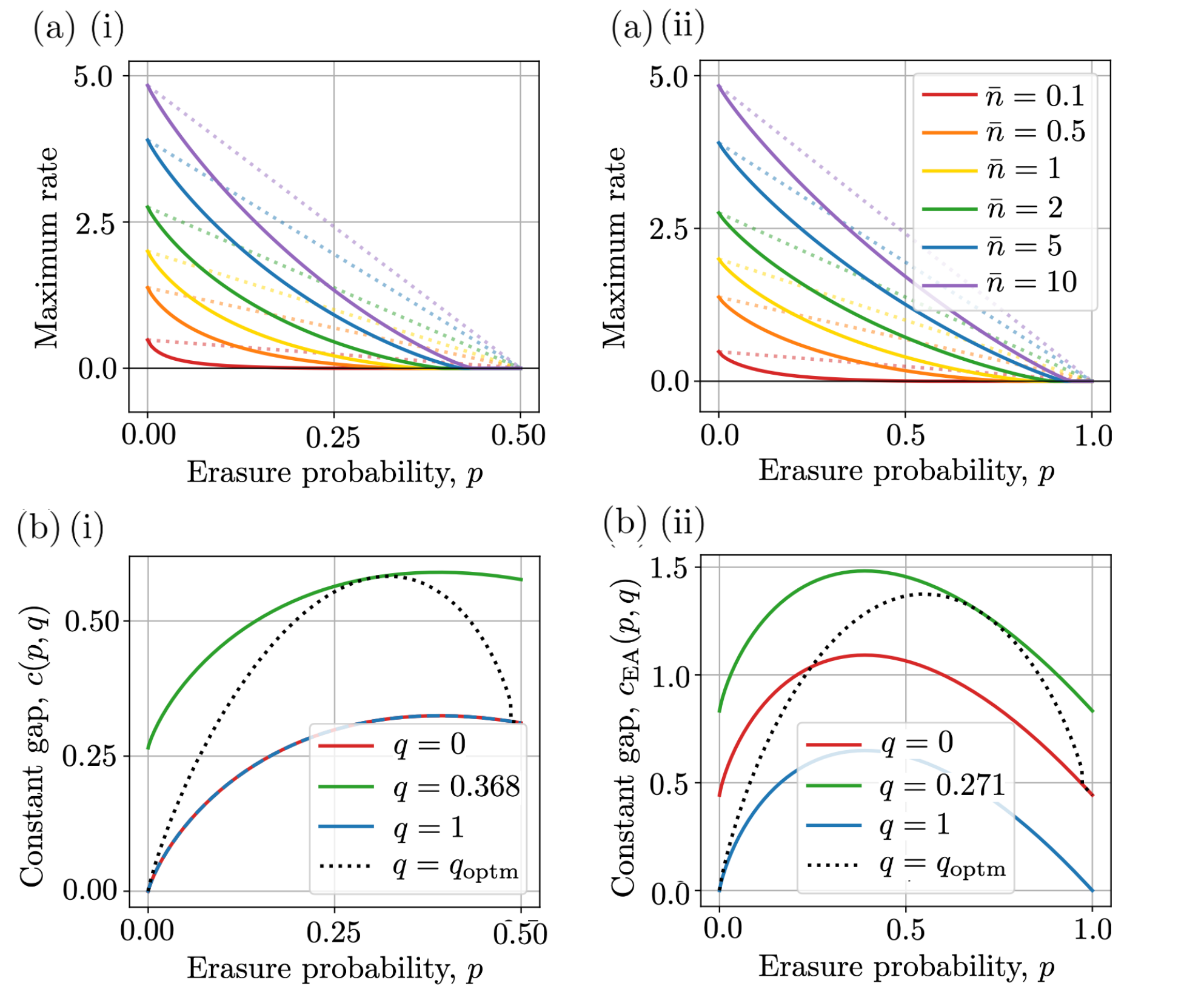}
    \caption{(a) Maximum rate of our random codes with energy constraints in (i) the standard case and (ii) the entanglement-assisted case. The dotted linear curves are the corresponding capacities. We see a nonlinear decrease in the maximum rate, arising fundamentally from the submultiplicity of dimensions within the typical subspace. The gap between the capacity and maximum rate is approximately equal to the constant functions plotted in (b). Here, the $q = 0.368$ and $q = 0.271$ curves are the values of $q$ that maximise the functions $c_{\textrm{stan}}$ and $c_{\textrm{EA}}$, while the dotted black $q_{\textrm{optm}}$ curve is the value of both $c$ functions given that the rate is maximised.}
    \label{sup_mat_fig:maximum_rate_and_constant}
\end{figure}

In the standard case, we have derived an upper bound on the variance of $\rho_{RE}(U, \bm{\gamma})$ in state space (Eq.~\eqref{sup_mat_eq:expected_distance_to_average_decoupled_state_full_result});
\begin{align}
\begin{split}
\label{sup_mat_eq:main_result_standard_in_terms_of_dimensions}
    \mathbb{E}_{U, \bm{\gamma}} ||\rho_{RE}(U, \bm{\gamma}) - \rho_R \otimes \bar{\rho}_E||_1 \le \sqrt{\frac{d_R d_E \alpha_B \zeta_{N-K} \Delta_K^2}{d_A^2}}
\end{split}
\end{align}
If we now insert the large $N$ expression for $d_A$ and $d_R d_E$ (Eq.~\eqref{sup_mat_eq:dimension_in_full_space_dA} and Eq.~\eqref{sup_mat_eq:dimension_in_reduced_space_dX} respectively), use the upper bounds $\zeta_{N-K}, \Delta_K^2 \le 1$ and include our upper bound on $\alpha_B$ which was derived in Sec.~\ref{sup_mat_sec:bounding_alpha_X_purity_reduced_state} (Eq.~\eqref{sup_mat_eq:alpha_X_bound_exact} with $X = (1-p)N$), then we derive the result
\begin{equation}
\label{sup_mat_eq:main_result_in_terms_of_xi_and_tau}
    \mathbb{E}_{U, \bm{\gamma}} ||\rho_{RE}(U, \bm{\gamma}) - \rho_R \otimes \bar{\rho}_E||_1 \le \textrm{poly}(N) \, \, 2^{-N \xi_{\textrm{stan}}(\bar{n}, p, q)}
\end{equation}
where $\xi_{\textrm{stan}}$ is defined as
\begin{align}
\begin{split}
    \xi_{\textrm{stan}}(\bar{n}, p) &= 2 H_{\textrm{therm}}(\bar{n}) - (\bar{n} + p) H_2\left(\frac{\bar{n}}{\bar{n} + p}\right) - (\bar{n} + q) H_2\left(\frac{\bar{n}}{\bar{n}+q}\right) \\
    & \quad + \, \bar{n} \log(r_{\textrm{opt}}^2) + (1 - p) \log(1 - r_{\textrm{opt}}^2) + 2p \log(1 - r_{\textrm{opt}})
\end{split}
\end{align}
with $r_{\textrm{opt}}$ from Eq.~\eqref{sup_mat_eq:r_opt_definition}.

In the high energy limit, it can be shown by using $\log(\tilde{m}_+ + y) = \log(\tilde{m}_+) + \frac{y}{\tilde{m}_+ \ln(2)} + \textrm{O}(\tilde{m}_+^{-2})$, and using $\log(\tilde{m}_+) = H_{\textrm{therm}}(\tilde{m}_+) - \frac{1}{\ln(2)} + \textrm{O}(\tilde{m}_+^{-1})$ that Eq.~\eqref{sup_mat_eq:main_result_in_terms_of_xi_and_tau} reduces to
\begin{align}
\begin{split}
    \mathbb{E}_{U, \bm{\gamma}} ||\rho_{RE}(U, \bm{\gamma}) - \rho_R \otimes \bar{\rho}_E||_1 \le \textrm{poly}(N) \, \, 2\left[ - \frac{1}{2} N \Big( (1 - 2p - q) H_{\textrm{therm}}(\bar{n}) - c_{\textrm{stan}}(p, q) + \textrm{O}(\bar{n}^{-1}) \Big)  \right]
\end{split}
\end{align}
where energy-independent function is given by
\begin{equation}
    c_{\textrm{stan}}(p, q) = 2p - p \log(p) - (1 + p) \log(1 + p) - q \log(q)
\end{equation}
and again we used the notation $2[\bullet] \coloneq 2^\bullet$.

The entanglement transmission rate of the code is given by $Q_{\textrm{rate}}(\bar{n}, q) = q H_{\textrm{therm}}(\bar{n})$. Similarly, the quantum capacity of the erasure channel for $p < 0.5$, presented in the main manuscript, is equal to $\mathcal{Q}(\Lambda_p; \bar{n}) = (1 - 2p) H_{\textrm{therm}}(\bar{n})$. Our final result for the standard random code thus is
\begin{equation}
\label{sup_mat_eq:final_result_standard_case}
     \mathbb{E}_{U, \bm{\gamma}} ||\rho_{RE}(U, \bm{\gamma}) - \rho_R \otimes \bar{\rho}_E||_1 \le \textrm{poly} (N) \, \, 2^{-\frac{1}{2}N \big( \mathcal{Q}(\Lambda_p; \bar{n}) - Q_{\textrm{rate}}(\bar{n}, q) - c_{\textrm{stan}}(p, q) + \textrm{O}(\bar{n}^{-1}) \big)}
\end{equation}
This bound is clearly exponentially small in $N$ whenever the rate is less than $\mathcal{Q} - c$. Therefore, Bob can reliably recover the $K = qN$ two-mode squeezed vacuum states for all rates up to the capacity minus a small gap, $c_{\textrm{stan}}$, in the high energy limit. This proves that the typical subspace random code is asymptotically optimal in the standard case, because it satisfies
\begin{equation}
    \lim_{N, \bar{n} \rightarrow \infty} \frac{Q_{\textrm{rate}}(\bar{n}, q_{\textrm{optm}})}{\mathcal{Q}(\Lambda_p; \bar{n})} = 1
\end{equation}
where $q_{\textrm{optm}}$ is the maximum value of $q$ for which $\xi_{\textrm{stan}}(\bar{n}, p, q) > 0$.
\\
\\
In the entanglement-assisted case, the upper bound on variance of $\rho_{RE}(U)$ in Eq.~\eqref{sup_mat_eq:expected_distance_to_average_decoupled_state_full_result} reads;
\begin{equation}
    \mathbb{E}_{U} ||\rho_{RE}(U) - \rho_R \otimes \bar{\rho}_E||_1 \le \sqrt{\frac{d_R d_E \alpha_B \beta_{N-K} \Delta_K^2}{d_A^2}}
\end{equation}
Taking the large $N$ limit of the dimensions, using $\Delta_K^2 \le 1$ and implementing our upper bound of $\alpha_B$ gives the same expression as in the standard case but multiplied by $\sqrt{\beta_{N-K}}$. Via submultiplicity, this itself is upper bounded by
\begin{align}
\begin{split}
    \beta_{N-K} &\le \left(\frac{1 - |z|^2}{1 + |z|^2}\right)^{N-K} \\
    &= \left(\frac{1}{2\bar{n} + 1}\right)^{N-K} \\
    &= (2\bar{n} + 1)^{-(1-q)N} \\
    &= 2^{-N (1 - q)\log(2\bar{n} + 1)}
\end{split}
\end{align}
Inserting this into the standard result gives the entanglement-assisted version $\xi$;
\begin{equation}
    \xi_{\textrm{EA}}(\bar{n}, p, q) = \xi_{\textrm{stan}}(\bar{n}, p, q) + \log(2\bar{n} + 1) -  q \log(2\bar{n} + 1)
\end{equation}
In the high energy limit, this leads to the bound 
\begin{equation}
    \mathbb{E}_{U} ||\rho_{RE}(U) - \rho_R \otimes \bar{\rho}_E||_1 \le \textrm{poly}(N) \, 2^{-N \Big( \mathcal{Q}_{\textrm{EA}}(\Lambda_p, \bar{n}) - Q_{\textrm{rate}}(\bar{n}, q) - c_{\textrm{EA}}(p, q) + \textrm{O}(\bar{n}^{-1})\Big)}
\end{equation}
with new energy-independent function
\begin{equation}
\label{sup_mat_eq:EA_constant_gap_definition}
    c_{\textrm{EA}}(p, q) = c_{\textrm{stan}}(p, q) - (1 - q) \left( 1 - \frac{1}{\ln(2)}\right).
\end{equation}
Once again, this bound is exponentially small for any rate $Q_{\textrm{rate}} < \mathcal{Q}_{\textrm{EA}} - c_{\textrm{EA}}$. This means that our entanglement-assisted random code is asymptotically optimal up to the energy-independent gap $c_{\textrm{EA}}(p, q)$.
\\
\\
This proves that in the large $N$, high energy limit, random codes that scramble information within the finite-dimensional typical subspace are asymptotically optimal for entanglement-transmission through the continuous-variable erasure channel. When away from this limit, the maximum rate decreases nonlinearly with $p$. In Fig.~\ref{sup_mat_fig:maximum_rate_and_constant} (a), we plot the maximum rate of the code for a different energy constraint in both the standard and entanglement-assisted cases. This is found by maximising $Q_{\textrm{rate}}(\bar{n}, q) = q H_{\textrm{therm}}(\bar{n})$ subject to $\xi(\bar{n}, p, q) > 0$. For $\bar{n} > 10$, the maximum rate plots all look like $\mathcal{Q} - c(p,q_{\textrm{optm}}) + \textrm{O}(<0.1)$, where we have plotted $c(p, q_{\textrm{optm}})$ as black dotted curves in Fig.~\ref{sup_mat_fig:maximum_rate_and_constant} (b). From (b), we also plot the maximum and minimum values of the constant and see that it really is of order unity.

\subsection{Dimensions not scaling with $N$}
In the above derivation, we took the reduced number of modes in $A_1$ and $B$ as scaling with the total number of modes, e.g., $A_1$ had $qN$ modes while $B$ had $(1-p) N$ modes. When looking at the Hayden-Preskill protocol \cite{hayden2007black}, the question is how many modes does Bob need to access in $B$ in order to recover the information initially encoded in $A_1$ in the entanglement-assisted case. The DV result shows for any constant $K$, the number of qudits in $B$ need just be a constant amount higher and Bob can recover the information. In this subsection, we show that our bound becomes ineffective when taking $K$ and the number of modes in $B$ to be constant, fundamentally due to added complexity from the submultiplicity of dimensions in the typical subspace. To do this, we take the result from Eq.~\eqref{sup_mat_eq:expected_distance_to_average_decoupled_state_full_result} but treat the expressions assuming $K$ is constant and $B$ consists of a constant number of modes.

First, we look at the expression for reduced dimensions from the first line of Eq.~\eqref{sup_mat_eq:dimension_in_reduced_space_dX} and again employ Stirling's formula. Let the subsystem $X$ consist of a constant set of $X$ modes. The reduced dimension reads
\begin{align}
\begin{split}
    d_X &= {\tilde{m}_+N + X \choose \tilde{m}_+ N}  \\
    &\approx \textrm{poly}(N) \, 2^{\tilde{m}_+ N \log( 1 + \frac{X}{\tilde{m}_+ N})} \\
    &= \textrm{poly}(N)
\end{split}
\end{align}
where we used $\log(1 + \varepsilon) = \varepsilon + \textrm{O}(\varepsilon^2)$ in the final line to show $d_X$ scales polynomially with $N$ (note that we take $2^{1/N} \in \textrm{poly}(N)$ when looking at scaling in the large $N$ limit). Now, let the subsystem $Y$ consist of a $N-Y$ modes for constant $Y$. This subsystem has a dimension
\begin{align}
\begin{split}
    d_Y &= {(\tilde{m}_+ + 1 )N - Y \choose \tilde{m}_+ N} \\
    &\approx \textrm{poly}(N) \, 2^{N H_{\textrm{therm}}(\tilde{m}_+)}
\end{split}
\end{align}
This is clearly equivalent to $d_A$ up to polynomial corrections.
Similarly, we can look at the bound on the reduced maximally mixed state purity $\alpha_X$ for constant $X$. The bound from Eq.~\eqref{sup_mat_eq:full_alphaX_bound_no_approx} implies that 
\begin{align}
\begin{split}
    \alpha_X & \le \frac{(rs)^{-\tilde{m}_+ N} (1 - rs)^{-X} ((1-r)(1-s))^{-(N-X)}}{(1-rs)(1-r)(1-s)} \\
    &\approx \textrm{poly}(N) \, 2^{-N \Big( \tilde{m}_+ \log(rs) + \log\big((1-r)(1-s)\big)  \Big)}
\end{split}
\end{align}
If we take $r=s$, then it can easily be shown that the bound is minimised when $r = \frac{\tilde{m}_+}{\tilde{m}_+ + 1}$. This then gives
\begin{equation}
    \alpha_X \le \textrm{poly}(N) \, 2^{N \big(2\tilde{m}_+ H_{\textrm{therm}}(\tilde{m}_+)\big)}
\end{equation}
Finally, we look at purity of some initial $N-Y$ thermal states in the typical subspace, denoted as $\beta_{N-Y}$ in Eq.~\eqref{sup_mat_eq:thermal_state_purity_typical_subspace}
\begin{align}
\begin{split}
    \beta_{N-Y} &\le \left(\frac{1}{2\bar{n}_z - 1}\right)^{N-Y} \\
    &= \textrm{poly}(N) \, 2^{-N \log(2\bar{n}_z + 1)}
\end{split}
\end{align}
We now bring these components together; in the entanglement-assisted case where $R$ ($A_2$) consists of $K$ modes ($N-K$ modes) and $B$ ($E$) consists of $M$ ($N-M$) modes for constant $K$ and $M$, our result from Eq.~\eqref{sup_mat_eq:expected_distance_to_average_decoupled_state_full_result} can be upper bounded  by
\begin{align}
\begin{split}
    \mathbb{E}_{U} || \rho_{RE}(U) - \rho_R \otimes \bar{\rho}_E ||_1 &\le \sqrt{\frac{d_R d_E \alpha_B \beta_{N-K}}{d_A^2}} \\
    &= \textrm{poly}(N) \, 2^{\frac{1}{2} N \left( 2(\bar{n} - 1) H_{\textrm{therm}}(\bar{n}) + \frac{1}{\ln(2)} + 1  \right)}
\end{split}
\end{align}
For any $\bar{n}$, the exponent is positive and hence in the large $N$ limit our bound is exponentially large. This means that we can not prove the exact analogous Hayden-Preskill result because we cannot prove that Bob can recover Alice's information if he has access to $>K$ modes.
Fundamentally, this stems from the submultiplicity of typical subspace dimensions leading to very complex analytic expressions that require new bounding techniques when compared with the finite-dimensional derivation. These new bounds bury the relationship between the number of modes in $A_1$ and the number of modes in $B$ under an exponential term. Instead, our CV construction of the Hayden-Preskill protocol requires that if  Alice inputs $K = qN$ modes then Bob requires access to $(q + \varepsilon_{\textrm{EA}}) N$ modes in the high-energy limit to recover her information, where $\varepsilon_{\textrm{EA}} = (c_{\textrm{EA}} + \textrm{O}(\bar{n}^{-1}))/H_{\textrm{therm}}(\bar{n})$ is given in terms of the constant from Eq.~\eqref{sup_mat_eq:EA_constant_gap_definition} and thermal state entropy. In other words, the CV Hayden-Preskill protocol changes the scaling from a constant number more modes to to a constant fraction more modes.
Nonetheless, it shows that we can extend these DV information theoretic toy models for black holes to CV thermal states, which are exactly the sort of optical states emitted as Hawking radiation.

%% file: PLON_averaged_state.tex
\section{Average output state from a passive linear optical network}
\label{sup_mat_sec:average_output_state_from_PLON}

In this section, we analytically calculate the average output state from the set of passive linear optical networks (PLONs) - this is the full set of energy-conserving operations that can be implemented with just beam splitters and phase shifters. 
We calculate this for the setup considered in Ref.~\cite{Zhong2023information}, namely one mode from each of $K$ two-mode squeezed vacuum states is input into an $N$-mode PLON with an additional $N-K$ vacua input into the remaining modes. In Ref.~\cite{Zhong2023information}, the authors argued that the average output state on the support of the PLON is diagonal in the number basis and takes a tensor product form; since the single mode reduced state of a TMSV $\ket{z}$ is a thermal state $\rho_z$, this would be equivalent to showing
\begin{equation}
\label{sup_mat_eq:ansatz_average_state}
    \mathbb{E}_{\hat{V}} \big[ \hat{V} (\rho_z^{\otimes K} \otimes \ketbra{0}^{\otimes N-K}) \hat{V}^\dag\big] \, \stackrel{\text{ansatz}}{=} \, (\bar{\rho}_1)^{\otimes N}, \quad \quad \bar{\rho}_1 = \sum_{n = 0}^\infty p(n) \ketbra{n}
\end{equation}
where $\hat{V}$ is a particular PLON, $\bar{\rho}_1$ is the reduced average state on a single mode and $p(n)$ is an unknown probability distribution. Since the set of passive linear optical networks acting on $N$ modes is isomorphic to the unitary group $\textrm{U}(N)$, the average is taken with respect to the Haar measure.
\\
\\
In Ref.~\cite{Zhong2023information}, the authors introduced the idea of typical subspace scrambling. The argued that the average output state from a PLON has a tensor product structure, and hence if one considered many modes then the average state would form a typical subspace in which they could scramble information. However, we will show that the true average output state does not take this tensor product form due to residual classical correlations that exist between modes.
These arise from the energy-preservation of PLONs which necessitates that the average state must have the same energy-spectrum of the initial state, and in turn requires global classical correlations.
This further implies that the true average output state does not form an exact typical subspace because it does not have the correct tensor product form. We then further show that the overlap between the true average output state and corresponding tensor product ansatz state is, at most, small when in the high-energy limit. Finally, while we do not know if the true average output we derive has significant overlap with the \emph{typical subspace} generated by the ansatz state $\bar{\rho}_1^{\, \otimes N}$, we can not think of a clear theoretical reason for believing so.

\subsection{Average output state derivation}

First, recall that the action of a passive linear optical network $\hat{V}$ on the creation and annihilation operators of an $N$ mode system is
\begin{equation}
\label{sup_mat_eq:PLON_action_on_creation_operators}
    \hat{V} \hat{a}_i^\dag \hat{V}^\dag = (U_{\hat{V}})_{ij} \hat{a}_j^\dag, \quad \quad \hat{V} \hat{a}_i \hat{V}^\dag = (U_{\hat{V}})^*_{ij} \hat{a}_j
\end{equation}
where $U_{\hat{V}} \in \textrm{U}(N)$ is the unitary matrix associated with $\hat{V}$ and the Einstein summation convention (where double indices are summed over) was used. In future, we will drop the $\hat{V}$ subscript from $U_{\hat{V}}$.

We now look at an input state consisting of $K$ two-mode squeezed vacuum states and store one mode from each in a reference Hilbert space $R$ and while sending the other into an $N$-mode PLON, given by $\hat{V}$, with an additional $N-K$ vacuum states. We denote the support of $\hat{V}$ with $A$, let $A_i$ ($R_i$) denote the $i^{\textrm{th}}$ mode in $A$ ($R$) and have the $K$ modes from the TMSV states occupy the first $K$ modes in $A$. We write the $i^{\textrm{th}}$ TMSV state as $\ket{z}_{R_i A_i} = (1-|z|^2)^{\frac{1}{2}} \exp[z \hat{b}_i^\dag \hat{a}_i^\dag] \ket{0,0}_{R_i A_i}$ for $i \in \{1, \dots, K\}$, where $\hat{a}_i^\dag$ acts on $A_i$ and $\hat{b}_i^\dag$ on $R_i$. The initial state is thus $\ket{\psi_0(z)}_{RA} = (\ket{z}^{\otimes K})_{R A_1 \dots A_K} \otimes (\ket{0}^{\otimes N-K})_{A_{K+1} \dots A_N}$. The action of a PLON $\hat{V}$ associated with a unitary $U$ can then be written as
\begin{align}
\begin{split}
    \hat{V} \ket{\psi_0} &= (1 - |z|^2)^{\frac{K}{2}} \exp[ \sum_{i = 0}^{K} z \hat{b}_i^\dag \left(\sum_{j = 1}^N U_{ij}\hat{a}_j^\dag\right)] \ket{0}_{RA} \\
    &= (1 - |z|^2)^{\frac{K}{2}} \sum_{n = 0}^\infty \frac{z^n}{n!} \big(U_{ij} \hat{b}_i^\dag \hat{a}_j^\dag \big)^n \ket{0}_{RA} \\
    &= (1 - |z|^2)^{\frac{K}{2}} \sum_{n = 0}^\infty \frac{z^n}{n!} \big(U_{i_1 j_1} \hat{b}_{i_1}^\dag \hat{a}_{j_1}^\dag \big) \big(U_{i_2 j_2} \hat{b}_{i_2}^\dag \hat{a}_{j_2}^\dag\big) \dots \big(U_{i_n j_n} \hat{b}_{i_n}^\dag \hat{a}_{j_n}^\dag\big) \ket{0}_{RA} \\
    &= (1 - |z|^2)^{\frac{K}{2}} \sum_{n = 0}^\infty \frac{z^n}{n!} U_{i_1 j_1} U_{i_2 j_2} \dots U_{i_n j_n} \hat{b}_{i_1}^\dag \hat{b}_{i_2}^\dag \dots \hat{b}_{i_n}^\dag \hat{a}_{j_1}^\dag \hat{a}_{j_2}^\dag \dots \hat{a}_{j_n}^\dag \ket{0}_{RA}
\end{split}
\end{align}
where we used Eq.~\eqref{sup_mat_eq:PLON_action_on_creation_operators} in the first line. The repeated $i$ indices run from $1$ to $K$ while the repeated $j$ indices run from $1$ to $N$ (recall that repeated indices are summed over). With this, the average output state from the PLON can be written as
\begin{align}
\begin{split}
\label{sup_mat_eq:average_PLON_state_step1}
    \mathbb{E}_{\hat{V}}\big[ \hat{V} \ketbra{\psi_0}\hat{V}^\dag  \big] &= (1 - |z|^2)^K \sum_{n = 0}^\infty \sum_{m = 0}^\infty \frac{z^n z^{*m}}{n! m!} \int_{\textrm{U}(N)} \textrm{d}\mu(U) \, U_{i_1 j_1} \dots U_{i_n j_n} U^*_{i'_1 j'_1} \dots U^*_{i'_m j'_m} \\
    & \quad \quad \quad \times \, \hat{b}^\dag_{i_1} \dots \hat{b}^\dag_{i_n} \hat{a}^\dag_{j_1} \dots \hat{a}^\dag_{j_n} \ketbra{0}_{RA} \hat{b}_{i'_1} \dots \hat{b}_{i'_m} \hat{a}_{j'_1} \dots \hat{a}_{j'_m}
\end{split}
\end{align}
where the integral over $U$ is with respect to the Haar measure. We now proceed to calculate this integral using the Weingarten moment result from \cite{collins2006integration} 
\begin{align}
\begin{split}
    \int_{\textrm{U}(N)} \textrm{d}\mu(U) \, U_{i_1 j_1} \dots U_{i_n j_n} U^*_{i'_1 j'_1} \dots U^*_{i'_m j'_m} &= \delta_{n, m} \sum_{\sigma, \tau \in \mathcal{S}_n} \delta_{i_1, i'_{\sigma(1)}} \delta_{i_2, i'_{\sigma(2)}} \dots \delta_{i_n, i'_{\sigma(n)}} \delta_{j_1 , j'_{\tau(1)}} \delta_{j_2, j'_{\tau(2)}} \dots \delta_{j_n j'_{\tau(n)}} \textrm{Wg} (\tau \sigma^{-1})
\end{split}
\end{align}
where $\mathcal{S}_n$ is the symmetric group of $n$ items.
Inserting this back into Eq.~\eqref{sup_mat_eq:average_PLON_state_step1} gives
\begin{align}
\begin{split}
    \mathbb{E}_{\hat{V}}\big[ \hat{V} \ketbra{\psi_0} \hat{V}^\dag  \big] &= (1 - |z|^2)^K \sum_{n = 0}^\infty \frac{|z|^{2n}}{(n!)^2} \sum_{\sigma, \tau \in \mathcal{S}_n} \delta_{i_1, i'_{\sigma(1)}} \delta_{i_2, i'_{\sigma(2)}} \dots \delta_{i_n, i'_{\sigma(n)}} \delta_{j_1 , j'_{\tau(1)}} \delta_{j_2, j'_{\tau(2)}} \dots \delta_{j_n j'_{\tau(n)}} \textrm{Wg}(\tau \sigma^{-1}) \\
    & \quad \quad \quad \times \, \big(\hat{b}^\dag_{i_1} \dots \hat{b}^\dag_{i_n} \ketbra{0}_R \hat{b}_{i'_1} \dots \hat{b}_{i'_n}\big) \times \big(\hat{a}^\dag_{j_1} \dots \hat{a}^\dag_{j_n} \ketbra{0}_A \hat{a}_{j'_1} \dots \hat{a}_{j'_n} \big)
\end{split}
\end{align}
We can now use the fact that the set $\{\tau(1), \tau(2), \dots, \tau(n)\}$ is equal to the set $\{1, 2, \dots, n\}$ and the bosonic commutation relations $[\hat{a}^\dag_{i}, \hat{a}^{\dag}_j] = [\hat{a}_i, \hat{a}_j] = 0$ to show that $\hat{a}^\dag_{j_{\sigma(1)}} \dots \hat{a}^\dag_{j_{\sigma(n)}} \ket{0} = \hat{a}^\dag_{j_1} \dots \hat{a}^\dag_{j_n} \ket{0}$ (and similarly for the creation/annihilation operators on $R$). This allows us to drop all the Kronecker delta terms and the primed indices, giving
\begin{align}
\begin{split}
    \mathbb{E}_{\hat{V}}\big[ \hat{V} \ketbra{\psi_0} \hat{V}^\dag  \big] &= (1 - |z|^2)^K \sum_{n = 0}^\infty \frac{|z|^{2n}}{(n!)^2} \big(\hat{b}^\dag_{i_1} \dots \hat{b}^\dag_{i_n} \ketbra{0}_R \hat{b}_{i_1} \dots \hat{b}_{i_n}\big) \times \big(\hat{a}^\dag_{j_1} \dots \hat{a}^\dag_{j_n} \ketbra{0}_A \hat{a}_{j_1} \dots \hat{a}_{j_n} \big) 
    \sum_{\sigma, \tau \in \mathcal{S}_n} \textrm{Wg}(\tau \sigma^{-1})
\end{split}
\end{align}
The sums over the Weingarten function now just acts as an $n$-dependent coefficient. We can use the closure of the group $\mathcal{S}_n$, the fact that $|\mathcal{S}_n| = n!$ and the analytic expression for the sum over Weingarten functions given in Ref.~\cite{collins2014integration} to show
\begin{align}
\begin{split}
    \sum_{\sigma, \tau \in \mathcal{S}_n} \textrm{Wg}(\tau \sigma^{-1}) &= n! \sum_{\tau' \in \mathcal{S}_n} \textrm{Wg}(\tau') \\
    &= {n + N - 1 \choose n}^{-1}
\end{split}
\end{align}
This gives the average state as
\begin{equation}
    \mathbb{E}_{\hat{V}} \big[ \hat{V} \ketbra{\psi_0} \hat{V}^\dag \big] = (1 - |z|^2)^K \sum_{n = 0}^\infty \frac{1}{(n!)^2} |z|^{2n} {n + N - 1 \choose n}^{-1} \big(\hat{b}^\dag_{i_1} \dots \hat{b}^\dag_{i_n} \ketbra{0}_R \hat{b}_{i_1} \dots \hat{b}_{i_n}\big) \times \big(\hat{a}^\dag_{j_1} \dots \hat{a}^\dag_{j_n} \ketbra{0}_A \hat{a}_{j_1} \dots \hat{a}_{j_n} \big) 
\end{equation}
We now just need to calculate the action of the creation/annihilation operators on $R$ and $A$. One can use simple counting arguments to find the answer, but we instead choose to introduce the notation $\big(f(\bm{\hat{a}^\dag})\big) \tilde{\circ} \ketbra{0}$ for functions $f(\bm{a^\dag})$ that have positive powers of $\bm{a^\dag} = (\hat{a}_1^\dag, \dots, \hat{a}_N^\dag)$. This $\tilde{\circ}$ operator has the following defining features
\begin{align}
\begin{split}
\label{sup_mat_eq:tilde_circ_functional_definition}
    \hat{a}^\dag_k \tilde{\circ} \ketbra{0} &= \hat{a}_k^\dag \ketbra{0} \hat{a}_k \\
    (c \,\hat{a}^\dag_k) \tilde{\circ} \ketbra{0} &= c \, (\hat{a}_k^\dag) \tilde{\circ} \ketbra{0} \\
    (c \, \hat{a}_{k_1}^\dag \hat{a}^\dag_{k_2}) \tilde{\circ} \ketbra{0} &= c \, \hat{a}^\dag_{k_1} \hat{a}^\dag_{k_2} \ketbra{0} \hat{a}_{k_1} \hat{a}_{k_2} \\
    (c_1 \hat{a}_{k_1}^\dag + c_2 \hat{a}_{k_2}^\dag) \tilde{\circ} \ketbra{0} &= (c_1 \hat{a}_{k_1}^\dag) \tilde{\circ} \ketbra{0} +  (c_2 \hat{a}_{k_2}^\dag) \tilde{\circ} \ketbra{0}
\end{split}
\end{align}
This is similar to the notation $X \circ Y = X \,Y X^\dag$, but diagonal in the creation / annihilation operators.

Using this notation, we can rewrite the iterated sum over creation/annihilation operators on the vacuum as
\begin{align}
\begin{split}
    \hat{a}_{j_1}^\dag \dots \hat{a}_{j_n}^\dag \ketbra{0} \hat{a}_{j_1} \dots \hat{a}_{j_n} &= \hat{a}_{j_1}^\dag \dots \hat{a}_{j_{n-1}}^\dag \left(\sum_{k = 1}^N \hat{a}^\dag_k \ketbra{0} \hat{a}_k \right) \hat{a}_{j_1} \dots \hat{a}_{j_{n - 1}} \\
    &= \hat{a}_{j_1}^\dag \dots \hat{a}_{j_{n - 1}} \left( \Big( \sum_{k = 1}^N \hat{a}_k^\dag \Big) \tilde{\circ} \ketbra{0} \right) \hat{a}_{j_1} \dots \hat{a}_{j_{n-1}} \\
    &= \hat{a}^\dag_{j_1} \dots \hat{a}^\dag_{j_{n - 2}} \left( \Big( \sum_{k = 1}^N \hat{a}_k^\dag \Big)^2 \tilde{\circ} \ketbra{0} \right) \hat{a}_{j_1} \dots \hat{a}_{j_{n - 2}} \\
    &= \left(\sum_{k = 1}^N \hat{a}_k^\dag \right)^n \tilde{\circ} \ketbra{0}
\end{split}
\end{align}
From this, we can use the multinomial theorem to expand the power
\begin{align}
\begin{split}
    \left( \sum_{k = 1}^N \hat{a}_k^\dag\right)^n \tilde{\circ} \ketbra{0} &= \sum_{\bm{x} \in \bm{P}(n, N)} {n \choose x_1, \dots, x_N} \left(\hat{a}_1^{\dag x_1} \dots \hat{a}_N^{\dag x_N}  \right) \tilde{\circ} \ketbra{0} \\
    &= n! \sum_{\bm{x} \in \bm{P}(n, N)} \ketbra{\bm{x}}
\end{split}
\end{align}
where $\bm{x} = (x_1, x_2, \dots, x_N)$ is a vector of photon numbers and $\bm{P}(n, N)$ is the set of ways of distributing $n$ indistinguishable items among $N$ bins, i.e, $\bm{P}(n, N) = \{\bm{x} : x_1, \dots, x_N \ge 0 , \, \sum_{k = 1}^N x_k = n\}$. Note that in the second line, we used $\hat{a}_k^{\dag x_k} \tilde{\circ} \ketbra{0} = x_k! \ketbra{x_k}$.
The full expression is proportional to the $n$ photon subspace on $N$ modes, giving a final expression for the sums over the creation/annihilation operators
\begin{equation}
    \hat{a}^\dag_{j_1} \dots \hat{a}^\dag_{j_n} \ketbra{0}_A \hat{a}_{j_1} \dots \hat{a}_{j_n} = n! \, \Pi_N(n), \quad \quad \hat{b}^\dag_{i_1} \dots \hat{b}^\dag_{i_n} \ketbra{0}_R \hat{b}_{i_1} \dots \hat{b}_{i_n} = n! \, \Pi_{K}(n)
\end{equation}
Bringing everything together, we derive that the average state is equal to 
\begin{equation}
    \mathbb{E}_{\hat{V}} \big[ \hat{V} \ketbra{\psi_0} \hat{V}^\dag \big] = (1 - |z|^2)^K \sum_{n = 0}^\infty |z|^{2n} {n + N - 1 \choose n}^{-1} \Pi_{K}(n)_R \otimes \Pi_{N}(n)_A
\end{equation}
The component of the state ${\tilde{n} + N - 1 \choose n}^{-1} \Pi_{N}(n)$ is the normalised, maximally mixed state on the $n$ photon subspace and has unit trace, and hence the spectrum is the same as in the initial state.
Intuitively, in each energy subspace (i.e, $n$ photon subspace) the averaging process transforms any state into the maximally mixed state. And because the PLON is passive, this averaging process takes place in each energy subspace individually. This leads to residual classical correlations that prevent an exact typical subspace forming from the average state on $A$.

\subsection{Reduced states}
The reduced state on $R$ correctly recovers a $K$-fold tensor product of single-mode thermal states, $\rho_z^{\otimes K} = (1 - |z|^2)^K \sum_{n = 0}^\infty |z|^{2n} \Pi_K(n)$. This follows from $\textrm{tr}[\Pi_N(n)] = {n + N - 1 \choose n}$. The reduced state on $A$, denoted $\bar{\rho}_A$, is equal to
\begin{equation}
    \bar{\rho}_A = (1 - |z|^2)^K \sum_{n = 0}^\infty |z|^{2n} \frac{{n + K - 1 \choose n}}{{n + N - 1 \choose n}} \Pi_N(n)
\end{equation}
We can rewrite this into a form that resembles a product state but with a kernel that introduces classical correlations between modes; let $f(n) = (1 - |z|^2)^K |z|^{2n} \frac{{n + K - 1 \choose n}}{{n + N - 1 \choose n}}$.
\begin{align}
\begin{split}
    \bar{\rho}_A &= \sum_{n = 0}^\infty f(n) \Pi_N(n) \\
    &= \sum_{n = 0}^\infty \sum_{m_1 = 0}^n f(n) \ketbra{m_1} \otimes \Pi_{N-1}(n - m) \\
    &= \sum_{n = 0}^\infty \sum_{m_1 = 0}^\infty f(n + m_1) \ketbra{m_1} \otimes \Pi_{N - 1}(n) \\
    &= \sum_{n = 0}^\infty \sum_{m_1 = 0}^\infty \sum_{m_2 = 0}^n f(n + m_1) \ketbra{m_1, m_2} \otimes \Pi_{N - 2}(n - m_2) \\
    &= \dots \\
    &= \sum_{m_1, m_2, \dots, m_N = 0}^\infty \left(\sum_{n = 0}^\infty f(n + m_1 + \dots + m_N) \right) \ketbra{m_1, m_2, \dots, m_N}
\end{split}
\end{align}
where we iteratively decomposed $\Pi_X(n) = \sum_{m = 0}^n \ketbra{m} \otimes \Pi_{X-1}(n-m)$ and applied the identity $\sum_{n = 0}^\infty \sum_{m = 0}^n f(n, m) = \sum_{n, m = 0}^\infty f(n + m, m)$. This resembles a product state $\sigma^{\otimes N}$ for $\sigma = \sum_{m = 0}^\infty p_n \ketbra{n}$, but with a kernel that leads to classical correlations between modes.

If $K = 1$, then $f(n + m_{\textrm{tot}}) = (1 - |z|^2) |z|^{2(n + m_{\textrm{tot}})} {n + m_{\textrm{tot}} + N - 1 \choose n + m_{\textrm{tot}}}^{-1}$ and the sum over $n$ can be given in terms of the Gaussian hypergeometric functions, $_2F_1(a, b; c; x) = \sum_{n = 0}^\infty \frac{(a)_n (b)_n}{(c)_n} \frac{x^n}{n!}$ where $(a)_n = \Gamma(a + n) / \Gamma(a)$ is the Pochhammer symbol;
\begin{equation}
    \sum_{n = 0}^\infty f(n + m_{\textrm{tot}}) = (1 - |z|^2) |z|^{2 m_{\textrm{tot}}} \frac{_2F_1(1, m_{\textrm{tot}}; m_{\textrm{tot}} + N; |z|^2)}{{m_{\textrm{tot}} + N - 1 \choose m_{\textrm{tot}} }}
\end{equation}
This gives the average state on $A$ for $K = 1$ the form
\begin{equation}
    \bar{\rho}_A = \sum_{\bm{m} \in \mathbb{N}_0^N} q(\bm{m}) \ketbra{\bm{m}}, \quad \quad q(\bm{m}) = (1 - |z|^2) |z|^{2 m_{\textrm{tot}}} \frac{_2F_1(1, m_{\textrm{tot}}; m_{\textrm{tot}} + N; |z|^2)}{{m_{\textrm{tot}} + N - 1 \choose m_{\textrm{tot}}}}
\end{equation}
for $\bm{m} = (m_1, m_2, \dots, m_N)$ and $m_{\textrm{tot}} = m_1 + \dots + m_N$.

In the $K = N$ case, the initial state on $A$ is an $N$-fold tensor product of thermal states, $\rho_z^{\otimes N}$. The resulting average state on $A$ is the exact same tensor product of thermal states, and hence this state is an eigenstate of the ``average PLON'' operator. So while it would be true that a typical subspace is formed in this case, there already existed the same typical subspace on the initial state so the passive linear optical network would be unnecessary.
\\
\\
If we consider $M$ modes within $A$ (in a Hilbert space denoted $A_M$), then we can use the identity $\Pi_N(n) = \sum_{m = 0}^n \Pi_{M}(m) \otimes \Pi_{N-M}(n-m)$ to derive the reduced state $\bar{\rho}_{A_M}$,
\begin{equation}
    \bar{\rho}_{A_M} = (1 - |z|^2)^K \sum_{n = 0}^\infty \sum_{m = 0}^\infty  |z|^{2n} \frac{{n + K - 1 \choose n} {n - m + N - M - 1 \choose n - m}}{{n + N - 1 \choose n}} \Pi_{M}(m)
\end{equation}
And the single-mode average state, denoted $\bar{\rho}_1$, can be written as
\begin{equation}
    \bar{\rho}_{1} = \sum_{n = 0}^\infty p(n) \ketbra{n}, \quad \quad p(n) = (1 - |z|^2)^K |z|^{2n} \sum_{m = 0}^\infty |z|^{2m} \frac{{m + N - 2 \choose m} {n + m + K - 1 \choose n + m}}{{n + m + N - 1 \choose n + m}} 
\end{equation}
In the $K = 1$ case, the coefficient $c(n)$ can again be given in terms of the Gaussian hypergeometric function $_2F_1(a, b; c; x)$;
\begin{align}
\begin{split}
    \label{sup_mat_eq:single_mode_average_state_prob_dist}
    p(n) &= \frac{(1 - |z|^2) |z|^{2n}} {{n + N - 1 \choose n}}  \,_2F_1(n + 1, N - 1; n + N; |z|^2)
\end{split}
\end{align}
If we take the $K=1$ case for the random code in Ref.~\cite{Zhong2023information}, arbitrary entanglement-transmission rates can be achieved by increasing the squeezing of the TMSV state as a function of the total number of modes. Therefore, the $K=1$ case with arbitrary squeezing generically captures the key features of their result and we can focus attention on this simpler case.

\begin{figure}
    \centering
    \includegraphics[width=0.85\linewidth]{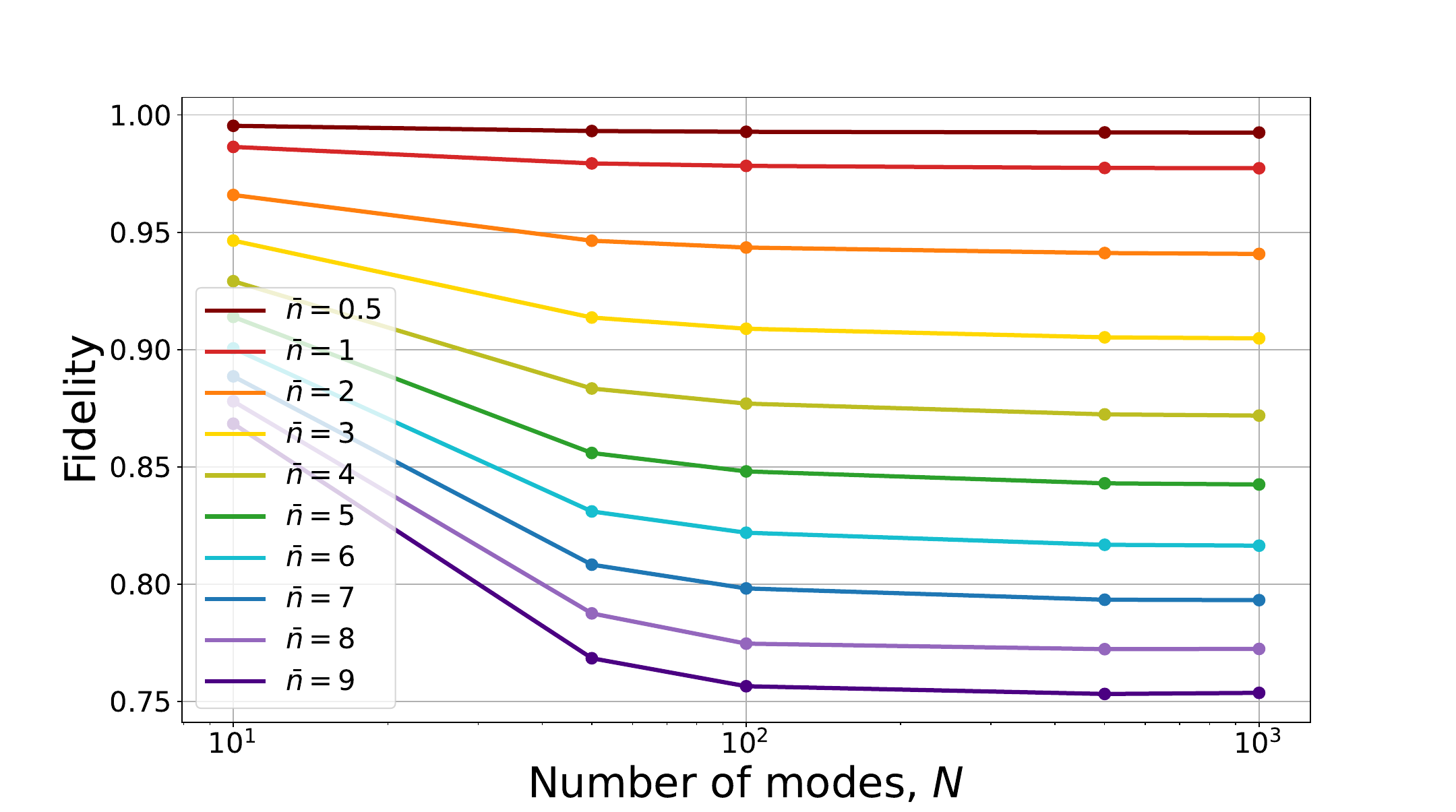}
    \caption{Fidelity between $\bar{\rho}_A$ and the ansatz tensor product state $\bar{\rho}_1^{\, \otimes N}$ for $K = 1$ and expected photon number $\bar{n}$ (i.e, Eq.~\eqref{sup_mat_eq:fidelity_to_ansatz_state_1}). As $\bar{n}$ increases, the fidelity decreases almost (but not quite) linearly. As the number of modes increases, the fidelity initially decreases before plateauing beyond $N \approx 100$. In the high-energy, large $N$ limit, this suggests that the fidelity becomes very low. We considered only up to $\bar{n} = 10$ because 1) the number of terms grows exponentially with the numerical cutoff, which itself grows with $\bar{n}$ and 2) the native hypergeometric function $_2F_1(a, b; c; |z|^2)$ in the Python package scipy.special returns $\infty$ when $\bar{n}$ and $N$ become too large.
    }
    \label{sup_mat_fig:fidelity_tensor_product_ansatz}
\end{figure}

\subsection{Implication on the product state ansatz}
We have proven that the product state ansatz in Eq.~\eqref{sup_mat_eq:ansatz_average_state} is not exactly right, but perhaps it holds approximately? To test this conjecture, we can calculate the fidelity between $\bar{\rho}_A$ and the tensor product of single-mode average states $\bar{\rho}_1^{\,\otimes N}$ for the $K=1$ case;
\begin{align}
\begin{split}
\label{sup_mat_eq:fidelity_to_ansatz_state_1}
    \textrm{Fid}[\bar{\rho}_A, \bar{\rho}_1^{\,\otimes N}] = \sum_{n = 0}^\infty \sum_{\bm{x} \in \bm{P}(n, N)} \sqrt{f(n) p(x_1) p(x_2) \dots p(x_N)}
\end{split}
\end{align}
where $f(n) = (1 - |z|^2) |z|^{2n} {n + N - 1 \choose n}^{-1}$ and $p(n)$ is given in Eq.~\eqref{sup_mat_eq:single_mode_average_state_prob_dist}.
There is a lot of degeneracy in the summands of these representations. This can be removed by noting that that summands with indices $\bm{x} = (x_1, \dots, x_N)$ and $\bm{x'} = (x_1', \dots, x_N')$ are identical if they are equivalent up to permutations. We can therefore decompose the sum over $\bm{x}$ into a sum over all partitions of the associated $n$ with length $\le N$. We denote $Q\vdash n$ as an integer partition of $n$, which we write as $Q = (Q_1^{m_1}, Q_2^{m_2}, \dots, Q_{|{Q}|}^{m_{|Q|}})$ where $Q_i$ is the $i^{\textrm{th}}$ unique integer of the partition and the superscript $m_i$ is the number of times that integer appears in the partition. The number of unique integers in $Q$ is denoted $|Q|$. For example, $Q = (4^1, 3^2, 1^3) = (4 + 3 + 3 + 1 + 1 + 1) \vdash 13$. With this, the fidelity is given by
\begin{equation}
\label{sup_mat_eq:fidelity_to_ansatz_state_2}
    \textrm{Fid}[\bar{\rho}_A, \bar{\rho}_1^{\,\otimes N}] = \sum_{n = 0}^\infty \sum_{\substack{Q \vdash n, \\
    \sum_{i = 1}^{|Q|} m_{Q_i} \le N}} d(Q) \sqrt{f(n) p(0)^{m_0} p(Q_1)^{m_1} p(Q_2)^{m_2} \dots p(Q_{|Q|})^{m_{|Q|}} }
\end{equation}
where $m_0 = N - \sum_{i = 1}^{|Q|} m_i$.
If applicable, the $p(0)^{m_0}$ term accounts for the modes that have index 0 (i.e, no photons) which are not included in the integer partition of $n$. The degeneracy factor $d(Q)$ is the number of permutationally equivalent ways the term in the square root appears in the sum - it is equal to the multinomial coefficient $d(Q) = {N \choose m_0, m_1, m_2, \, \dots,\, m_{|Q|}}$.

When numerically calculating the capacity, a finite cutoff $n_{\textrm{cutoff}}$ is chosen. We can only study the low-to-intermediate energy regime because the number of terms increases exponentially with $n_{\textrm{cutoff}}$, although using high powered computing and/or focusing only on the partitions that represent the majority of $\bm{P}(n, N)$ (e.g, the partitions centered around $\bm{x} = (n/N, n/N, \dots, n/N)$ or $\bm{x} = (1, 1, \dots, 1, 0, \dots, 0)$ depending on if $n/N >1$ or $n/N<1$) would allow the fidelity to be calculated for significantly higher energies.

In Fig.\ref{sup_mat_fig:fidelity_tensor_product_ansatz}, we plot the fidelity $\textrm{Fid}[\bar{\rho}_A, \bar{\rho}_1^{\,\otimes N}]$ for a single TMSV state input with expected photon numbers from $\bar{n}=0.5$ to $\bar{n} = 9$ with between $N = 50$ and $N = 1000$ modes. We choose a cutoff such that our fidelity calculation includes at least $99.9\%$ of the state. In the figure, we see two clear phenomena; first, we see that increasing $N$ leads to a decrease in the fidelity until it plateaus beyond $N \approx 100$ modes. Second, the fidelity decreases with respect to $\bar{n}$ at an almost linear rate, dropping from $98\%$ at $\bar{n} = 1$ to just $75\%$ at only $\bar{n} = 9$. These two results suggest that as we continue increasing the energy, the fidelity will continue to worsen. Since the effective state space increases with energy, we would conjecture that the fidelity monotonically decreases as $\bar{n}$ grows. However, it is not clear whether the fidelity tends to zero or some non-zero constant in the very high energy limit.

If the random coding scheme from Ref.~\cite{Zhong2023information} wants a non-negligible rate, then for $K=1$ it requires that $H_{\textrm{therm}}(\bar{n}) \sim Q_{\textrm{rate}} N$. To achieve a rate $Q_{\textrm{rate}}$, the expected photon number of the TMSV state must scale exponentially as $\bar{n} \sim 2^{Q_{\textrm{rate}}N}$, and so it is the high energy limit where they would want the tensor product ansatz to be most accurate. However, we have presented evidence above that in this regime their ansatz likely performs the worst. Therefore, we argue that the average state $\bar{\rho}_A$ does not form the tensor product structure used to justify the typical subspace expansion, a result that undermines components of the derivation in Ref.~\cite{Zhong2023information}. 
While there is still a chance that the state $\bar{\rho}_A$ could have a significant overlap with the typical subspace induced by $\bar{\rho}_1^{\, \otimes N}$, at this time we are not aware of any theoretical justification for such a claim.